\documentclass[preprint,12pt]{elsarticle}

\usepackage{lineno,hyperref}
\modulolinenumbers[5]
\usepackage[margin=1.0in]{geometry}

\journal{}

\usepackage{amsmath}
\usepackage{color}
\usepackage{graphicx}
\usepackage{multirow}
\usepackage{siunitx}
\usepackage{xcolor}
\usepackage{empheq}
\usepackage{tabularx}
\usepackage{subcaption}
\usepackage{longtable}
\usepackage{makecell}

\graphicspath{ {./images/} }

\usepackage{latexsym,amsfonts,amscd,amssymb,amsthm}
\usepackage{graphics}
\usepackage{caption}
\usepackage{epstopdf}
\usepackage{changebar}
\usepackage{indentfirst}
\usepackage{verbatim}
\usepackage{array}
\numberwithin{equation}{section}

\newcommand{\revised}[1]{\textcolor{black}{#1}}

\begin{document} 

\begin{frontmatter}


\title{Modeling blood flow in networks of viscoelastic vessels\\ with the \revised{1-D} augmented fluid–structure interaction system}

\author[mymainaddress]{Francesco Piccioli\corref{mycorrespondingauthor}}
\cortext[mycorrespondingauthor]{Corresponding author}
\ead{francesco.piccioli@unife.it}
\author[Giuliaaddress,Giuliamainaddress]{Giulia Bertaglia}
\author[mymainaddress]{Alessandro Valiani}
\author[mymainaddress]{Valerio Caleffi}
\address[mymainaddress]{Department of Engineering, University of Ferrara, Via G. Saragat 1, 44122 Ferrara, Italy}
\address[Giuliaaddress]{Istituto Nazionale di Alta Matematica “Francesco Severi", P.le Aldo Moro 5, 00185 Roma, Italy}
\address[Giuliamainaddress]{Department of Mathematics and Computer Science, University of Ferrara, Via Machiavelli 30, 44121 Ferrara, Italy}

\begin{abstract}
A noteworthy aspect in blood flow modeling is the definition of the mechanical interaction between the fluid flow and the biological structure that contains it, namely the vessel wall. Particularly, it has been demonstrated that the addition of a viscous contribution to the mechanical characterization of vessels brings positive results when compared to \textit{in-vivo} measurements. 
In this context, the \revised{numerical} implementation of boundary conditions able to keep memory of the viscoelastic contribution of vessel walls assumes an important role, especially when dealing with large circulatory systems. In this work, viscoelasticity is taken into account in entire networks via the Standard Linear Solid Model. The implementation of the viscoelastic contribution at boundaries (inlet, outlet and junctions), is carried out considering the hyperbolic nature of the mathematical model. Specifically, a non-linear system is established based on the definition of the Riemann Problem at junctions, characterized by rarefaction waves separated by contact discontinuities, among which the mass and the total energy are conserved. 
Basic junction tests are analyzed, such as a trivial 2--vessels junction, for both a generic artery and a generic vein, and a simple 3--vessels junction, considering an aortic bifurcation scenario. The chosen \revised{asymptotic-preserving} IMEX Runge-Kutta Finite Volume scheme is demonstrated to be second-order accurate in the whole domain and well-balanced, even when including junctions. 
\revised{Two different benchmark models of the arterial network are then implemented}, differing in number of vessels and in viscoelastic parameters. Comparison of the results obtained in the two networks underlines the high sensitivity of the model to the chosen viscoelastic parameters. In these numerical tests, \revised{the conservation of the contribution provided by the viscoelastic characterization of vessel walls is assessed in the whole network, including junctions and boundary conditions}. 
\end{abstract}


\end{frontmatter}

\section{Introduction}
\label{sect:introduction}
Robust and efficient numerical schemes able to solve fluid mechanics balance laws are at the basis of the analysis of wave propagation phenomena in human cardiovascular networks. One-dimensional (1-D) models have been broadly used to study how blood pressure and flow rate affects the biological structure that contains the fluid, namely arterial and venous walls \cite{Quarteroni2003,Mynard2014b,Quarteroni2016,Reymond2009a}. In fact, it is commonly acknowledged that the use of computational models provides a great assistance to the clinical research, having the possibility to supply data that, otherwise, would require invasive techniques \cite{Ambrosi2012,Formaggia2009,Willemet2016}. Furthermore, computational hemodynamics represents a valuable tool to foresee possible occurrences of diseases and the development of pathologies related to the arterial system, as in the case of arterial hypertension or atherosclerosis \cite{Alastruey2007,Alastruey2012,Alastruey2012b,Liang2011,Liang2018,Muller2019}, or to the venous system \cite{Muller2015b,Toro2016}. 

Thus, to properly and accurately investigate the propagation of blood in circulatory systems, mathematical models need to consider that blood mechanically interacts with vessel walls and tissues. Vessel walls are indeed deeply affected by internal pressure, undergoing strong deformations, even collapsing in the case of veins under specific circumstances \cite{Murillo2019,Toro2013,Spiller2017}. The fluid-structure interaction (FSI) in blood propagation phenomena
requires the introduction of a constitutive law which defines the transfer of energy between \revised{blood and vessel wall} \cite{Bertaglia2020,Fung1997,Leguy2019,Shapiro1977}. 
Mechanically speaking, rheological properties of arteries and veins are well-described by viscoelastic typical features \cite{Armentano1995,Valdez2009,Zocalo2008}. Although neglected in some works in favor of simpler elastic models \cite{Boileau2015,Murillo2019,Siviglia2012,Xiao2014}, viscoelasticity adds a valuable contribution to the representation of the problem, being able to capture damping effects related to a partial loss of energy occurring during the deformation of the vessel \cite{Alastruey2011a,Alastruey2012,Bertaglia2020,Bertaglia2020a,Blanco2020,Blanco2014,Muller2016,Reymond2009a,Valdez2009}. 
This fundamental feature is also observable from hysteresis loops obtained by means of stress and strain measurements in different vessels \cite{Bertaglia2020a,Raghu2011}. 

In this work, the viscoelasticity of blood vessels is introduced in the mathematical model recurring to the Standard Linear Solid Model (SLSM), consisting of a spring in series with a Kelvin-Voigt (KV) unit, the latter composed by a spring and a dash-pot in parallel \cite{Bertaglia2020,Bertaglia2018,Lemini2014,Lakes2009}. Most of the models dealing with viscous effects of vessels make use only of the KV unit \cite{Alastruey2011a,Ghigo2017,Liang2018,Montecinos2014,Mynard2015,Wang2013}. The employment of the more complex 
SLSM allows a better description of the real mechanical behavior of biological tissues \revised{because, in contrast with the KV model, it permits to include} the definition of the exponential relaxation of the stress (pressure) over time \revised{\cite{Westerhof1970,Westerhof2019}}. This \revised{ability of the model is of great importance since stress relaxation is} one of the characteristic features of viscoelastic materials, besides creep and hysteresis \cite{Gurtin1961,Lakes2009}. 

A viscoelastic constitutive tube law, relating pressure to cross-sectional area, is defined through the SLSM and added in partial differential equation (PDE) form to the main system of governing equations, composed by the equations of conservation of mass and momentum, giving rise to the so-called augmented fluid-structure interaction (a-FSI) system \cite{Bertaglia2020,Bertaglia2020a}. The a-FSI blood flow model, solved through an Asymptotic-Preserving (AP) IMEX Runge-Kutta scheme in time and a Finite Volume (FV) method in space, has already been validated in previous single--vessel studies, demonstrating its capability to correctly simulate flow rate and pressure trends in patient-specific simulations \cite{Bertaglia2020,Bertaglia2020a} and also assessing its sensitivity to uncertainties inherent the viscoelastic parameters involved \cite{Bertaglia2021}.

This work aims at further developing the recently proposed a-FSI blood flow model, investigating the treatment of networks of viscoelastic vessels and thus \revised{the junction numerical modeling}. Different approaches have been discussed in literature to deal with this topic: in \cite{Fullana2009a} the junction is conceived as an elastic tank; in \cite{Mynard2015} the junction is modeled via a control volume analysis; in \cite{Alastruey2012b,Reymond2009a,Sherwin2003a} the \revised{implementation of the model representing the junction} relies on the Method of Characteristics (MOC). However, none of these works take into account the viscoelastic contribution within the \revised{numerical} implementation of the node, neglecting it in favor of a local elastic approach. To the authors' knowledge, only in \cite{Muller2016} a \revised{numerical} implementation of viscoelastic bifurcations is proposed, via the solution of a Riemann Problem (RP), but modeling viscoelasticity through the simple KV unit. The innovative contribution of this work lays in a detailed treatment of viscoelastic \revised{boundary conditions (BCs)}, accounting for the viscoelastic rheology of vessels through the more appropriate SLSM,
presenting an original methodology to be computed at the boundaries of the 1-D physical \revised{domain. Inspired by the work in \cite{Muller2016}, the proposed methodology, specifically developed for junctions, represents a feasible tool also for inlet and outlet boundaries and naturally derives from the hyperbolic eigenstructure of the a-FSI system}. 

The rest of the manuscript is structured as follows: in Section \ref{sect:mathematical_model} the mathematical model is presented, hence the a-FSI system is defined and the hyperbolic eigenstructure is analyzed. In Section \ref{sect:numerical_model} the numerical model is discussed and the \revised{numerical treatment} of \revised{BCs}  -- at inlet, outlet and junctions -- is presented in detail for both the elastic and the viscoelastic cases. The procedure here proposed to implement junctions is valid for both bifurcation and confluence cases. \revised{Bifurcations are mostly related to arteries, whereas confluences are mostly related to veins}. 
In Section \ref{sect:results} various numerical results are shown and discussed. At first, the trivial 2--vessels junction test is presented for arteries and veins, in both elastic and viscoelastic regimes. In these tests, accuracy analysis of the model is performed as well as well-balancing analysis. Then, a 3--vessels junction problem is analyzed concerning a benchmark elastic aortic bifurcation case and performing the same test also with a viscoelastic configuration of the vessels.  
At last, results concerning the simulation of two benchmark arterial networks are presented, again for both the mechanical characterizations of vessel walls. 
Both the arterial networks deal with arterial vessels with diameters typical of the human macro-circulation. Blood flow within these vessels is characterized by high Womersley numbers indicating that the respective flow is pulsating \cite{Fung1997}, strengthening the choice of using the sophisticated a-FSI model to obtain realistic results.
The first network replicates an \textit{in-vitro} model of arterial tree representing the largest central systemic arteries of the human vascular system, whereas the second one is a reduced version of an anatomically detailed arterial network. 
Final conclusions are drawn in Section \ref{sect:conclusions}.

\section{Mathematical model}
\label{sect:mathematical_model}

The mathematical model here discussed recalls the blood flow model presented in \cite{Bertaglia2021,Bertaglia2020,Bertaglia2020a}, to which the reader is referred for further details.

\subsection{1-D a-FSI blood flow model}
\label{sect:a-FSI}

The well established system of governing equations defining blood flow circulation in compliant vessels is composed of the standard equations of balance of mass and momentum of the fluid mechanics. The 1-D formulation of the model is obtained averaging the incompressible 3-D Navier-Stokes equations over the vessel cross-section, under the assumption of axial symmetry of the geometry and of the flow \cite{Formaggia2009}, and hypothesizing large wavelengths with respect to the radius of the vessels \cite{Wang2013}. To close the resulting system, a constitutive relationship relating the cross-sectional area to the internal pressure is needed, which is the so-called tube law. This third equation can be expressed both in elastic and viscoelastic formulations: the elastic formulation consists in a first reasonable approximation of the mechanical behavior of vessels \cite{Muller2014a,Muller2014,Mynard2008,Sherwin2003,Toro2013,Willemet2014}, whereas the viscoelastic one allows for a more accurate characterization of the FSI occurring between blood flow and vessel walls, considering damping effects \cite{Alastruey2012b,Alastruey2012,Bertaglia2021,Bertaglia2020,Bertaglia2020a,Mynard2015,Raghu2011}. In the proposed model, viscoelasticity is taken into account recurring to the SLSM, which is the simplest viscoelastic model able to exhibit all the three primary features of a viscoelastic material: creep, stress relaxation and hysteresis. Finally, to ensure a formally correct numerical treatment of possible longitudinal discontinuities of vessels properties, either geometrical or mechanical -- i.e.\ equilibrium cross-sectional area $A_{0}(x)$, instantaneous Young modulus $E_{0}(x)$ and external pressure $p_{ext}(x)$ -- three additional closure equations are added to the system, imposing these quantities to be constant in time \cite{Bertaglia2020,Castro2008,Muller2013,Toro2013}. Thus, the a-FSI system is obtained:
\begin{subequations} {\label{a-FSI}}
\begin{eqnarray}
\label{eq:mass}
\partial_{t} A +\partial_{x} (Au) = 0, \\ 
\label{eq:momentum}
\partial_{t} (Au) + \partial_{x} (Au^{2}) + \frac{A}{\rho} \partial_{x} p = \frac{f}{\rho}, \\ 
\label{eq:tubelaw}
\partial _{t} p + d\, \partial _{x} (Au) = S, \\ 
\label{eq:A0}
\partial_{t}A_{0} = 0, \\ 
\label{eq:E0}
\partial_{t}E_{0} = 0, \\
\label{eq:pe} 
\partial_{t}p_{ext} = 0.
\end{eqnarray}
\end{subequations}
Here $A(x,t)$ is the cross-sectional area of the vessel, $u(x,t)$ is the cross-sectional averaged blood velocity, $p(x,t)$ is the internal blood pressure, $\rho$ is the blood density, and $x$ and $t$ are space and time, respectively. 

In \eqref{eq:momentum} $f$ represents the friction loss term. \revised{The velocity profile is assumed self-similar and axisymmetric, and $f$ is defined as follows} \cite{Bertaglia2020a,Quarteroni2016,Xiao2014}:
\begin{equation}
f = -2(\zeta+2)\mu\pi u,
\end{equation}
where $\mu$ is dynamic blood viscosity and $\zeta = (2-\alpha_c) / (\alpha_c-1)$ is a parameter depending on $\alpha_c$, whose definition relies on the velocity profile. \revised{Generally}, the velocity profile is assumed to be close to flat ($\alpha_c \approx 1.0$) for the computation of the friction loss term. 

In \eqref{eq:tubelaw} the parameter $d(x,t)$ accounts for the elastic contribution of the vessel wall 
\cite{Bertaglia2021,Bertaglia2020,Bertaglia2020a}:
\begin{equation}
\label{eq:d_el}
d = \frac{K}{A}(m\alpha^{m} - n\alpha^{n}),
\end{equation}
where $\alpha = A/A_{0}$ is the dimensionless cross-sectional area, while \revised{the wall stiffness coefficient $K(x)$ and parameters} $m$ and $n$ account for the specific mechanical properties of the material (therefore depending on the vessel type). The wave speed $c(x,t)$ results \cite{Bertaglia2020,Bertaglia2020a}:
\begin{equation}
\label{eq:celerity}
c = \sqrt{\frac{A\,d}{\rho}} .
\end{equation}
For a discussion on the choice of parameters $K$, $m$, and $n$, for both arteries and veins, the reader is referred to \cite{Bertaglia2020a,Muller2014}.
In the same equation, the source term $S(x,t)$ accounts for the viscous contribution of the vessel wall behavior and reads:
\begin{equation}
\label{eq:sourceterm}
S = \frac{1}{\tau_{r}} \left[ \frac{E_{\infty}}{E_{0}} \, K(\alpha^{m} - \alpha^{n}) - (p - p_{ext}) \right].
\end{equation}\\
Eq.~\eqref{eq:tubelaw}, together with \eqref{eq:sourceterm}, directly derives from the constitutive law of the chosen SLSM, which is indeed characterized by three main parameters: the instantaneous Young modulus $E_0(x)$, the asymptotic Young modulus $E_{\infty}(x)$ and the relaxation time $\tau_r(x)$. The interplay between these three parameters is defined through the following equation \cite{Bertaglia2020,Bertaglia2020a}:
\begin{equation}
\label{eq:SLSMclosingeq}
\tau_r = \eta \, \frac{E_0 - E_{\infty}}{E_0^2},
\end{equation} 
where $\eta(x)$ is the viscosity coefficient of the SLSM.

Finally, it is possible to write the non-linear non-conservative system \eqref{a-FSI} in the general compact form:
\begin{equation}
\label{hypsys1}
\partial_{t} \boldsymbol{Q} + \partial_{x} \boldsymbol{f}(\boldsymbol{Q}) + \boldsymbol{B}(\boldsymbol{Q}) \partial_{x} \boldsymbol{Q}=\boldsymbol{S}(\boldsymbol{Q}),
\end{equation}
and, furthermore, in the following quasi-linear form:
\begin{equation}
\label{hypsys2}
\partial_{t} \boldsymbol{Q} + \boldsymbol{A}(\boldsymbol{Q}) \partial_{x} \boldsymbol{Q}=\boldsymbol{S}(\boldsymbol{Q}),
\end{equation}
with $\boldsymbol{A}(\boldsymbol{Q}) = \partial\boldsymbol{f} / \partial\boldsymbol{Q} + \boldsymbol{B}(\boldsymbol{Q})$. Here, $\boldsymbol{Q}$ is the vector of the state variables, $\boldsymbol{f}(\boldsymbol{Q})$ represents the vector of the analytical fluxes related to the conservative part of the system, while $\boldsymbol{B}(\boldsymbol{Q})$ identifies the non-conservative matrix of the problem, and $\boldsymbol{S}(\boldsymbol{Q})$ is the vector of the source terms. For further details the reader is invited to refer to \ref{appendix}.

\subsubsection{Asymptotic limits}
\label{sect:asymptotic_lim}

The two Young moduli, $E_0$ and $E_{\infty}$, are associated to the initial and final phases of the material's deformation, respectively. In a simple relaxation experiment, the law defining how the apparent Young modulus $E(t)$ of the material gradually changes in time, from the instantaneous value to the asymptotic one, is the so-called \textit{relaxation function} \cite{Lakes2009}:
\begin{equation}
\label{eq:relaxationeq}
E(t) = E_0 \, \exp\left(-\frac{t}{\tau_r}\right)+E_{\infty}\left[ 1- \exp\left(-\frac{t}{\tau_r}\right) \right].
\end{equation}
By definition, when dealing with a simple elastic behavior, the Young modulus is constant in time  \cite{Lakes2009}: $E(t) = E_{\infty} = E_0$. In fact, analyzing the asymptotic
limit of Eq.~\eqref{eq:relaxationeq}, if $\tau_r \rightarrow 0 \Rightarrow E(t) \to E_{\infty}$, thus leading to an elastic mechanism. It is important to notice that in this limit system \eqref{a-FSI} becomes \textit{stiff}. On the other hand, also an excessively long relaxation time, $\tau_r \to \infty$, does not allow the development of viscoelastic features, resulting $E(t) \to E_0$. In both cases, the source term of Eq.~\eqref{eq:tubelaw}, defined in Eq.~\eqref{eq:sourceterm}, tends to zero, $S \to 0$, which confirms that the formulation of $S$ is coherent with the assumed mechanical behavior and consistent with the equilibrium limit \cite{Bertaglia2021,Bertaglia2020,Bertaglia2020a}. 
From this analysis, we can deduce that the more $E_{\infty} \rightarrow E_0$, the more the behavior of the material becomes purely elastic. Therefore, in the SLSM the ratio $z=E_{\infty}/E_0$ can be seen as an index of the viscoelasticity of the material. In the a-FSI model here discussed, this ratio \revised{is computed by an} empirical formula presented in \cite{Bertaglia2020a}, which is the result of a calibration process carried out considering \revised{various hysteresis loops taken from literature \cite{Giannattasio2008,Salvi2012}:
\begin{equation}
\label{eq:Einf2E0}
\frac{E_{\infty}}{E_0} = e^{-1.3\cdot 10^{-5}\eta}.
\end{equation}}
\subsubsection{Riemann Invariants}
\label{sect:eigenstructure}
\revised{As discussed in \cite{Bertaglia2020}, system \eqref{a-FSI} is hyperbolic and, in contrast with other approaches proposed in literature (e.g. \cite{Alastruey2011a,Montecinos2014}), this mathematical model preserves its hyperbolicity even when considering the viscoelastic contribution of vessel walls, without the need for any mathematical reformulation. Nevertheless, we highlight here that system \eqref{a-FSI} is not \emph{strictly} hyperbolic. In this work, we}
evaluate the nature of its characteristic fields, as well as the quantities that are kept constant across these fields, namely the Riemann Invariants (RIs) \cite{Toro2009}, \revised{considering only sub-critical flow cases}.

The study of the eigenstructure of the a-FSI system and the detailed derivation of its RIs is presented in \ref{appendix}. From the analysis, system \eqref{a-FSI} results characterized by two genuinely non-linear fields, associated with either shocks or rarefactions, and by four linearly degenerate (LD) fields, associated with contact discontinuity waves. 
The RIs associated with the LD fields are:
\begin{equation}
\label{eq:RILD}
\Gamma_{1}^{LD} = Au, \qquad \Gamma_{2}^{LD} = p + \frac{1}{2}\rho u^{2}.
\end{equation}
On the other hand, RIs associated with the genuinely non-linear fields read:
\begin{equation}
\label{eq:RIgenNL}
\Gamma_{1} = u+\int \frac{c(A)}{A}\mathrm{d}A, \quad \Gamma_{2} = u-\int \frac{c(A)}{A}\mathrm{d}A, \quad \Gamma_{3} = p-\int d(A)\mathrm{d}A.
\end{equation}
It is worth noticing that the integral in $\Gamma_{3}$ can be analytically computed, resulting:
\begin{equation}
\Gamma_{3} = p - K \left( \alpha^{m} - \alpha^{n} \right).
\label{eq:Gamma3}
\end{equation}
The same holds true for $\Gamma_{1}$ and $\Gamma_{2}$, but only when dealing with arteries:
\begin{equation}
\Gamma_{1} = u + 4c, \qquad \Gamma_{2} = u - 4c. 
\end{equation}

When available, the analytical expressions of the three RIs associated with the genuinely non-linear fields are employed in the model, within their range of applicability, to reduce the computational cost.
Through the use of $\Gamma_{1}$, $\Gamma_{2}$ and $\Gamma_{3}$ it is possible to implement inlet and outlet \revised{BCs}, when having, for instance, a prescribed flow rate/velocity at the inlet and a prescribed pressure at the outlet of the domain, as presented in Section \ref{sect:in_out_BC}. Moreover, these RIs are used also for the computation of junctions for both elastic and viscoelastic cases, as further discussed in Section \ref{sect:junction}.
\subsection{0-D lumped-parameter model}
\label{sect:RCR}
At the outlet of peripheral arteries, the 0-D lumped-parameter model, named either 3-element Windkessel model or RCR model, is used to simulate the effects of both terminal resistance and terminal compliance on the propagation of pulse waves. The RCR model is built in analogy with an electric circuit composed of a first resistor, with resistance $R_{1}$, which has the effect of absorbing incoming waves and reducing artificial backwards reflections (therefore fixed to match the characteristic impedance of the terminal 1-D vessel \cite{Alastruey2008}), connected in series with a second resistor, $R_{2}$, which is in parallel combination with a capacitor of compliance $C$ \cite{Bertaglia2020a,Boileau2015,Reymond2009a,Willemet2014,Xiao2014}. In this work, the peripheral inductance is neglected since it has a minor effect on reflected waves under standard conditions \cite{Alastruey2008,Alastruey2012b}.
The resulting system of equation for the 0-D lumped-parameter model reads \cite{Bertaglia2020a}:
\begin{subequations} {\label{RCR}}
\begin{eqnarray}
\label{RCR1}
C \, \frac{\mathrm{d} p_{C}}{\mathrm{d} t} = q^{*} - q_{out}, \\
\label{RCR2}
R_{1}\, q^{*} = p(A^{*})-p_{C}, \\
\label{RCR3}
R_{2}\, q_{out} = p_{C}-p_{out},
\end{eqnarray}
\end{subequations}
where $p(A^{*})$ and $q^{*}=A^{*}u^{*}$ are the unknown variables at the final boundary of the 1-D domain, $p_{C}$ is the pressure at the capacitor and $p_{out}$ and $q_{out}$ are the variables at the outlet of the RCR unit. 

\section{Numerical method}
\label{sect:numerical_model}

As discussed in Section \ref{sect:asymptotic_lim}, the a-FSI blood flow system \eqref{a-FSI} is a hyperbolic system presenting a \textit{stiff} relaxation term in Eq.~\eqref{eq:tubelaw}. Under physiological conditions, the relaxation time can vary from values of order one to very small values if compared to the time scale determined by the characteristic speeds of the system. Therefore, to solve the problem, a robust numerical scheme that works efficiently for all ranges of the relaxation time is needed. This represents a challenging task if one aims at maintaining high order of accuracy and stability of the scheme \cite{Ascher1997,Pareschi2001,Pareschi2005}. In this work, the following discussed asymptotic-preserving (AP) Implicit-Explicit (IMEX) Runge-Kutta (RK) Finite Volume approach is employed.

\subsection{AP-IMEX Runge-Kutta Finite Volume scheme}
\label{sect:IMEX}

For the time discretization of system \eqref{a-FSI}, the stiffly accurate IMEX-SSP2(3,3,2), characterized by 3 stages for both the implicit and explicit parts and second order of accuracy, is adopted \cite{Pareschi2005}. This scheme is AP and asymptotic accurate in the zero relaxation limit, meaning that the consistency of the scheme with the equilibrium system is guaranteed and the order of accuracy is preserved in the stiff limit \cite{Bertaglia2021}. 
As typical for IMEX schemes, an L-stable diagonally implicit Runge-Kutta (DIRK) method is used for the treatment of the stiff part \revised{(here represented by the source terms)}, ensuring elevated robustness; while an explicit strong-stability-preserving (SSP) scheme is \revised{provided} for all the non-stiff components of the system, to maximize the efficiency \cite{Pareschi2005}. 

For the space discretization, \revised{a second-order FV approach is considered, with a uniform grid of length $l_v$ and $n_{c,v}$ number of cells for each \textit{v-th} vessel, a mesh spacing $\Delta x_{v} = l_v / n_{c,v}$}, and a \revised{global} time step size $\Delta t = t^{n+1}-t^{n}$ satisfying the $\mathsf{CFL}$ condition \cite{Toro2009}\revised{:
\begin{equation}
\label{eq:CFL}
\Delta t \leq \mathsf{CFL}\, \frac{\mathrm{min}\{\Delta x_{v}\}}{\mathrm{max}\{\lambda_1,\ldots,\lambda_6\}} \,, \quad v=1,\ldots,n_v\,,
\end{equation}
where $\lambda_1,\ldots,\lambda_6$ are the eigenvalues of the system (see \ref{appendix}) \revised{and $n_v$ identifies the number of vessels considered}. We highlight that a local time stepping (LTS) procedure might also be considered to increment computational efficiency \cite{Muller2016a}. For ease of reading, the subscript $v$ is omitted in the following. On the \textit{i-th} cell, $I_i=[x_{i-\frac{1}{2}},x_{i+\frac{1}{2}}]$ with $\Delta x = x_{i+\frac{1}{2}}-x_{i-\frac{1}{2}}$,} the chosen numerical technique leads to the following final discretization of system \eqref{a-FSI}:
\begin{subequations} \label{IMEX-RKscheme}
\small
\begin{align}
\label{IMEX-RK1}
\mathit{\boldsymbol{Q}}_i^{(k)} = \mathit{\boldsymbol{Q}}_i^n - \frac{\Delta t}{\Delta x}\sum\limits_{j=1}^{k-1} \tilde a _{kj} \left[ \left( \mathit{\boldsymbol{F}}_{i+\frac{1}{2}}^{(j)} - \mathit{\boldsymbol{F}}_{i-\frac{1}{2}}^{(j)} \right) + \left( \mathit{\boldsymbol{D}}_{i+\frac{1}{2}}^{(j)} + \mathit{\boldsymbol{D}}_{i-\frac{1}{2}}^{(j)} \right) + \mathit{\boldsymbol{B}} \left(  \mathit{\boldsymbol{Q}}_i^{(j)} \right) \Delta \mathit{\boldsymbol{Q}}_i^{(j)} \right] 
+ \Delta t \sum\limits_{j=1}^{k} a _{kj}\mathit{\boldsymbol{S}} \left(  \mathit{\boldsymbol{Q}}_i^{(j)} \right), \\
\label{IMEX-RK2}
\mathit{\boldsymbol{Q}}_i^{n+1} = \mathit{\boldsymbol{Q}}_i^n - \frac{\Delta t}{\Delta x}\sum\limits_{k=1}^{s} \tilde{\omega}_{k} \left[ \left( \mathit{\boldsymbol{F}}_{i+\frac{1}{2}}^{(k)} - \mathit{\boldsymbol{F}}_{i-\frac{1}{2}}^{(k)} \right) + \left( \mathit{\boldsymbol{D}}_{i+\frac{1}{2}}^{(k)} + \mathit{\boldsymbol{D}}_{i-\frac{1}{2}}^{(k)} \right) + \mathit{\boldsymbol{B}} \left(  \mathit{\boldsymbol{Q}}_i^{(k)} \right) \Delta \mathit{\boldsymbol{Q}}_i^{(k)} \right] 
+ \Delta t \sum\limits_{k=1}^{s} \omega_{k}\mathit{\boldsymbol{S}} \left(  \mathit{\boldsymbol{Q}}_i^{(k)} \right) .
\end{align}
\normalsize
\end{subequations}
Here, $\boldsymbol{Q}^n_i$ is the vector of the cell-averaged variables at time $t^{n}$. Matrices $\boldsymbol{\tilde{a}}=(\tilde a_{kj})$ and  $\boldsymbol{a}=( a_{kj})$ are $s\times s$ matrices characterizing the explicit and implicit stages, respectively, of the chosen IMEX RK scheme, while the coefficient vectors $\boldsymbol{\tilde{\omega}} = (\tilde{\omega}_1,\ldots,\tilde{\omega}_s)$ and $\boldsymbol{\omega} = (\omega_1,\ldots,\omega_s)$ represent the explicit and implicit weights, respectively, with $s$ identifying the number of the RK stages \cite{Pareschi2001,Pareschi2005}. 
$\boldsymbol{F}$ is the vector of the numerical fluxes and $\boldsymbol{D}$ is the vector of the non-conservative jumps, evaluated at each cell boundary through the path-conservative Dumbser-Osher-Toro (DOT) Riemann solver \cite{Dumbser2011a,Dumbser2011,Leibinger2016}:
\begin{equation}
\label{eq:flux}
	\boldsymbol{F}_{i\pm\frac{1}{2}} = \frac{1}{2} \left[ \boldsymbol{f}\left(\boldsymbol{Q}_{i\pm\frac{1}{2}}^{+}\right) + \boldsymbol{f}\left(\boldsymbol{Q}_{i\pm\frac{1}{2}}^{-}\right)\right] - \frac{1}{2} \int_{0}^{1} \left \lvert \boldsymbol{A}\left(\Psi\left(\boldsymbol{Q}_{i\pm\frac{1}{2}}^{-},\boldsymbol{Q}_{i\pm\frac{1}{2}}^{+},s\right)\right)\right \rvert \frac{\partial \Psi}{\partial s}\mathrm{d} s ,
\end{equation}
\begin{equation}
\label{eq:fluct}
	\boldsymbol{D}_{i\pm \frac{1}{2}} = \frac{1}{2} \int_{0}^{1} \boldsymbol{B}\left(\Psi\left(\boldsymbol{Q}_{i\pm \frac{1}{2}}^{-},\boldsymbol{Q}_{i\pm\frac{1}{2}}^{+},s\right)\right)\frac{\partial \Psi}{\partial s}\mathrm{d} s .
\end{equation}
To achieve second-order accuracy also in space and to avoid spurious oscillations near discontinuities, boundary extrapolated values $\boldsymbol{Q}_{i\pm\frac{1}{2}}^{\mp}$ are computed using a Total-Variation-Diminishing (TVD) method, recurring, \revised{at each \textit{k-th} RK step,} to the classical minmod slope limiter \cite{Toro2009}\revised{:
\begin{equation}\label{eq:minmod}
    \Delta \boldsymbol{Q}_i^{(k)} = \mathrm{minmod}(\boldsymbol{Q}_i^{(k)} - \boldsymbol{Q}_{i-1}^{(k)},\, \boldsymbol{Q}_{i+1}^{(k)} - \boldsymbol{Q}_i^{(k)}) ,
\end{equation}
hence:
\begin{equation}\label{eq:reconstruction}
    \boldsymbol{Q}_{i\pm\frac{1}{2}}^{(k),\mp} = \boldsymbol{Q}_i^{(k)} \pm  \Delta \boldsymbol{Q}_i^{(k)} .
\end{equation}}
Integrals in Eqs. \eqref{eq:flux} and \eqref{eq:fluct} are approximated by a 3-points Gauss-Legendre quadrature formula after that a simple linear path $\Psi$, connecting left to right boundary values in the phase-space, is chosen \cite{Bertaglia2018}.

It is noted that the numerical method here presented has been demonstrated to be exactly well-balanced \cite{Bertaglia2020}, hence exactly conserves a rest initial condition (C-property as defined in \cite{Bermudez1994}). Finally, another characteristic worth to be remarked is that\revised{, consistently with the AP propriety,} the chosen scheme can be reformulated to obtain a totally explicit algorithm. \revised{This permits to avoid the adoption of iterative procedures, like Newton-Raphson method, to solve the implicit terms, leading to an additional consistent reduction of the computational cost with respect to standard semi-implicit or fully implicit methods. 
We underline here that this capability of IMEX schemes, for which the implicit part can be rearranged to compute the involved component of the state variable equation by equation, is valid for many applications and represents a consistent advantage of these methods}.
For further details on these aspects, the reader is referred to \cite{Bertaglia2021,Bertaglia2020,Bertaglia2020a,Pareschi2005}. 

\subsection{Boundary conditions}
\label{sect:BC}
The model presented in this work is characterized by three types of \revised{BCs}: inlet BC, outlet BC and internal BC, i.e.\ at junctions. All types of \revised{BCs} have been implemented to account for the viscous contribution of the vessel wall. Particularly, in the previous works related to the a-FSI blood flow model \cite{Bertaglia2021,Bertaglia2020,Bertaglia2020a}, the pressure at boundary sections was always evaluated neglecting the viscoelastic contribution of the tube law, in favor of a local elastic approach. On the contrary, the here proposed new approach employs an additional RI, namely $\Gamma_{3}$, defined in Eq.~\eqref{eq:Gamma3}, to consider the variation of pressure in the 1-D domain induced by the wall viscoelasticity at boundaries. At first, inlet and outlet \revised{BCs} are presented in Section \ref{sect:in_out_BC}, whereas in Section \ref{sect:junction} the treatment of junction \revised{BCs} is analyzed. A final \revised{section} is left to discuss the problem of maintaining the high-order of accuracy of the scheme even at boundaries.

\subsubsection{Inlet and outlet \revised{BCs}}
\label{sect:in_out_BC}
Inflow BC is implemented imposing at each RK step an inlet flow rate $q_{in}$, 
recurring to the $\Gamma_{2}$ RI associated with the genuinely non-linear field, defined in Eq.~\eqref{eq:RIgenNL}. To take into account viscoelasticity at the boundary, once the inlet cross-sectional area $A_{in}$ is computed from the inlet prescription, the $\Gamma_{3}$ RI is used in its analytical form, presented in Eq.~\eqref{eq:Gamma3}, for the evaluation of $p_{in}$, pressure at the inlet section.

At outlet sections, namely at the last cell of each terminal vessel of the network, the 3-element Windkessel/RCR model presented in Section \ref{sect:RCR} is coupled with the 1-D model through the solution of the problem at the interface. A null outlet pressure $p_{out}=0$ is assigned in these sections to simulate the blood pressure when the flux reaches the venous system \cite{Boileau2015,Xiao2014}. Following, the variable $p_C$ presented in system \eqref{RCR} must be computed. This system has two algebraic equations, Eqs. \eqref{RCR2} and \eqref{RCR3}, and one ordinary differential equation (ODE), Eq.~\eqref{RCR1}. Hence, it can be rewritten as a unique ODE problem:
\begin{equation} 
\label{RCR3rewritten}
\frac{\mathrm{d}p_C}{\mathrm{d}t}= \frac{1}{C} \left[ \frac{1}{R_1} \left( p(A^*) - p_C \right)- \frac{1}{R_2} \left( p_C - p_{out} \right) \right].
\end{equation}
In this work, this ODE is integrated in time following the IMEX RK method previously discussed in Section \ref{sect:IMEX}, treating the equation explicitly since it does not involve \textit{stiff} terms. 
Thus, the explicit RK discretization of Eq.~\eqref{RCR3rewritten} reads:
\begin{subequations}
\begin{align} \label{pCkth}
p_C^{(k)} =  p_C^n + \Delta t \sum\limits_{j=1}^{k-1} \frac{\tilde a _{kj}}{C} \left[ \frac{1}{R_1} \left( p(A^*)^{(j)} - p_C^{(j)} \right)- \frac{1}{R_2} \left( p_C^{(j)} - p_{out} \right) \right],
\\
\label{pCRK}
p_C^{n+1} = p_C^n + \Delta t \sum\limits_{k=1}^{s} \frac{\tilde{\omega}_{k}}{C} \left[ \frac{1}{R_1} \left( p(A^*)^{(k)} - p_C^{(k)} \right)- \frac{1}{R_2} \left( p_C^{(k)} - p_{out} \right) \right] .
\end{align}
\end{subequations}
It is worth to underline that the application of the IMEX RK scheme at the outlet BC increases the consistency of the model, maintaining an homogeneous approach among physical domain and boundary sections. 
Following the same approach used at the inlet interface, in order to couple the 1-D domain with the RCR model, the $\Gamma_{1}$ RI associated with the genuinely non-linear field, defined in Eq.~\eqref{eq:RIgenNL}, is employed. This yields to a non-linear equation in $A^*$, solved with Newton's method (see \cite{Bertaglia2020a} for more details). Once $A^*$ is obtained, the velocity $u^{*}$ is computed via $\Gamma_{1}$ and the pressure $p^{*}$ is computed through the elastic tube law, when the problem is elastic, or using $\Gamma_{3}$ to account also for the viscous contribution of the vessel wall, when considering the full viscoelastic problem.
\subsubsection{Junction \revised{BCs}}
\label{sect:junction}

\begin{figure}[t]
\centering
\includegraphics[scale=0.9]{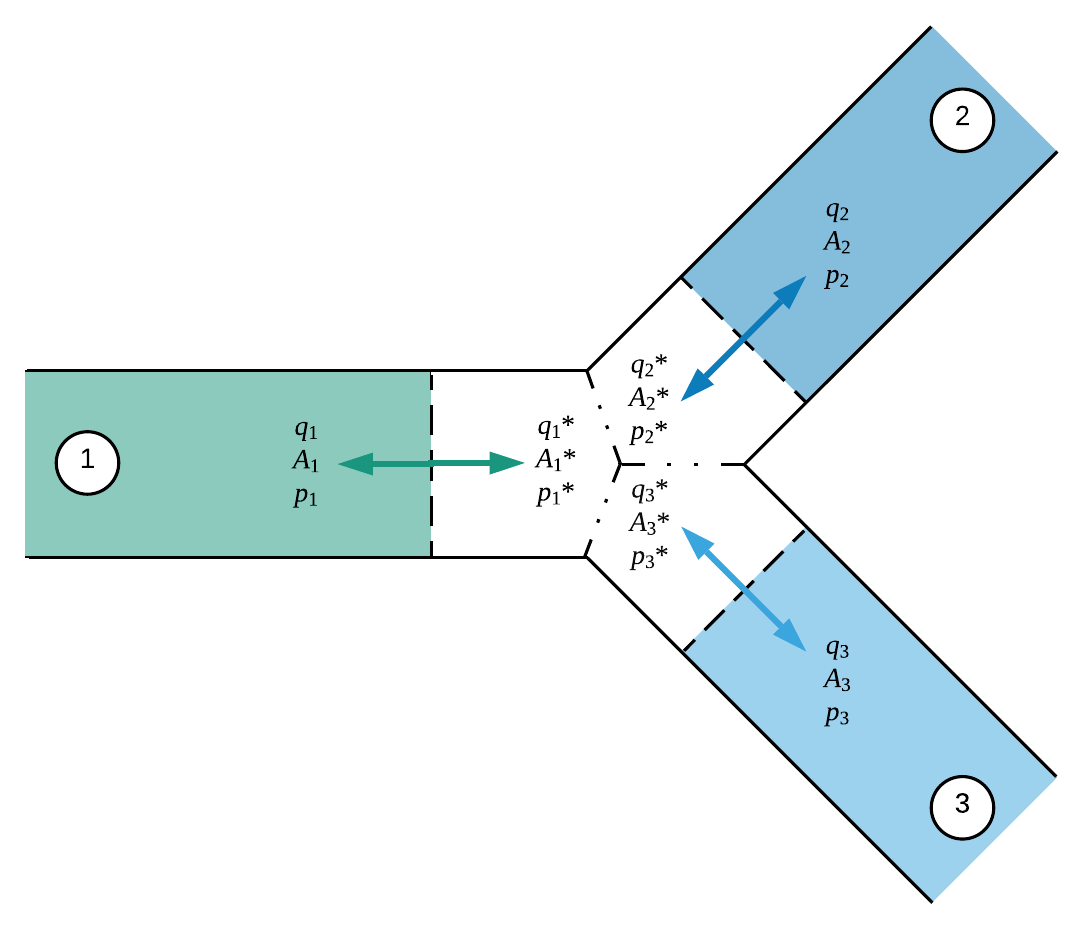}
\caption{3--vessels junction scheme. Reading the figure from left to right, the scheme shows a bifurcation, whereas reading the figure from right to left, it shows a confluence. With respect to the interpretation related to the Junction Riemann Problem, dashed lines represent rarefaction waves, whereas dot-dashed lines represent contact discontinuity waves. The initial conditions of the problem are given by the vector of averaged state variables of the afferent cells of vessel domains, i.e.~ $q_j=A_j u_j$, $A_j$ and $p_j$, with $j = 1,2,3$, whereas the intermediate constant states are $q_j^*=A_j^* u_j^*$, $A_j^*$ and $p_j^*$, again with $j = 1,2,3$.}
\label{bifurcation}
\end{figure}

Besides inlet and outlet sections, junctions among two or more blood vessels involve internal boundaries where specific conditions must be imposed to \revised{couple} the 1-D domains of the branching vessels. The present work proposes a method able to face both the simplest 2--vessels junction, consisting in two linked vessels with either the same or different geometrical and mechanical characteristics, and junctions consisting in more branching vessels (even more than three). In this last case, the model is able to deal with both the bifurcation case, in which a parent vessel separates in two or more daughter vessels (occurring \revised{mostly} in arterial networks) and the confluence case, in which two or more parent vessels merge in one daughter (\revised{generally} in venous networks). In Fig.~\ref{bifurcation} the scheme of a generic 3--vessels junction is shown. 

Generally, 3-D analyses of flow at either bifurcations or confluences show that blood flow becomes very complex at junctions, with the possible development of transient separation and of secondary flows \cite{Quarteroni2017,Xiao2014}. However, these issues are commonly confined in the proximity of the junction and their effects on pulse wave propagation are neglected due to the \textit{long wavelength approximation} \cite{Alastruey2012b}. Moreover, energy losses at junctions are neglected in this work, due to their little contribution \cite{Matthys2007a,Wang2013}, although some studies account for them (for example considering branching angles \cite{Quarteroni2003,Mynard2015,Steele2003,Stettler1981}). The \revised{inclusion} of  coefficients of pressure losses at junctions is beyond the scope of this work and will be considered for future \revised{developments}. 

Junction BC is \revised{numerically} implemented through the solution of a non-linear system of equations that derives from the RP arising at these internal boundaries, in the following called Junction Riemann Problem (JRP) \revised{\cite{Colombo2008,Colombo2008b}}.

The solution of elementary one dimensional RPs involving a set of two balance laws, one for the mass and one \revised{for} the momentum, is characterized by a single intermediate constant state (the so called \emph{star region}), separated from the initial imposed constant states by non-linear waves, such as shock waves or rarefactions. If the addition of further equations to the set of governing equations results in the enrichment of the eigenstructure of the system with null eigenvalues, as it happens for the augmented system \eqref{a-FSI}, stationary contact waves become part of the solution. Restricting our analysis to sub-critical flows (i.e.\ $u<c$) the non-linear waves are directed from the center towards the periphery and the intermediate states become two, separated from each other by the new contact discontinuity waves \cite{Toro2013}. The discontinuity between the intermediate states is stationary and remains located at the initial discontinuity of the initial conditions.
Now, conceiving the position of the initial discontinuity as a junction section between branches, it becomes clear that the RP partial solution related to each branch consists only of an initial state separated from a single intermediate state by a non-linear wave, while the intermediate states of the two adjacent branches are separated from each other by the contact discontinuities.

The extension of this interpretation to the JRP for a confluence of $N$ branches into a node is straightforward. Again, the initial state of each branch is separated from the \emph{star region} of the same branch by a non-linear wave, while the $N$ intermediate states of the $N$ branches, which are adjacent to the node, are separated from each other by contact waves. 
\revised{For an application of the same approach to different balance laws, the reader is addressed to \cite{Bertaglia2021a,Elshobaki2018b,Elshobaki2018,Muller2015,Muller2016}}.

To properly define a simplified solution of the JRP suitable to act as a junction mathematical model, two assumptions are made \textit{a priori}: 
\begin{enumerate}
\item The flux is sub-critical, namely the speed index, defined as the ratio between velocity of the fluid and celerity \cite{Shapiro1977}, is always less than one and therefore the non-linear waves \revised{joining 1-D domain states and star region states} are directed from the central node towards the periphery \cite{Toro2009};
\item Both the resulting non-linear waves are rarefactions, hence the JRP is treated following a Two-Rarefactions Riemann Solver (TRRS) \cite{Toro2009}. 
\end{enumerate}
As an example, referring to Fig.~\ref{bifurcation}, which clearly illustrates the so-conceived JRP, the initial conditions of the problem are given by the vector of averaged state variables of the afferent cells of vessel domains, i.e.~ $q_j=A_j u_j$, $A_j$ and $p_j$, here with $j = 1,2,3$, whereas the intermediate constant states are $q_j^*=A_j^* u_j^*$, $A_j^*$ and $p_j^*$, again with $j = 1,2,3$. 

Given a general junction connecting $N$ vessels, $\mathit{\boldsymbol{Q}_{j}^{\mathrm{1D}}}$, $j = 1,\ldots,N$, constant states are identified at time $t$ at the junction section, provided by the vectors of averaged state variables at the interface, one for each adjacent cell of the $N$ afferent vessels.   
If variables are discontinuous across the interface, $N$ new intermediate states originate at the node at time $t+\Delta t$. This set is formed by the starred constant states, $\mathit{\boldsymbol{Q}_{j}^{*}}$, $j = 1,\ldots,N$, which identify the $3N$ unknowns of the JRP: $q_j^*$, $A_j^*$ and $p_j^*$.
As introduced in Section \ref{sect:eigenstructure}, assuming that the non-linear waves are always rarefactions, it is possible to write the non-linear system at junctions recurring to the RIs related to both the genuinely non-linear fields, $\Gamma_{1,2,3}$, defined in  Eq.~\eqref{eq:RIgenNL}, which identify the variables conserved on rarefaction waves, and the RIs related to the LD fields, $\Gamma^{LD}_{1,2}$, defined in  Eq.~\eqref{eq:RILD}, which indicate, instead, variables conserved across contact discontinuities. Notice that, as direct consequence of the inclusion of the RIs of the LD fields in the system, the conservation of mass ($Au$) and the conservation of total energy ($p+\frac{1}{2}\rho u^2$) at junctions is granted. 
Thus, the resulting non-linear system arising at each viscoelastic junction, for both arterial and venous systems, reads as follows:
\begin{subequations} {\label{junct-sys}}
\begin{eqnarray}
\label{junc_vis_flowrate}
\sum\limits_{j=1}^{N} \Theta_{n_{j}}A_{j}^{*}u_{j}^{*} = 0, \\
\label{junc_vis_press}
\left( p_{1}^{*} + \frac{1}{2}\rho u_{1}^{*^2} \right) - \left( p_{j}^{*} + \frac{1}{2}\rho u_{j}^{*^2} \right) = 0, \qquad &j = 2,\ldots,N, \\
\label{junc_vis_RI_Au}
u_{j}^{*} - u_{j}^{1D} + \Theta_{n_{j}}\int_{A_{j}^{1D}}^{A_{j}^{*}} \frac{c(A)}{A}\, \mathrm{d}A = 0, \qquad &j = 1,\ldots,N, \\
\label{junc_vis_RI_Ap}
p_{j}^{*} - p_{j}^{1D} - \int_{A_{j}^{1D}}^{A_{j}^{*}} d(A) \,\mathrm{d}A = 0, \qquad &j = 1,\ldots,N,
\end{eqnarray}
\end{subequations}
where $\Theta_{n_{j}}$ is an auxiliary function such that:
\begin{eqnarray}
\Theta_{n_{j}} =
\begin{cases}
+1, & \mbox{if } \xi_{n_{j}}=l_{j}, \\
-1, & \mbox{if } \xi_{n_{j}}=0,
\end{cases}
\end{eqnarray}
with $\xi_{n_{j}}$ local coordinate of the $j$-$th$ vessel, evaluated at node $n_j$, and $l_{j}$ vessel length. Notice that Eq.~\eqref{junc_vis_RI_Au} derives from $\Gamma_{1}$ and $\Gamma_{2}$, whereas Eq.~\eqref{junc_vis_RI_Ap} derives from $\Gamma_{3}$. The non-linear system is solvable applying a Newton-Raphson iterative method, using $\mathit{\boldsymbol{Q}_{j}^{\mathrm{1D}}}$ as initial guess for $\mathit{\boldsymbol{Q}_{j}^{*}}$ \cite{Alastruey2012}.

It is worth stressing that the mechanical behavior attributed to vessel walls affects the formulation of the junction non-linear system, namely if the vessel is chosen to behave in an elastic or in a viscoelastic way. Indeed, when choosing a simple elastic vessel wall, system \eqref{junct-sys} can be simplified as \cite{Alastruey2012,Muller2014a}:
\begin{subequations}{\label{junct-sys-el}}
\begin{eqnarray}
\label{junc_el_flowrate}
\sum\limits_{j=1}^{N} \Theta_{n_{j}}A_{j}^{*}u_{j}^{*} = 0, \\
\label{junc_el_press}
\left( p(A_{1}^{*}) + \frac{1}{2}\rho u_{1}^{*^2} \right) - \left( p(A_{j}^{*}) + \frac{1}{2}\rho u_{j}^{*^2} \right) = 0, \qquad &j = 2,\ldots,N, \\
\label{junc_el_RI_Au}
u_{j}^{*} - u_{j}^{1D} + \Theta_{n_{j}}\int_{A_{j}^{1D}}^{A_j^{*}} \frac{c(A)}{A} \,\mathrm{d}A = 0, \qquad &j = 1,\ldots,N.
\end{eqnarray}
\end{subequations}
Hence, in the elastic case the only unknown variables are the area $A_{j}^{*}$ and the flow rate $q_{j}^{*}=A_{j}^{*}u_{j}^{*}$, being the pressure $p(A_{j}^{*})$ calculated \textit{a posteriori} via the elastic tube law \cite{Bertaglia2020}:
\begin{equation}
p = p_{ext} + K \left[ \left(\frac{A^*}{A_0}\right)^m - \left(\frac{A^*}{A_0}\right)^n \right].
\end{equation}

\subsubsection{High order of accuracy at boundaries}
As discussed in Section \ref{sect:IMEX}, the proposed AP-IMEX RK FV scheme is second-order accurate both in time and space. To maintain the second-order of accuracy of the scheme in the whole domain, slope values $\Delta \boldsymbol{Q}_i$, \revised{$i=1,\ldots,n_c$,} are computed in every cell of the physical grid, even in those cells that are adjacent to either the external or the internal boundary sections of the network, namely inlet/outlet or junction interface, respectively. Following the minmod slope limiter \revised{\eqref{eq:minmod}}, at each \revised{\textit{k-th} RK step}, for each cell $i$, the slope $\Delta \mathit{\boldsymbol{Q}}_{i}^{(k)}$ is computed requiring the vector of averaged variables of both adjacent cells: $\mathit{\boldsymbol{Q}}_{i-1}^{(k)}$ and  $\mathit{\boldsymbol{Q}}_{i+1}^{(k)}$ \cite{Toro2009}.
However, cells adjacent to the inlet (resp.\ outlet) present a complication since $\boldsymbol{Q}_{i-1}^{(k)}$ (resp. $\boldsymbol{Q}_{i+1}^{(k)}$) is missing. The same applies at junctions, where a daughter (resp. parent) vessel misses $\boldsymbol{Q}_{i-1}^{(k)}$ (resp.\ $\boldsymbol{Q}_{i+1}^{(k)}$). This issue is overcame by computing the vector of variables at boundaries as presented in Sections \ref{sect:in_out_BC} and \ref{sect:junction} and assuming that these values are constant within the missing boundary cells, which are conceived as ghost cells. 
The so-obtained ghost cell vector of averaged variables, \revised{$\boldsymbol{Q}_g^{(k)}$}, compensates for the missing one in the slope computation. \revised{Hence, for the evaluation of the slope of the first physical cell of the inlet vessel of the network as well as for the first cell of daughter vessels:
\begin{equation*}
    \Delta \boldsymbol{Q}_1^{(k)} = \mathrm{minmod}(\boldsymbol{Q}_1^{(k)} - \boldsymbol{Q}_{g}^{(k)},\, \boldsymbol{Q}_{2}^{(k)} - \boldsymbol{Q}_1^{(k)});
\end{equation*}
while the formulation of the minmod slope limiter in the last physical cell of each peripheral vessel coupled with the RCR model and in the last cell of parent vessels reads:
\begin{equation*}
    \Delta \boldsymbol{Q}_{n_c}^{(k)} = \mathrm{minmod}(\boldsymbol{Q}_{n_c}^{(k)} - \boldsymbol{Q}_{n_c-1}^{(k)},\, \boldsymbol{Q}_{g}^{(k)} - \boldsymbol{Q}_{n_c}^{(k)}) .
\end{equation*}
}
Once \revised{$\Delta \mathit{\boldsymbol{Q}}_{i}^{(k)}$} is computed, boundary extrapolated values within cell $i$ are obtained \revised{through Eq.~\eqref{eq:reconstruction}} and fluxes and non-conservative jumps can be evaluated with Eqs.~\eqref{eq:flux} and \eqref{eq:fluct}, respectively. 
\revised{We remark that the proposed methodology can be extended to higher-order accuracy in space recurring to different spatial reconstruction, for instance adopting WENO approaches \cite{Cavallini2008}}.
\section{Numerical results}
\label{sect:results}

\revised{Following \cite{Muller2016}}, the \revised{numerical implementation of the} junction is first validated via a trivial 2--vessels junction case, which consists in two subsequent compliant vessels linked by a single junction BC, for which the reference solution is given by the simulation of the single continuous vessel not interrupted by the junction. This case is implemented for both vessel types, i.e.\ artery and vein, prescribing a pulse wave as inlet BC. 
An accuracy analysis is performed with these two tests to verify the conservation of the second order of accuracy of the scheme in the whole domain, even when junctions are considered. Moreover, \textit{dead-body} simulations\revised{, i.e.\ the steady case with null velocity and uniform pressure,} are performed to confirm that no numerical errors, such as spurious velocities, arise during the simulation, therefore respecting the well-balancing of the scheme. Subsequently, a 3--vessels junction test of an ideal human aortic bifurcation is \revised{performed}.
Finally, two examples of human arterial networks are \revised{considered}, accounting for both elastic and viscoelastic behavior of the vessels to depict differences deriving from the choice of the mechanical characterization of the vessel wall. When considering the viscoelastic rheological behavior, \revised{BCs} at each junction composing the network, as well as at inlet and outlet sections, \revised{are} implemented as described in Sections \ref{sect:in_out_BC} and \ref{sect:junction}. 

\revised{In all the performed simulations, if not otherwise stated, $\mathsf{CFL}$ = 0.9 is imposed. Moreover, for ease of reading, information regarding spatial and temporal discretization is generally given in the caption of figures showing results of the specific tests.
Numerical data} of viscoelastic simulations of the aortic bifurcation test (Section \ref{sect:bif}) and of the two arterial networks (Sections \ref{37AN} and \ref{ADAN56}) are made publicly available as Supporting Material, in order to facilitate and encourage further analysis.

\begin{table}[b]
\caption{Parameters of the 2--vessels junction test, valid for a generic artery (AA) and a generic vein (VV), with $l$ vessel length, $R_0$ equilibrium internal radius, $h_0$ wall thickness, $p_0$ equilibrium internal pressure, $E_0$ instantaneous Young modulus, $E_\infty$ asymptotic Young modulus, $\tau_r$ relaxation time and $\hat q, \hat \sigma, \hat t$ additional parameters of the test used for the inlet condition, defined in Eq.~\eqref{eq:test2vess}. Note that parameters are equal for both parent and daughter vessels in this trivial test.}
\label{tbl:2-vessel}
\centering
\begin{tabular}{c c c c c c c c c c c}
\hline
\rule{0pt}{3ex}&
$l$&
$R_0$&
$h_0$&
$p_0$&
$E_{0}$&
$E_{\infty}$&
$\tau _{r}$&
$\hat q$&
$\hat \sigma$&
$\hat t$\\
Test  &
$\mathrm{[cm]}$ &
$\mathrm{[cm]}$ &
$\mathrm{[mm]}$ &
$\mathrm{[mmHg]}$ &
$\mathrm{[MPa]}$ &
$\mathrm{[MPa]}$ &
$\mathrm{[ms]}$&
$[\mathrm{cm}^3 \mathrm{s}^{-1}]$ &
$[\mathrm{s}^{-2}]$ &
$\mathrm{[s]}$ \\
\hline
AA & 20 & 1.0 & 0.5 & 80 & 2.3 & 1.6 & 3.5 & 100 &10000 &0.025\\
VV & 20 & 0.5 & 0.5 & 10 & 3.0 & 2.6 & 0.022 & 10 &1000 &0.05\\
\hline
\end{tabular}
\end{table}


\subsection{2--vessels junction}

\begin{figure}[t]
\centering
\includegraphics[width=1\linewidth]{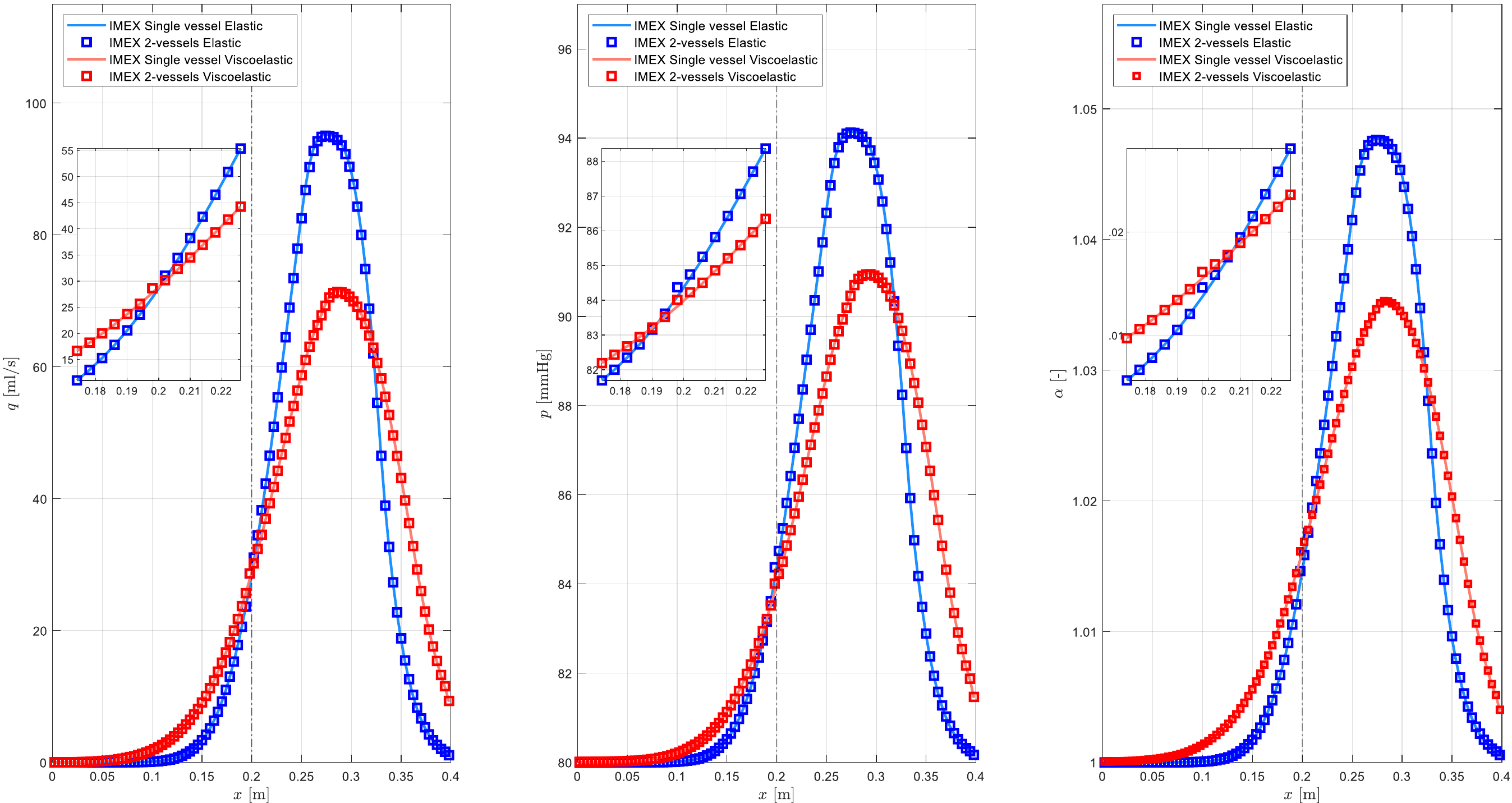}
\caption{Results of the 2--vessels junction test performed for a generic artery (AA case). Comparison between the reference solution (single artery) and the numerical solution (2 identical arteries separated by a trivial junction) when considering either an elastic or a viscoelastic wall mechanical behavior. Results are presented in terms of flow rate (left figure), pressure (center figure) and dimensionless cross-sectional area (right figure) at time $t = 0.068~ \mathrm{s}$, using $n_c= 50$ cells for each vessel. }
\label{fig:2vesAA}
\end{figure}
\begin{figure}[t]
\includegraphics[width=1\linewidth]{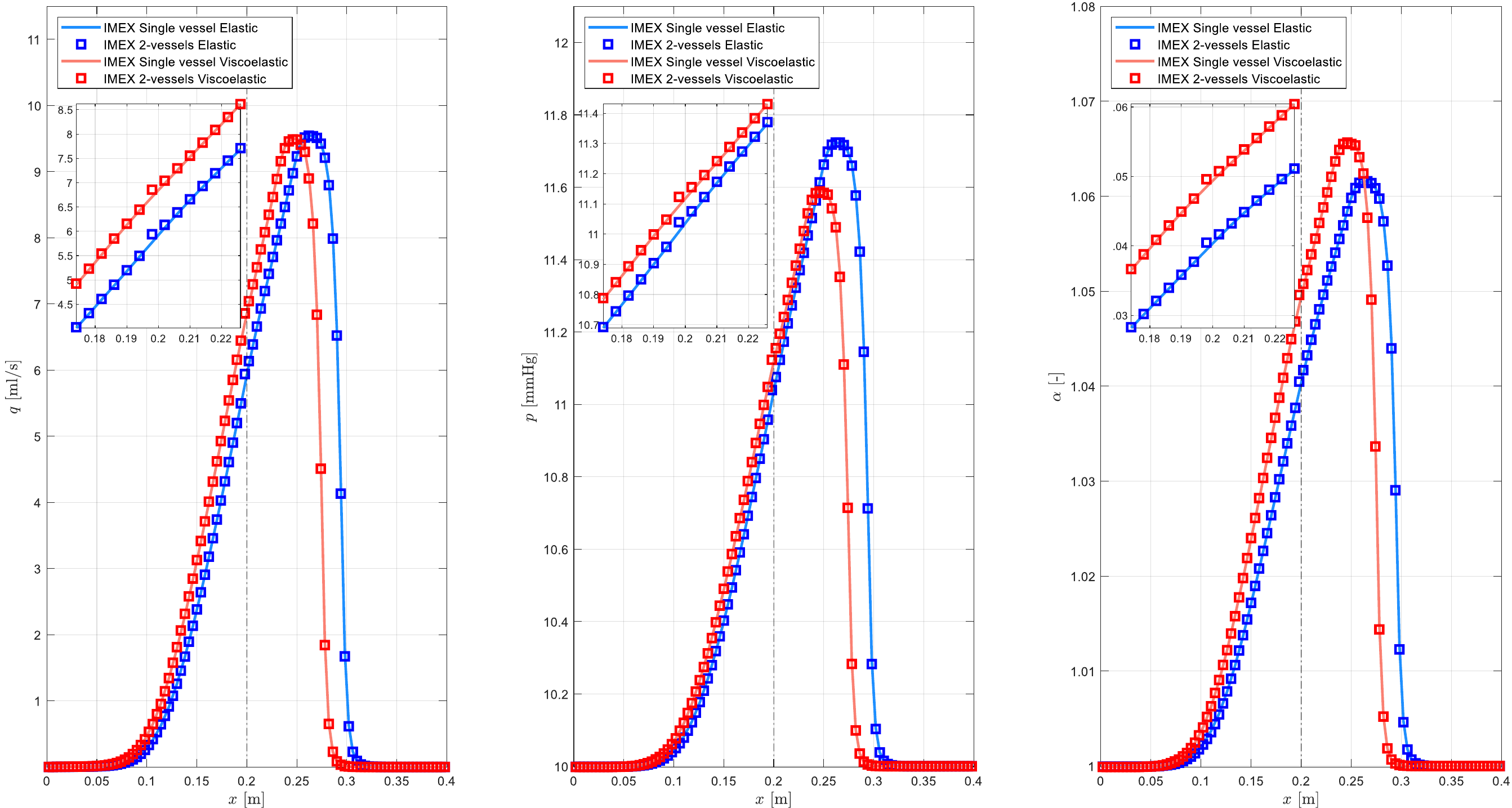}
\centering
\caption{Results of the 2--vessels junction test performed for a generic vein (VV case). Comparison between the reference solution (single vein) and the numerical solution (2 identical veins separated by a trivial junction) when considering either an elastic or a viscoelastic wall mechanical behavior. Results are presented in terms of flow rate (left figure), pressure (center figure) and dimensionless cross-sectional area (right figure) at time $t = 0.17~ \mathrm{s}$, using $n_c= 50$ cells for each vessel.}
\label{fig:2vesVV}
\end{figure}

The 2--vessels junction test has been performed for both blood vessel types, i.e.\ artery and vein, and for both wall mechanical behaviors, i.e.\ elastic and viscoelastic. The reference solution is given by the corresponding case of single vessel. To simulate the reference and the numerical solution, each parameter remains the same with the only difference that in the latter a trivial junction is inserted in the middle of the vessel domain. The 2--vessels junction test is purely abstract and specifically aimed at analyzing the \revised{numerical approach treating the} junction as described in Section \ref{sect:junction}. In the following tests, blood density, $\rho= 1050\, \mathrm{kg/m^3}$, and blood viscosity, $\mu = 4\, \mathrm{mPa \cdot s}$, are assumed as constant for both blood vessel types. Geometrical and mechanical parameters for the arterial case (AA) are taken from \cite{Muller2016}, whereas parameters for the venous case (VV) are chosen within physiological ranges, as presented in \cite{Physiology}. 
Regarding the SLSM parameters, the estimation procedure presented in \cite{Bertaglia2020a} is followed.
The relaxation time $\tau_r$ used for the artery is greater than the one used for the vein because the latter is characterized by a more restrained viscoelastic behavior, which leads to less dissipation of energy, since the venous walls are \revised{more deformable}, indeed tending to collapse when relatively high external pressures act \cite{Murillo2019,Spiller2017}. It is worth underlining that few works have been proposed concerning viscoelasticity applied to veins. However, this test is important to highlight the applicability of the junction non-linear system as internal BC for both vessel types, without claiming to be physiologically or clinically relevant. 

While at the outlet a simple transparent, reflection-free, BC is imposed, the following flow rate is prescribed at the inlet:
\begin{equation}
\label{eq:test2vess}
q_{in} = \hat{q} \, \exp\left[{-\hat{\sigma}\left(t - \hat{t}\right)^2}\right],
\end{equation}
where the parameters $\hat{q}$, $\hat{\sigma}$ and $\hat{t}$ are indicated in Table \ref{tbl:2-vessel}, together with the rest of the parameters used in these tests. Also parameters characterizing the inlet flow rate are taken from \cite{Muller2016} for the AA case, whereas, due to the lack of reference data, for the VV case they are chosen consistently with physiological values with the main purpose of testing the model.

Elastic and viscoelastic results presented in Figs.~\ref{fig:2vesAA} and \ref{fig:2vesVV} show a \revised{very good} agreement between reference and numerical solutions in the junction sections in both AA and VV tests, confirming the correct and robust \revised{numerical} implementation of the junction even in the innovative viscoelastic case. \revised{A better agreement between reference and numerical solutions would only be possible increasing the order of accuracy of the model, as presented in \cite{Muller2015}.} It is worth underlining that these \revised{satisfactory} results \revised{derive from the accurate numerical treatment} of the \revised{BCs} described in Section \ref{sect:BC}. Without the discussed combination of numerical techniques for treating internal boundaries, the numerical solution would be characterized by strong perturbations. This is true in these trivial 2--vessels cases, as well as in more complex networks \revised{(examples are not reported here for sake of brevity)}. As already discussed, the treatment of internal boundaries is a delicate task, especially when dealing with very wide networks. Indeed, even a small variation from the presented numerical model set up could provoke cumbersome problems to the network solution. In Figs.~\ref{fig:2vesAA} and \ref{fig:2vesVV}, it is also possible to appreciate the viscoelastic contribution given by the SLSM model. 
It can be noticed that for the VV case the viscoelastic effect is less evident than in the AA case due to the different viscoelastic parameters assigned to the two vessel walls. Furthermore, different characteristics between cases AA and VV can be observed in terms of dimensionless cross-sectional area $\alpha$, due to the diverse mechanical behavior associated to arteries and veins. Indeed, the tube law used for veins defines a more complex behavior with respect to the one used for arteries, having different coefficients $K$, $m$, and $n$ in Eq.~\eqref{eq:d_el}, which, for veins, account also for collapsing states.

\subsubsection{Accuracy analysis and well-balancing property}
\label{sect:accuracy_analysis}

\begin{table}[t!]
\caption{Parameters of the accuracy analysis performed for the 2--vessels junction test, considering a generic artery (AA) and a generic vein (VV), referring to Eq. \eqref{sys:sinICaccuracy}. Values are chosen within standard ranges found in the human body.}
\label{tbl:sinICaccuracy}
\centering
\begin{tabular}{c c c c c c c c c c c c}
\hline 
\rule{0pt}{3ex}&
$l$&
$T_0$&
$\tilde A$&
$\tilde{a}$&
$\tilde P$  &
$\tilde{p}$  &
$\tilde E_0$ &
$\tilde{e}$&
$h_0$&
$E_{\infty}$ &
$\tau _{r}$\\
Test &
[cm]&
[s]&
$\mathrm{[mm^2]}$ & 
$\mathrm{[mm^2]}$ & 
$\mathrm{[kPa]}$ &
$\mathrm{[kPa]}$ &
$\mathrm{[MPa]}$ & 
$\mathrm{[MPa]}$ & 
[mm]&
$\mathrm{[MPa]}$ & 
$\mathrm{[ms]}$\\
\hline
AA & 20& 1.0&  400 & 40 & 10 & 2.0 & 2.0 & 0.2 &1.5 & 1.6 & 0.36 \\
VV & 20& 1.0& 40 & 4.0 & 1.5 & 0.3 & 1.76 & 0.1 &0.3& 1.5 & 0.06 \\
\hline
\end{tabular}
\end{table}

\begin{table}[t!]
\caption{Results of the accuracy analysis performed for the 2--vessels junction test, concerning a generic artery (AA) and a generic vein (VV). Errors are computed for variables $A$, $Au$ and $p$ in terms of norms $L^1$, $L^2$ and $L^{\infty}$, showing the corresponding order of accuracy. \revised{The final time of the simulation is $t_{end}$ = 0.025 s for the AA test, and $t_{end}$ = 0.10 s for the VV test.}}
\label{tbl:orderofacc}
\centering
\begin{tabular}{c c c c c c c c c}
\hline
\rule{0pt}{3ex}
Test & 
Variable & 
$n_c$ & 
$L^{1}$ & 
$\mathcal{O}(L^{1})$ & 
$L^{2}$ & 
$\mathcal{O}(L^{2})$ & 
$L^{\infty}$ & 
$\mathcal{O}(L^{\infty})$ \\
\hline
\rule{0pt}{2.5ex}
\multirow{13}{*}{AA} & \multirow{4}{*}{$A$} & 33 & \num{3.47e-08} & - & \num{6.03e-08} & - & \num{1.36e-07} & - \\
& & 99 & \num{3.75e-09} & 2.02 & \num{6.58e-09} & 2.02 & \num{1.48e-08} & 2.02 \\
& & 297 & \num{4.13e-10} & 2.01 & \num{7.25e-10} & 2.01 & \num{1.64e-09} & 2.00 \\
& & 891 & \num{4.52e-11} & 2.01 & \num{7.95e-11} & 2.01 & \num{1.80e-10} & 2.01 \\
& & 2673 & \num{4.52e-12} & 2.10 & \num{7.94e-12} & 2.10 & \num{1.80e-11} & 2.10 \\
\rule{0pt}{3ex}
& \multirow{4}{*}{$Au$} & 33 & \num{1.70e-07} & - & \num{2.98e-07} & - & \num{7.51e-07} & - \\
& & 99 & \num{2.38e-08} & 1.79 & \num{4.29e-08} & 1.77 & \num{1.16e-07} & 1.70 \\
& & 297 & \num{2.82e-09} & 1.95 & \num{5.10e-09} & 1.94 & \num{1.40e-08} & 1.92 \\
& & 891 & \num{3.16e-10} & 2.00 & \num{5.74e-10} & 2.00 & \num{1.58e-09} & 2.00 \\
& & 2673 & \num{3.18e-11} & 2.10 & \num{5.78e-11} & 2.10 & \num{1.60e-10} & 2.10 \\
\rule{0pt}{3ex}
& \multirow{4}{*}{$p$} & 33 & \num{8.21e+00} & - & \num{1.43e+01} & - & \num{3.42e+01} & - \\
& & 99 & \num{9.14e-01} & 2.00 & \num{1.60e+00} & 2.00 & \num{3.75e+00} & 2.01 \\
& & 297 & \num{1.00e-01} & 2.01 & \num{1.76e-01} & 2.01 & \num{4.12e-01} & 2.01 \\
& & 891 & \num{1.10e-02} & 2.01 & \num{1.93e-02} & 2.01 & \num{4.52e-02} & 2.01 \\
& & 2673 & \num{1.10e-03} & 2.10 & \num{1.93e-03} & 2.10 & \num{4.51e-03} & 2.10 \\

\hline
\rule{0pt}{2.5ex}
\multirow{13}{*}{VV} & \multirow{4}{*}{$A$} & 33 & \num{2.91e-08} & - & \num{9.49e-08} & - & \num{7.55e-07} & - \\
& & 99 & \num{3.26e-09} & 2.00 & \num{1.11e-08} & 1.96 & \num{1.15e-07} & 1.71 \\
& & 297 & \num{3.57e-10} & 2.01 & \num{1.21e-09} & 2.01 & \num{1.38e-08} & 1.93 \\
& & 891 & \num{3.88e-11} & 2.02 & \num{1.34e-10} & 2.01 & \num{1.54e-9} & 2.00 \\
& & 2673 & \num{3.87e-12} & 2.10 & \num{1.34e-11} & 2.10 & \num{1.55e-10} & 2.09 \\
\rule{0pt}{3ex}
& \multirow{4}{*}{$Au$} & 33 & \num{1.61e-08} & - & \num{7.34e-08} & - & \num{6.33e-07} & - \\
& & 99 & \num{2.15e-09} & 1.83 & \num{9.46e-09} & 1.86 & \num{1.02e-07} & 1.59 \\
& & 297 & \num{2.59e-10} & 1.93 & \num{1.15e-09} & 1.91 & \num{1.35e-08} & 1.91 \\
& & 891 & \num{2.94e-11} & 1.99 & \num{1.32e-10} & 1.97 & \num{1.58e-9} & 1.95 \\
& & 2673 & \num{2.97e-12} & 2.10 & \num{1.34e-11} & 2.10 & \num{1.61e-10} & 2.08 \\
\rule{0pt}{3ex}
& \multirow{4}{*}{$p$} & 33 & \num{6.55e-01} & - & \num{2.12e+00} & - & \num{1.69e+01} & - \\
& & 99 & \num{8.89e-02} & 1.82 & \num{3.25e-01} & 1.71 & \num{3.54e+00} & 1.42 \\
& & 297 & \num{1.19e-02} & 1.83 & \num{4.77e-02} & 1.75 & \num{5.20e-01} & 1.75 \\
& & 891 & \num{1.33e-03} & 2.00 & \num{5.44e-03} & 1.98 & \num{6.15e-02} & 1.94 \\
& & 2673 & \num{1.37e-04} & 2.07 & \num{5.69e-04} & 2.05 & \num{6.48e-03} & 2.05 \\
\hline
\end{tabular}
\end{table}

The accuracy analysis is performed with the 2--vessels junction test for both the arterial (AA) and the venous (VV) case. In contrast with the above discussed tests, for these analyses periodic BCs are chosen. Furthermore, a sinusoidal initial condition is imposed as follows:
\begin{eqnarray} {\label{sys:sinICaccuracy}}
\mathit{\boldsymbol{Q}}_{IC} =
\begin{pmatrix}
A_{IC} \\  q_{IC} \\ p_{IC} \\ A_{0,\, {IC}} \\ E_{0,\, {IC}} \\ p_{ext,\, {IC}}
\end{pmatrix} =
\begin{pmatrix}
\tilde A + \tilde{a}\sin \left( \frac{2\pi x}{l} \right) \\
- \frac{\tilde{a} \, l}{T_{0}}\cos \left( \frac{2\pi x}{l} \right) \\
\tilde P + \tilde{p}\cos \left( \frac{2\pi x}{l} \right) \\
\tilde A+\tilde{a}\sin \left( \frac{2\pi x}{l} \right) \\
\tilde E_0+\tilde{e}\sin \left( \frac{2\pi x}{l}\right) \\
\tilde P +\tilde{p}\cos \left( \frac{2\pi x}{l} \right) \\
\end{pmatrix}.
\end{eqnarray}
Parameters for these analyses are listed in Table \ref{tbl:sinICaccuracy}. 
The reference solution is simulated with $n_c = 8019$ number of cells, and for each state variable (area $A$, flow rate $Au$ and pressure $p$) $L^1$, $L^2$ and $L^{\infty}$ norms are evaluated according to \cite{Caleffi2006}. 
Results accounting for a viscoelastic vessel wall behavior are presented in Table \ref{tbl:orderofacc}. The second-order accuracy is achieved for all the evolutionary variables in both the cases, AA and VV. Similar results, not reported here for brevity, confirm the expected order of accuracy also for the elastic case.

 \revised{The \textit{dead-body} simulation is first performed with periodic BCs, initial null velocity and uniform pressure on the geometry of the 2-vessels junction case. Successively, different configurations, including junctions with multiple branches in which the joined vessels exhibit mechanical and geometrical characteristics different from each other, are considered}. Results, here omitted, confirm that neither spurious velocities nor different numerical perturbations arise during the simulation. Therefore, the model is verified to be well-balanced (or, with the same connotation, satisfies the exact conservation property, i.e.\ C-property \cite{Bermudez1994}) even when involving junctions.

\begin{figure}[t!]
\centering
\begin{subfigure}{0.32\textwidth}
\centering
\includegraphics[width=\linewidth]{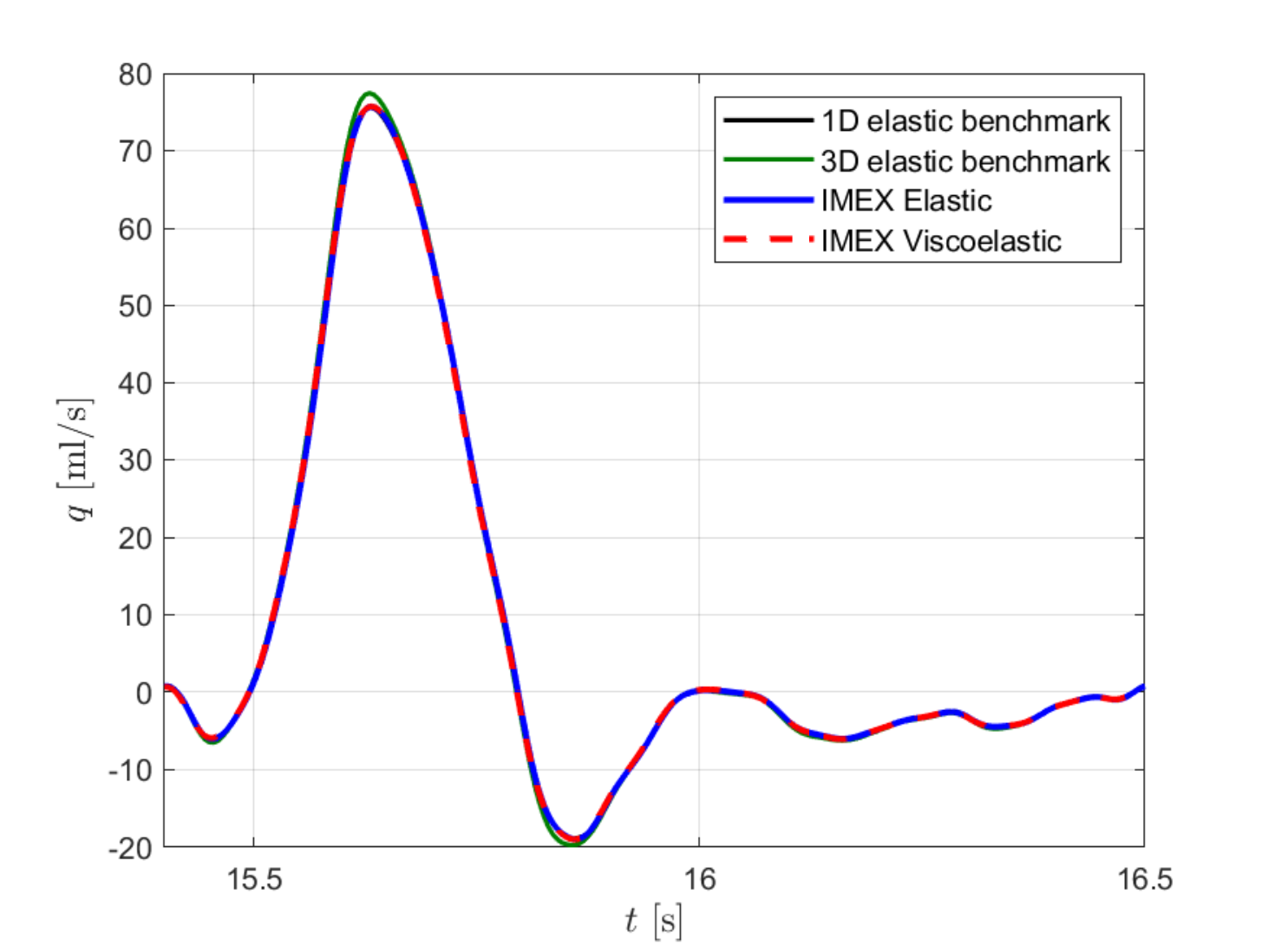}
\vspace*{ - 5mm}
\label{aorta_flowrate}
\end{subfigure}
\begin{subfigure}{0.32\textwidth}
\centering
\includegraphics[width=\linewidth]{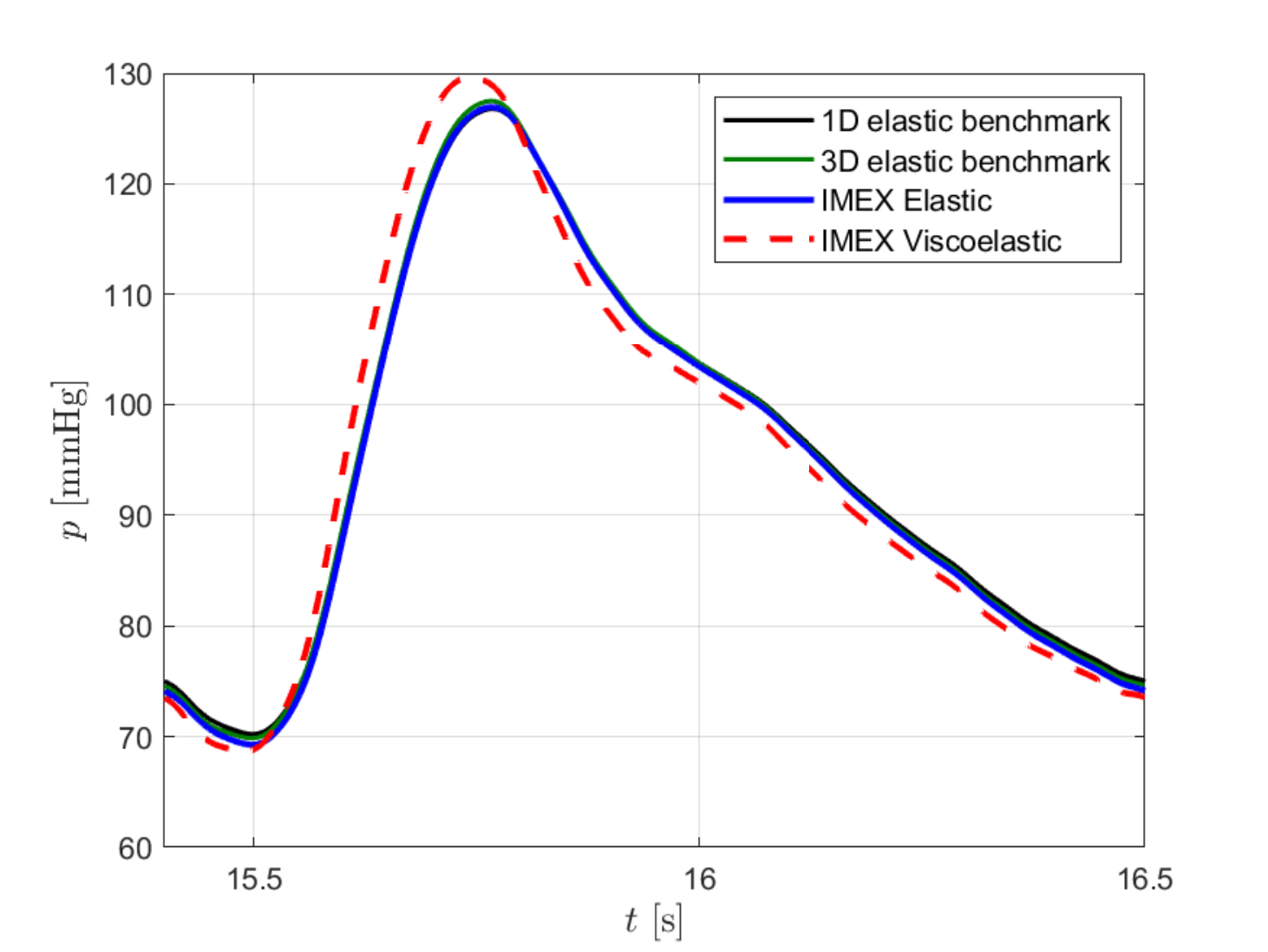}
\vspace*{ - 5mm}
\label{aorta_pressure}
\end{subfigure}
\begin{subfigure}{0.32\textwidth}
\centering
\includegraphics[width=\linewidth]{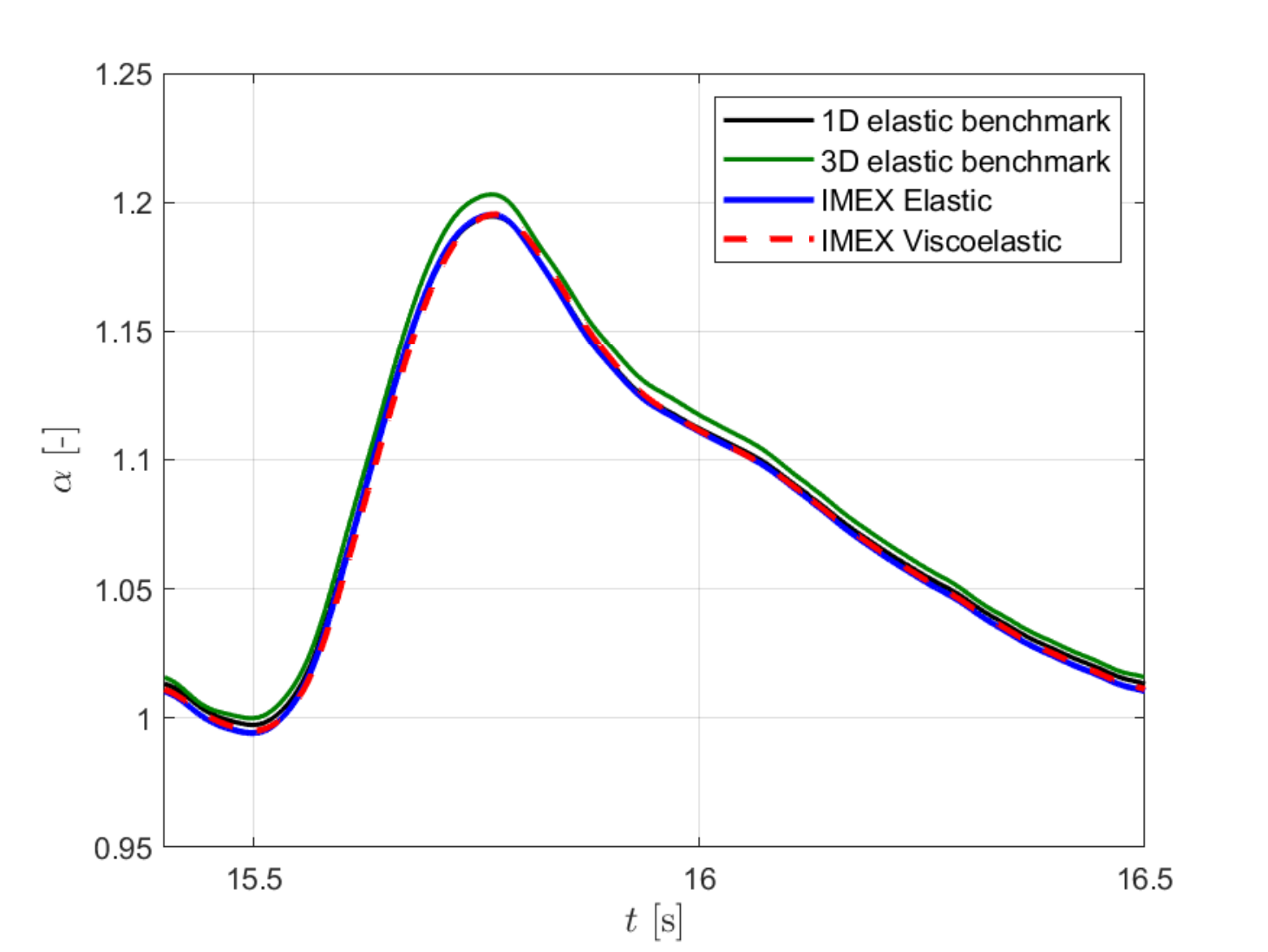}
\vspace*{ - 5mm}
\label{aorta_alpha}
\end{subfigure} 
\begin{subfigure}{0.32\textwidth}
\centering
\includegraphics[width=\linewidth]{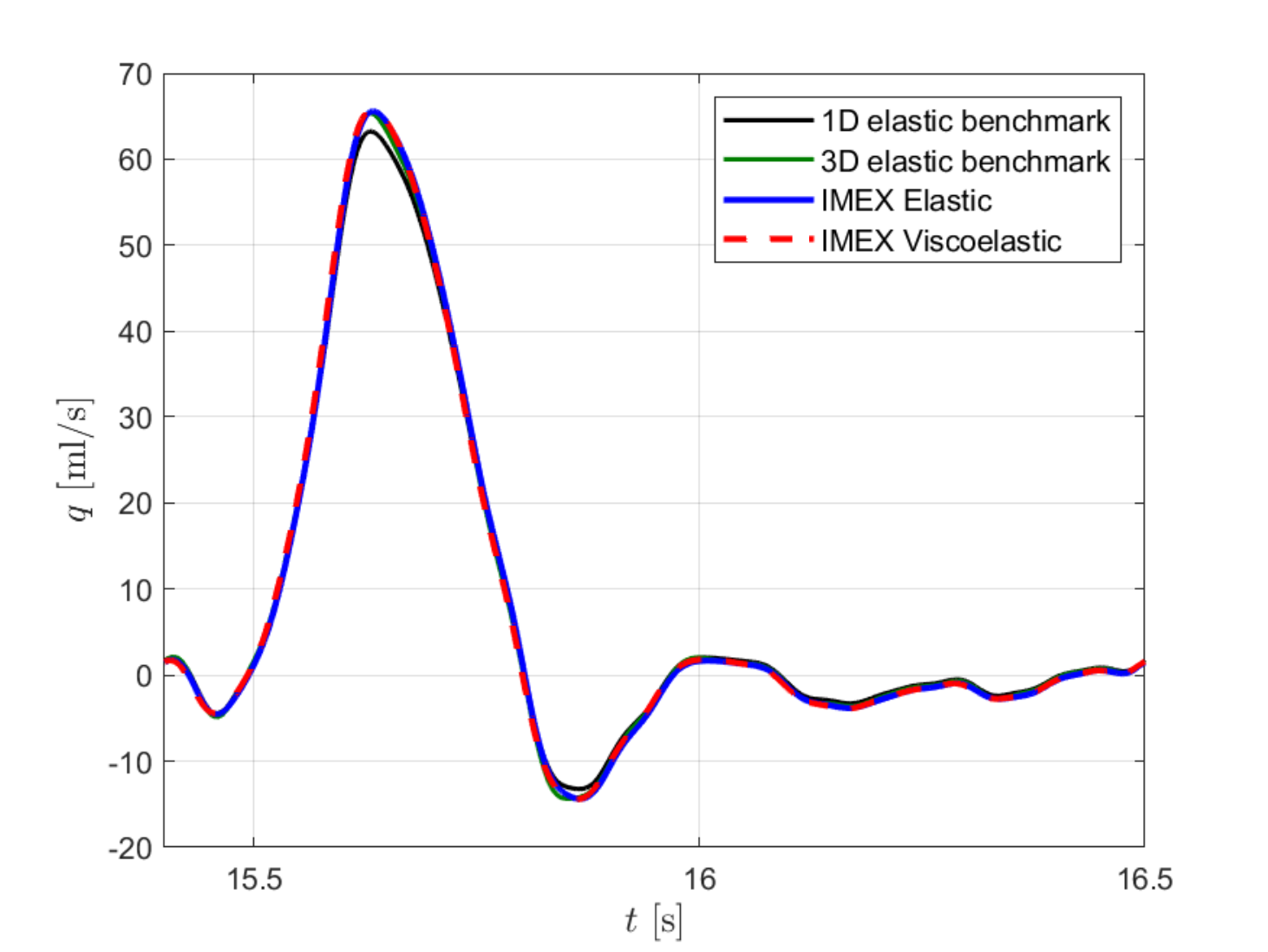}
\vspace*{ - 5mm}
\label{junc_flowrate}
\end{subfigure}
\begin{subfigure}{0.32\textwidth}
\centering
\includegraphics[width=\linewidth]{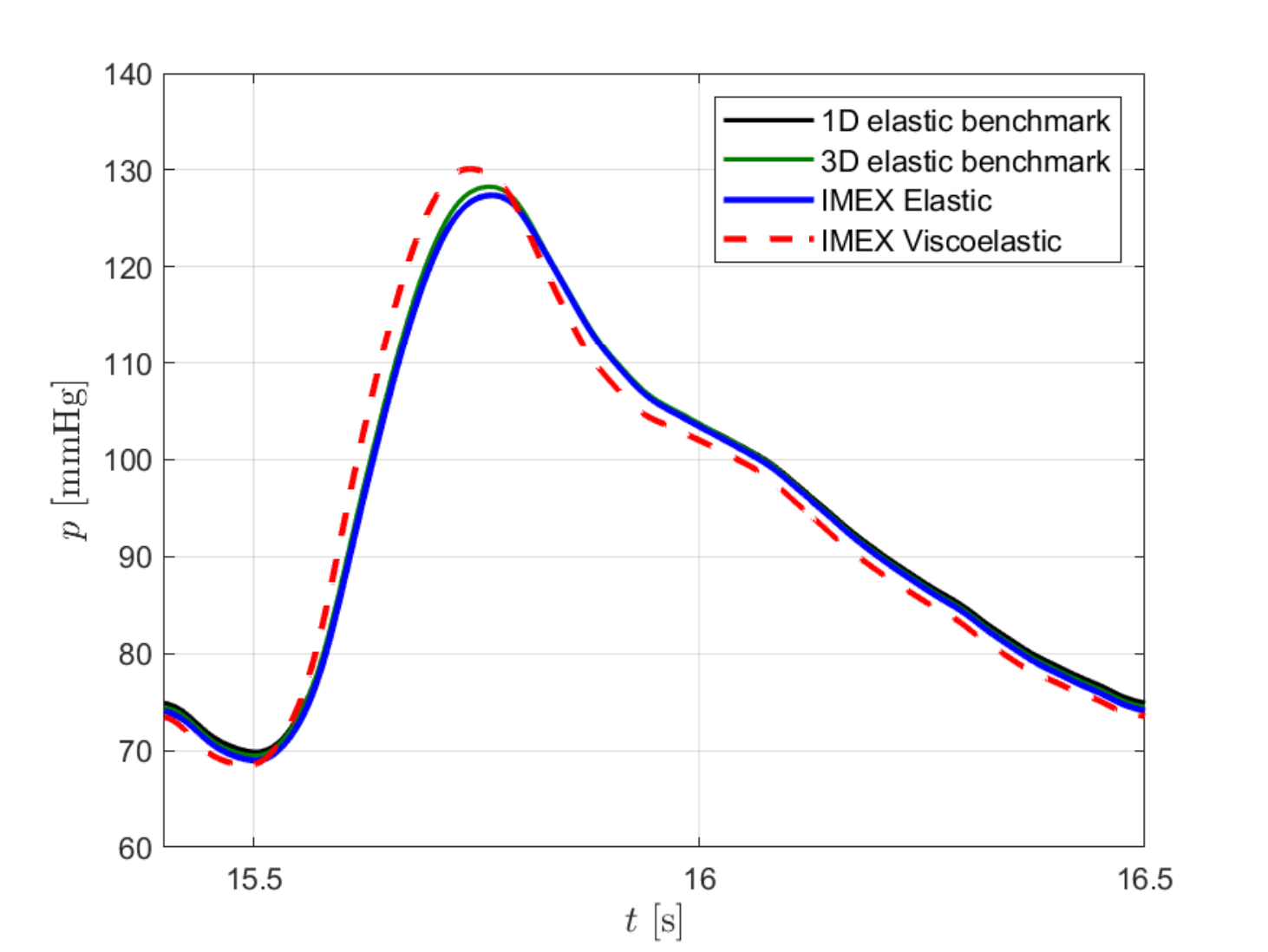}
\vspace*{ - 5mm}
\label{junc_pressure}
\end{subfigure}
\begin{subfigure}{0.32\textwidth}
\centering
\includegraphics[width=\linewidth]{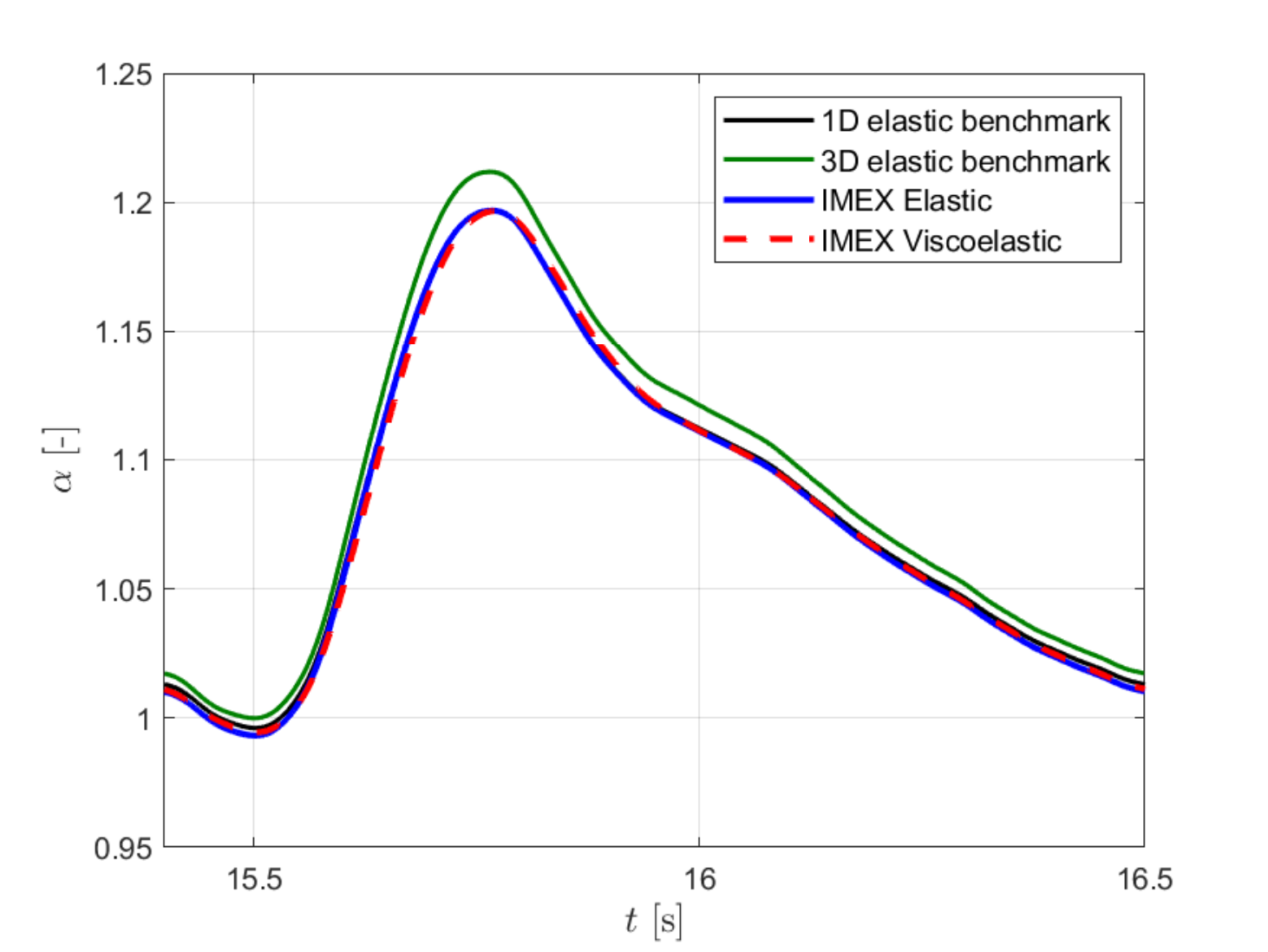}
\vspace*{ - 5mm}
\label{junc_alpha}
\end{subfigure} 
\begin{subfigure}{0.32\textwidth}
\centering
\includegraphics[width=\linewidth]{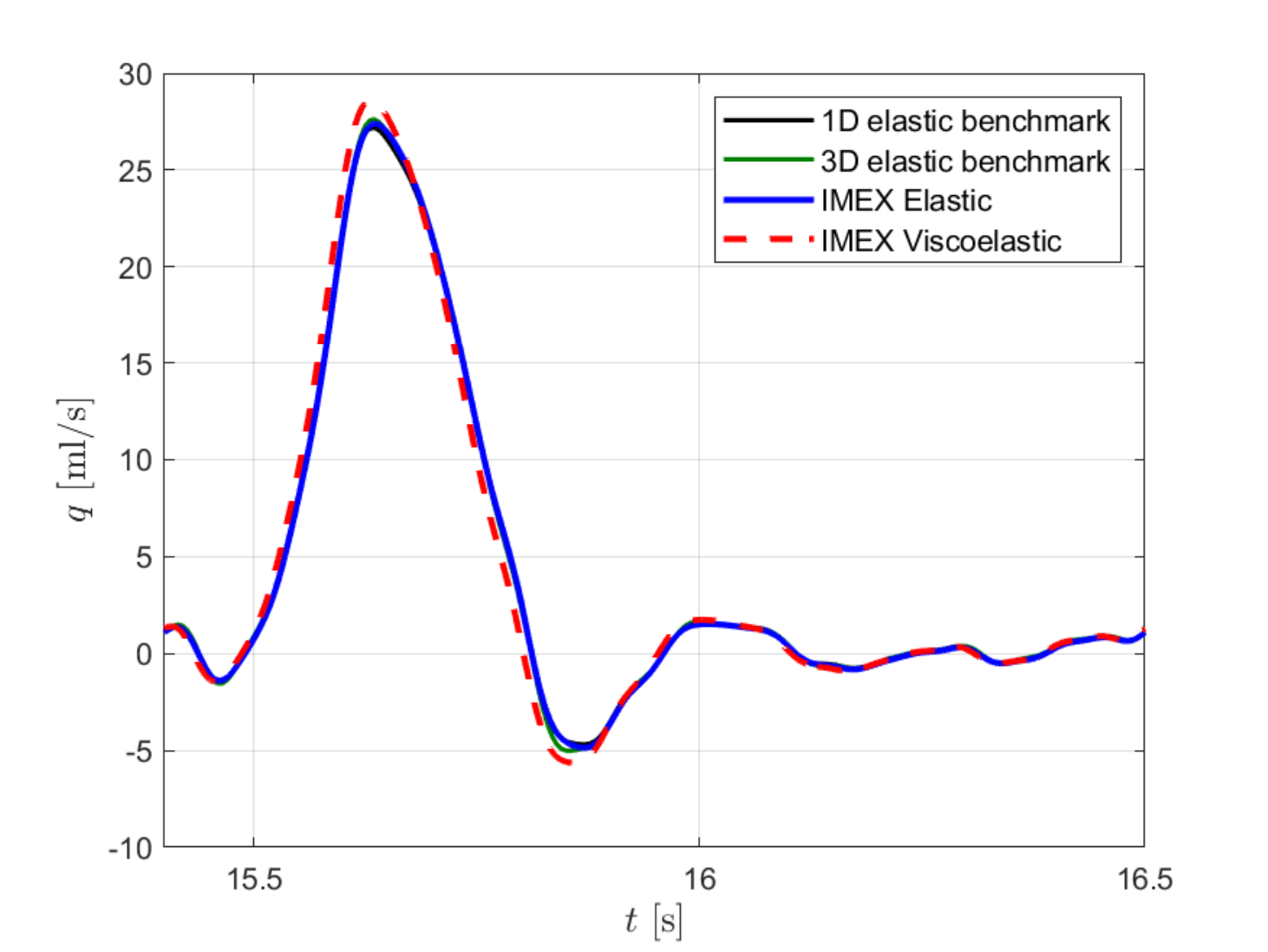}
\vspace*{ - 5mm}
\label{iliac_flowrate}
\end{subfigure}
\begin{subfigure}{0.32\textwidth}
\centering
\includegraphics[width=\linewidth]{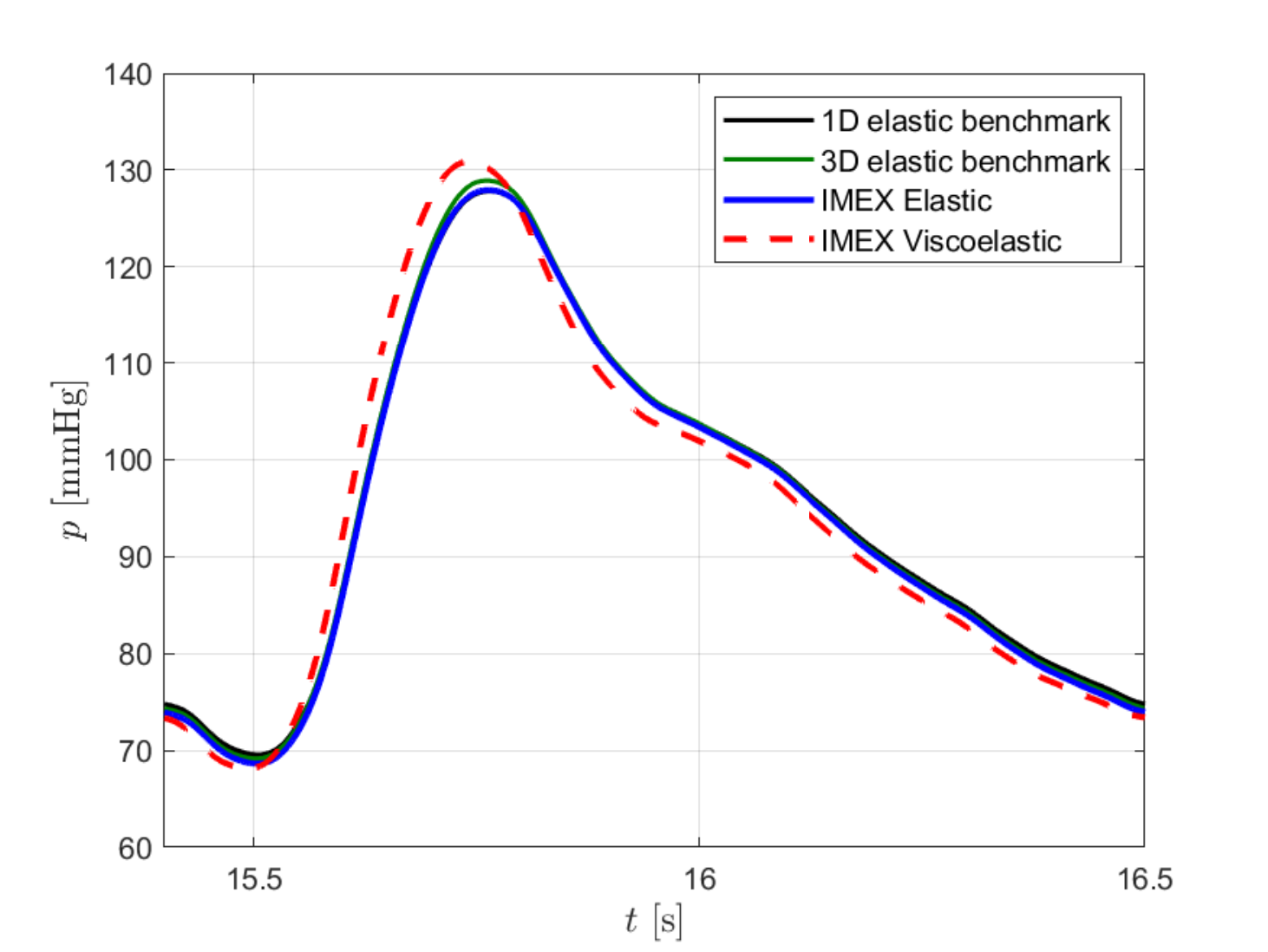}
\vspace*{ - 5mm}
\label{iliac_pressure}
\end{subfigure}
\begin{subfigure}{0.32\textwidth}
\centering
\includegraphics[width=\linewidth]{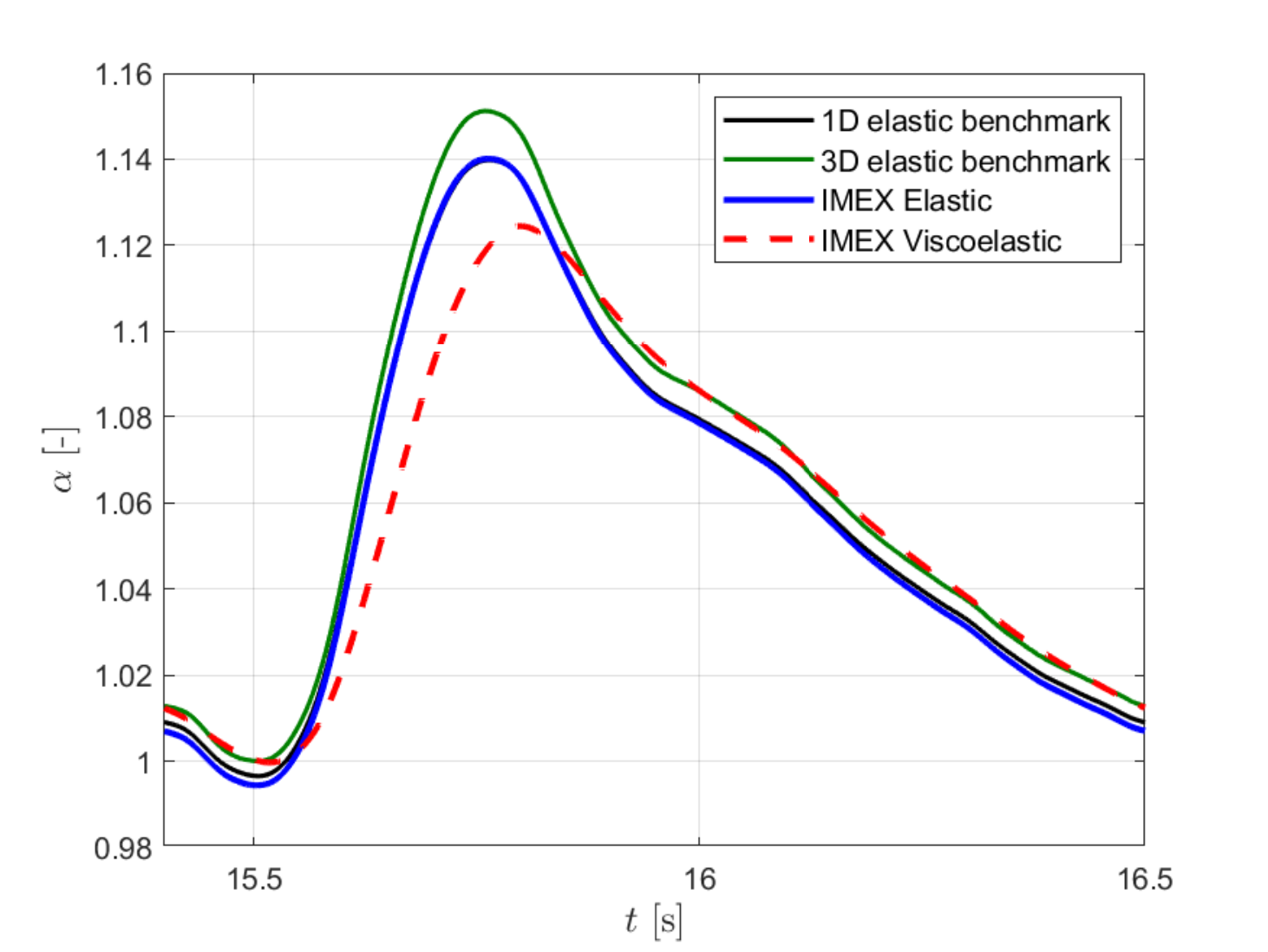}
\vspace*{ - 5mm}
\label{iliac_alpha}
\end{subfigure}
\caption{Results of the 3--vessels junction test, concerning an aortic bifurcation problem. Elastic and viscoelastic results obtained with the here proposed IMEX RK FV scheme are compared with \revised{six different 1-D and one 3-D elastic benchmark solutions} taken from \cite{Boileau2015}. Results at mid-section of the abdominal aorta (upper row), at junction section (center row) and at mid-section of the left iliac artery (lower row) are shown in terms of flow rate (left column), pressure (center column) and dimensionless cross-sectional area (right column), using $n_c= 5$ cells for each vessel. }
\label{fig:bif}
\end{figure}
\begin{figure}[t]
\begin{subfigure}{0.5\textwidth}
\centering
\includegraphics[width=1\linewidth]{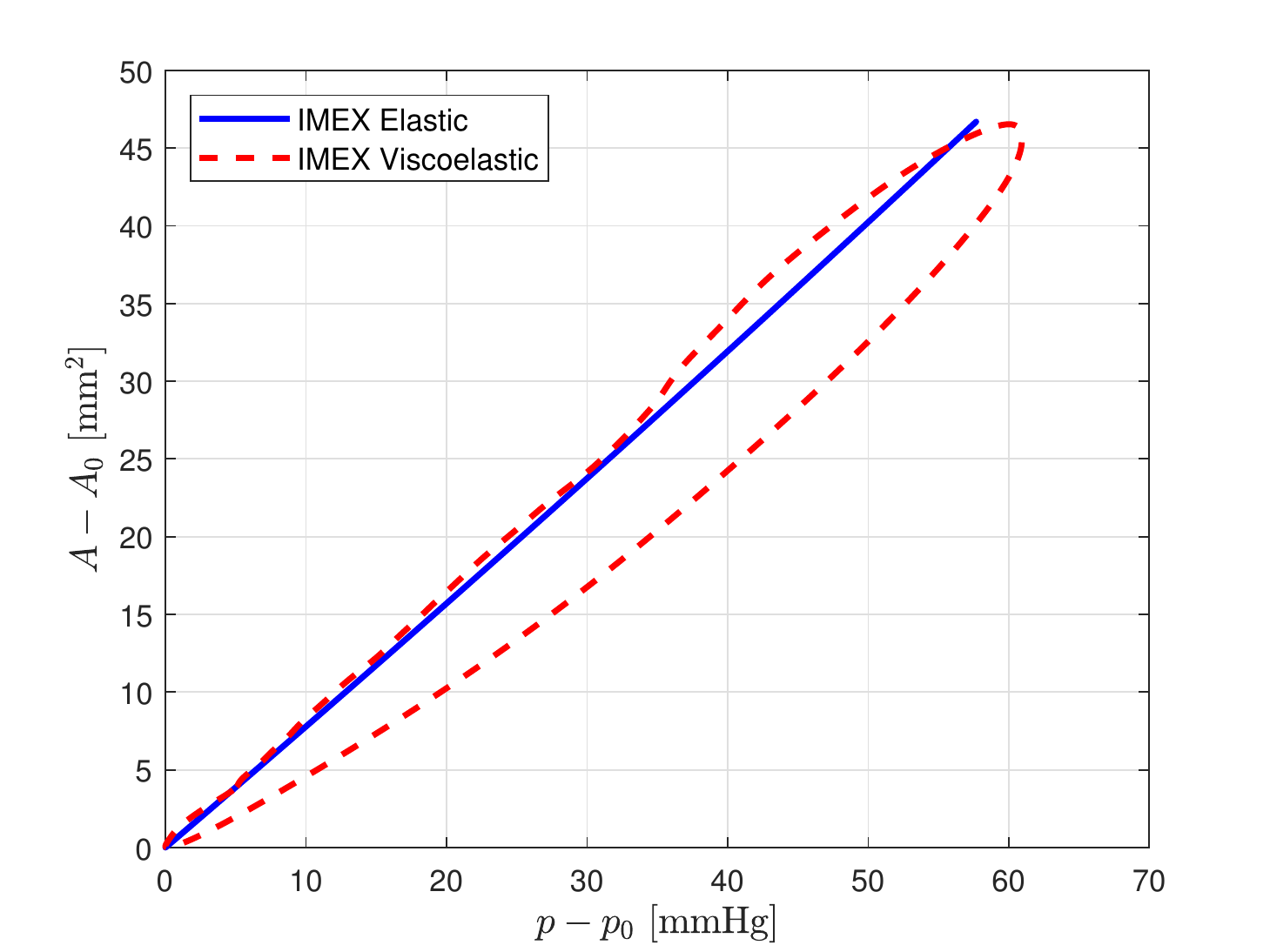}
\vspace*{ - 5mm}
\caption{Abdominal aorta}
\label{HL_AA}
\end{subfigure}
\begin{subfigure}{0.5\textwidth}
\centering
\includegraphics[width=1\linewidth]{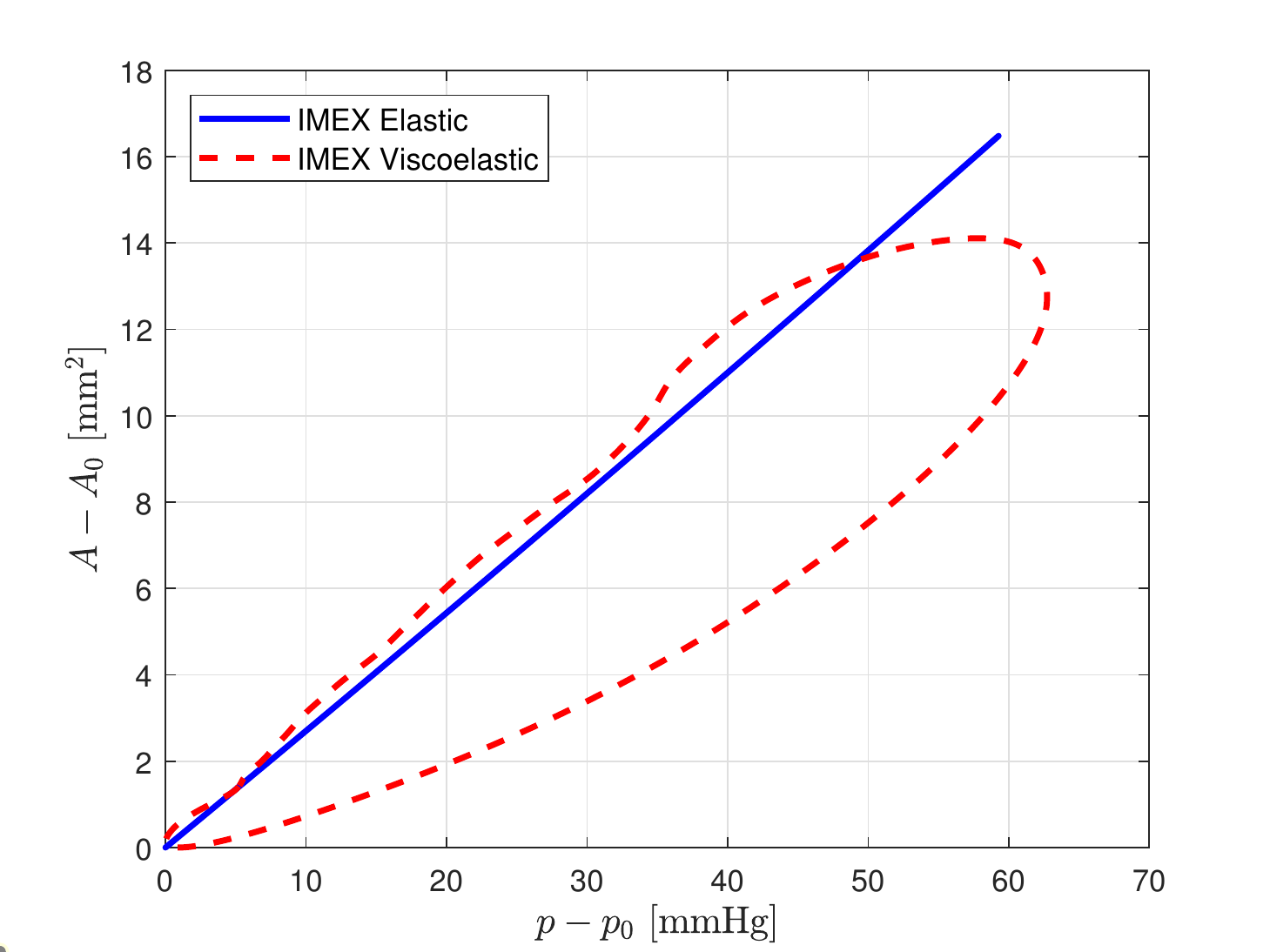}
\vspace*{ - 5mm}
\caption{Iliac artery}
\label{HL_IA}
\end{subfigure}
\caption{Hysteresis loops of the 3--vessels test, concerning an aortic bifurcation, in abdominal aorta (a) and iliac artery (b). The elastic line confirms the absence of dissipative effects, in contrast with the wide viscoelastic curve that shows energy dissipation. The loops evolve in counterclockwise. $A_0$ and $p_0$ are the equilibrium cross-sectional area and pressure, respectively, whose value coincides with that of the diastolic phase.}
\label{fig:hysteresis_loops}
\end{figure}

\subsection{3--vessels junction}
\label{sect:bif}
A 3--vessels junction test is performed accounting for a bifurcation case (3 arteries) in both the elastic and viscoelastic configuration. The model has been tested also for a venous confluence, but due to the lack of reference solutions to compare with, results are not reported in this paper.
The 3--vessels junction test consists in the bifurcation of the terminal part of the abdominal aorta (AbA) into the two iliac arteries (IAs) that perfuse the legs. Benchmark solutions, as well as geometrical and \revised{elastic} mechanical parameters, are taken from \cite{Boileau2015,Xiao2014}, to which the reader is referred. In fact, the bifurcation test proposed in \cite{Boileau2015,Xiao2014} only considers an elastic vessels wall behavior. At the inlet, a flow rate resembling the human flow rate waveform at this section of the AbA is prescribed as supplied by \cite{Murgo1980}\revised{, allowing to simulate a physiologically-based test}. Flow rate and pressure are simulated at the outlet sections via the RCR model. Resistances and compliance of the two outlets are given by \cite{Boileau2015}, as well as the pressure at the outlet of the RCR unit. It is worth to underline that the problem is symmetric, since left and right IAs have the same characteristics. 
Finally, the velocity profile is characterized by coefficients $\alpha_c = 1.1$ and $\zeta = 9$.

In this test, a comparison between the elastic and viscoelastic \revised{numerical approaches} is carried out, to investigate how much the viscous terms affect results with respect to a purely elastic bifurcation case. 
\revised{Due to the lack of reference viscoelastic parameters for this specific test, the procedure proposed in \cite{Bertaglia2020a} is considered to simulate viscoelasticity, i.e.\ to fix parameters of the SLSM. In particular:
\begin{itemize}
    \item the asymptotic Young moduli $E_{\infty}$ characterizing the three vessels considered in this numerical experiment are taken coincident with the elastic Young moduli used in the corresponding test in \cite{Boileau2015};
    \item since normally $\tau_r$ behaves almost like a biological constant \cite{Bertaglia2020a,Ghigo2017a}, it is fixed equal to the one obtained in the following discussed ADAN56 test (see Section \ref{ADAN56}) for all the three vessels.
\end{itemize}
Fixed these two parameters, $E_0$ and $\eta$ can be evaluated recurring to the following equations:
\begin{itemize}  
    \item Eq. \eqref{eq:SLSMclosingeq}, which gives a mathematical relationship among $E_0$, $E_{\infty}$ and $\tau_r$;
    \item Eq. \eqref{eq:Einf2E0} to compute the ratio $E_{\infty}/E_0$.
\end{itemize}}
This estimation procedure allows to have in this bifurcation test a viscoelastic behavior of vessels similar to that considered in the ADAN56 arterial network, further presented in Section \ref{ADAN56}. 
The reader is referred to \ref{appendixB} for the complete set of SLSM parameters of this test.

Fig.~\ref{fig:bif} shows numerical results of both elastic and viscoelastic simulations of the aortic bifurcation compared with benchmark solutions, which consist of one 3-D numerical solution and six 1-D numerical solutions performed with six different numerical methods (for further details regarding benchmark solutions see \cite{Boileau2015}). Concerning the elastic simulation, Fig.~\ref{fig:bif} demonstrates an excellent agreement between IMEX results and benchmark solutions. Hence, \revised{the capability of the proposed methodology, in the elastic formulation, to simulate in detail the multiple wave reflections generated at the bifurcation, which shape pressure, flow rate and area waveforms in the AbA and the two IAs is here confirmed. }
Regarding viscoelastic results, Fig.~\ref{fig:bif} reveals a modest viscous contribution \revised{of vessel walls, observing} a rather small difference between elastic and viscoelastic solutions.
Despite this modest difference, viscoelastic features are clearly observable in Fig.~\ref{fig:hysteresis_loops}, where hysteresis loops at the AbA mid-section and at the left IA mid-section are shown. The viscous dissipative effect can be observed comparing the wide loop of the viscoelastic simulation to the straight line obtained with the elastic configuration. Indeed, this result agrees with the one presented in \cite{Raghu2011}, where elastic-viscoelastic comparisons and consequent hysteresis loops are reported for the same aortic bifurcation test (the reader is referred to the \textit{Rest} case). Also in \cite{Raghu2011} there is no evident discrepancy between waveforms obtained with the two different mechanical behaviors of the vessel wall. Therefore, we can conclude that the IMEX FV method here proposed is well-conceived and robust enough to deal even with viscoelastic junction cases. 
\subsection{Human arterial networks}
\label{sect:network}
In this section, results obtained simulating two different networks of the largest central systemic arteries of the human vascular system are presented and discussed. The first network simulates an \textit{in-vitro} model composed of 37 main arteries, labeled as AN37 hereinafter, for which experimental measurements of flow rate and pressure are available \cite{Alastruey2011a,Matthys2007a}. The second network is a reduced version of the anatomically detailed arterial system, composed of the 56 largest arteries of the human arterial circulation, hence labeled as ADAN56 as in \cite{Blanco2014,Blanco2015}. For both networks, the simulated waveforms of the system variables are compared, for selected arteries, to reference solutions (experimental and numerical elastic benchmark) taken from \cite{Boileau2015}. Moreover, a viscoelastic simulation is carried out for the same two networks to assess how the SLSM \revised{application} affects the results. 
{All SLSM viscoelastic parameters, estimated following \cite{Bertaglia2020a}, are reported in \ref{appendixB} for both networks.}

\begin{figure}[t!]
\begin{subfigure}{1.0\textwidth}
\centering
\includegraphics[width=1\linewidth]{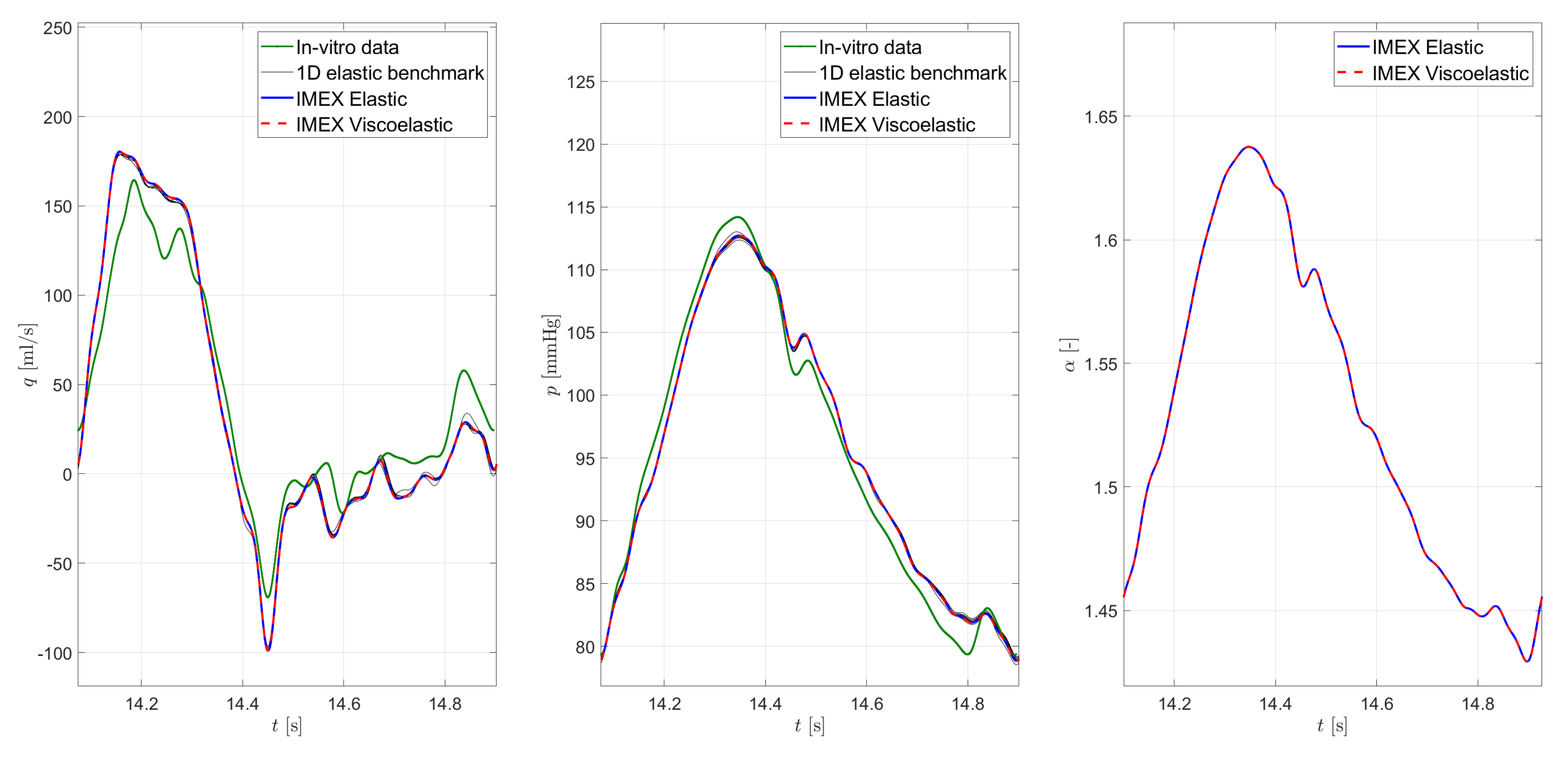}
\vspace*{ - 8mm}
\caption{Aortic arc II}
\label{37_AAII}
\end{subfigure}
\begin{subfigure}{1.0\textwidth}
\centering
\includegraphics[width=1\linewidth]{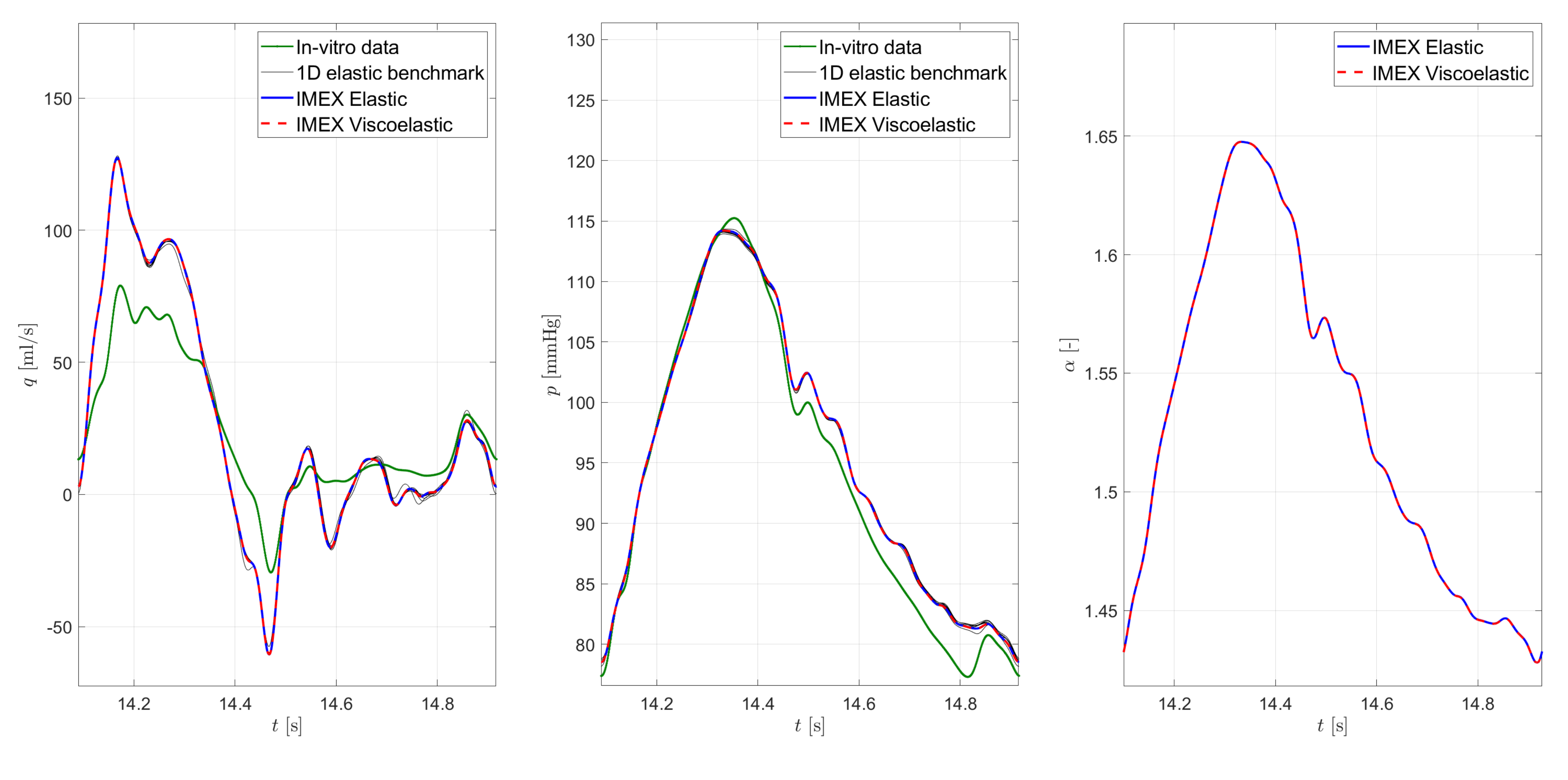}
\vspace*{ - 8mm}
\caption{Thoracic aorta II}
\label{37_TAII}
\end{subfigure}
\caption{Results of the AN37 network for 2 selected arteries. Results are obtained with the IMEX RK FV scheme considering both the elastic and the viscoelastic tube law to characterize the mechanical behavior of vessel walls. Comparison with \textit{in-vitro} experimental data and \revised{six different} numerical elastic benchmark are presented in terms of flow rate (left column) and pressure (center column), while for the dimensionless cross-sectional area (right column) reference results are not available.}
\label{fig:37AN_1}
\end{figure}
\begin{figure}[t!]
\begin{subfigure}{1.0\textwidth}
\centering
\includegraphics[width=1\linewidth]{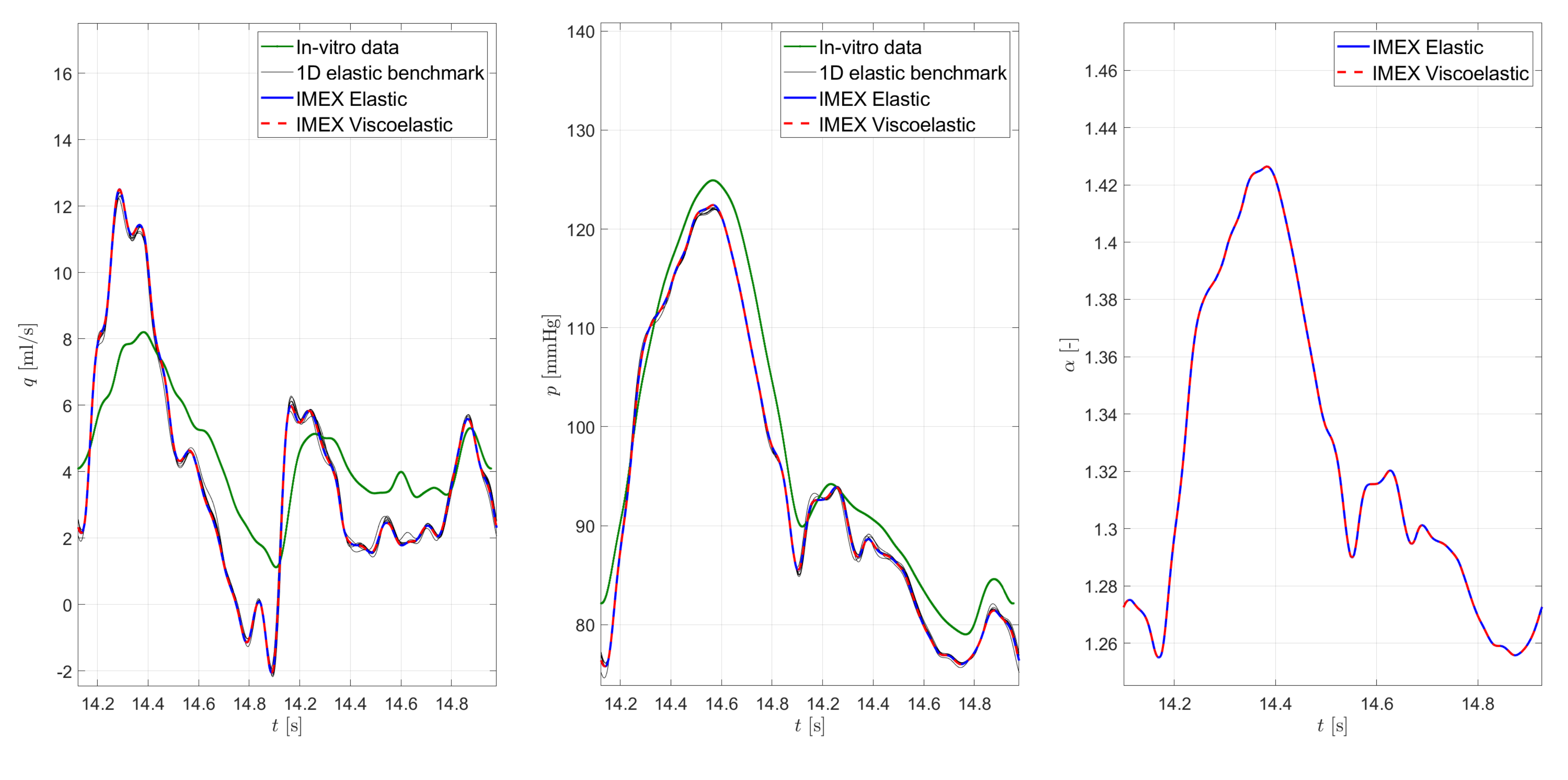}
\vspace*{ - 8mm}
\caption{Right iliac femoral artery II}
\label{37_IFRII}
\end{subfigure}
\begin{subfigure}{1.0\textwidth}
\centering
\includegraphics[width=1\linewidth]{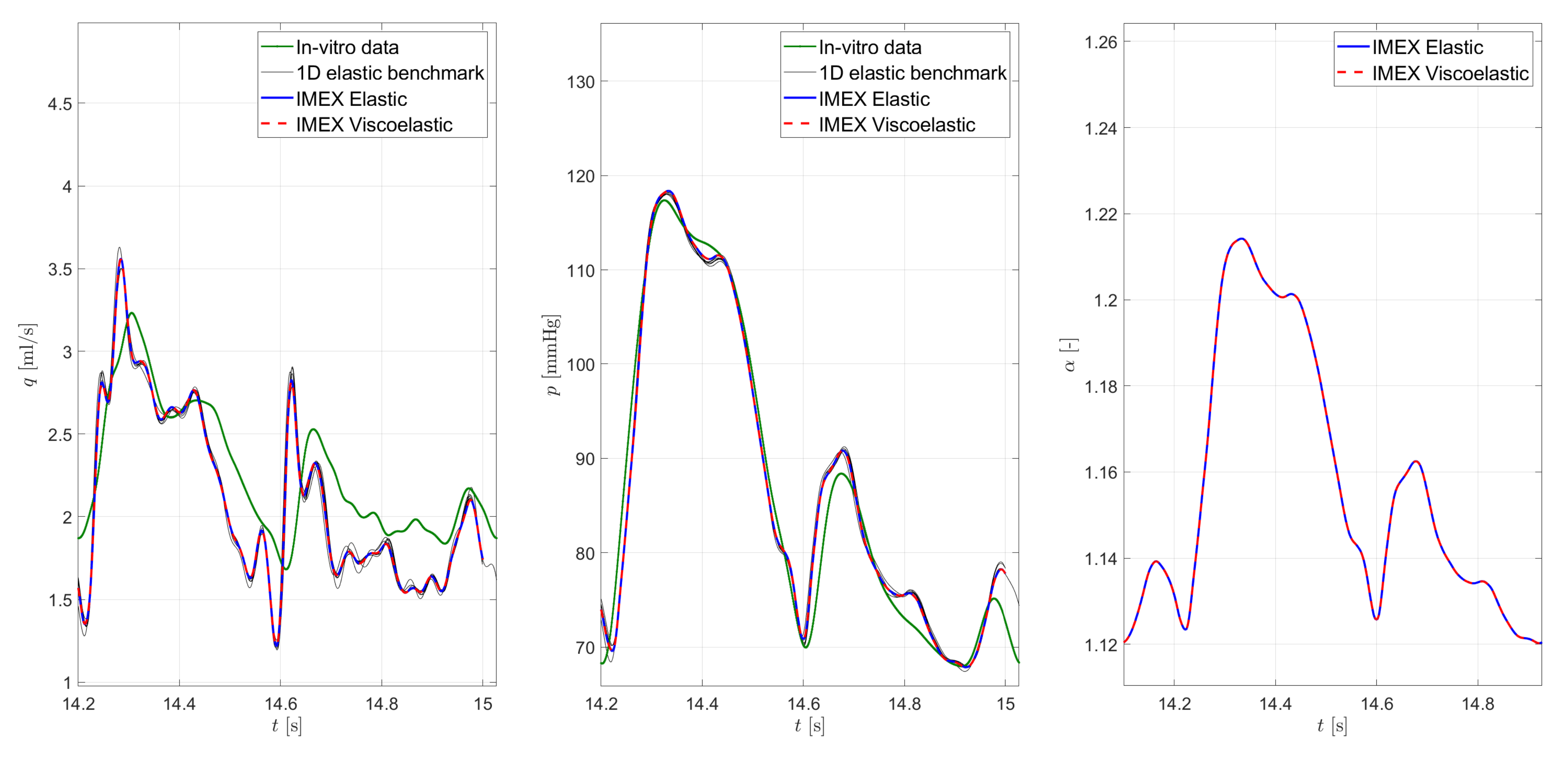}
\vspace*{ - 8mm}
\caption{Right anterior tibial artery}
\label{37_ATR}
\end{subfigure}
\caption{Results of the AN37 network for 2 selected arteries. Results are obtained with the IMEX RK FV scheme considering both the elastic and the viscoelastic tube law to characterize the mechanical behavior of vessel walls. Comparison with \textit{in-vitro} experimental data and \revised{six different} numerical elastic benchmark are presented in terms of flow rate (left column) and pressure (center column), while for the dimensionless cross-sectional area (right column) reference results are not available.}
\label{fig:37AN_2}
\end{figure}
\begin{figure}[t!]
\begin{subfigure}{1.0\textwidth}
\centering
\includegraphics[width=1\linewidth]{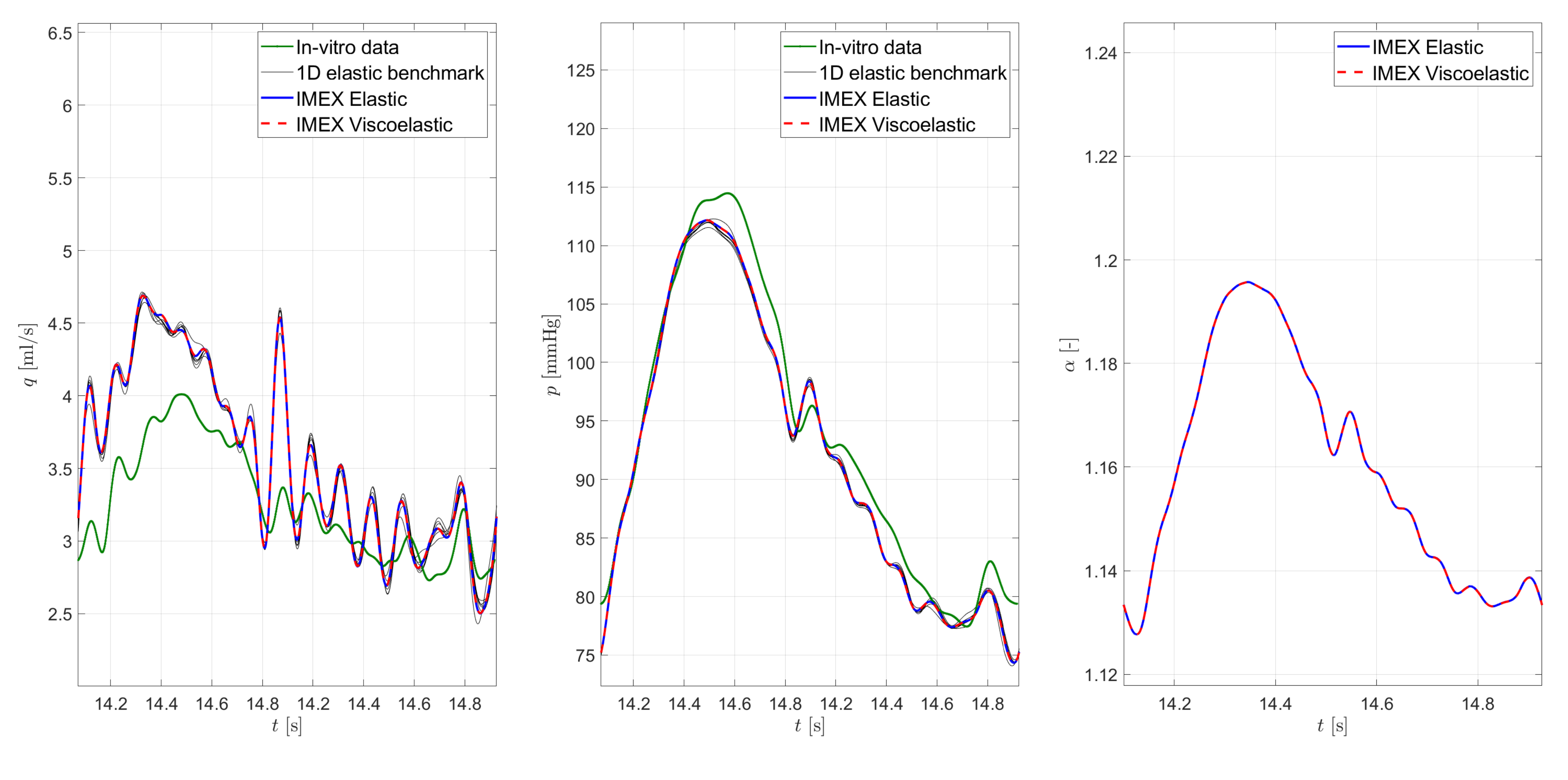}
\vspace*{ - 8mm}
\caption{Right ulnar artery}
\label{37_UR}
\end{subfigure}
\begin{subfigure}{1.0\textwidth}
\centering
\includegraphics[width=1\linewidth]{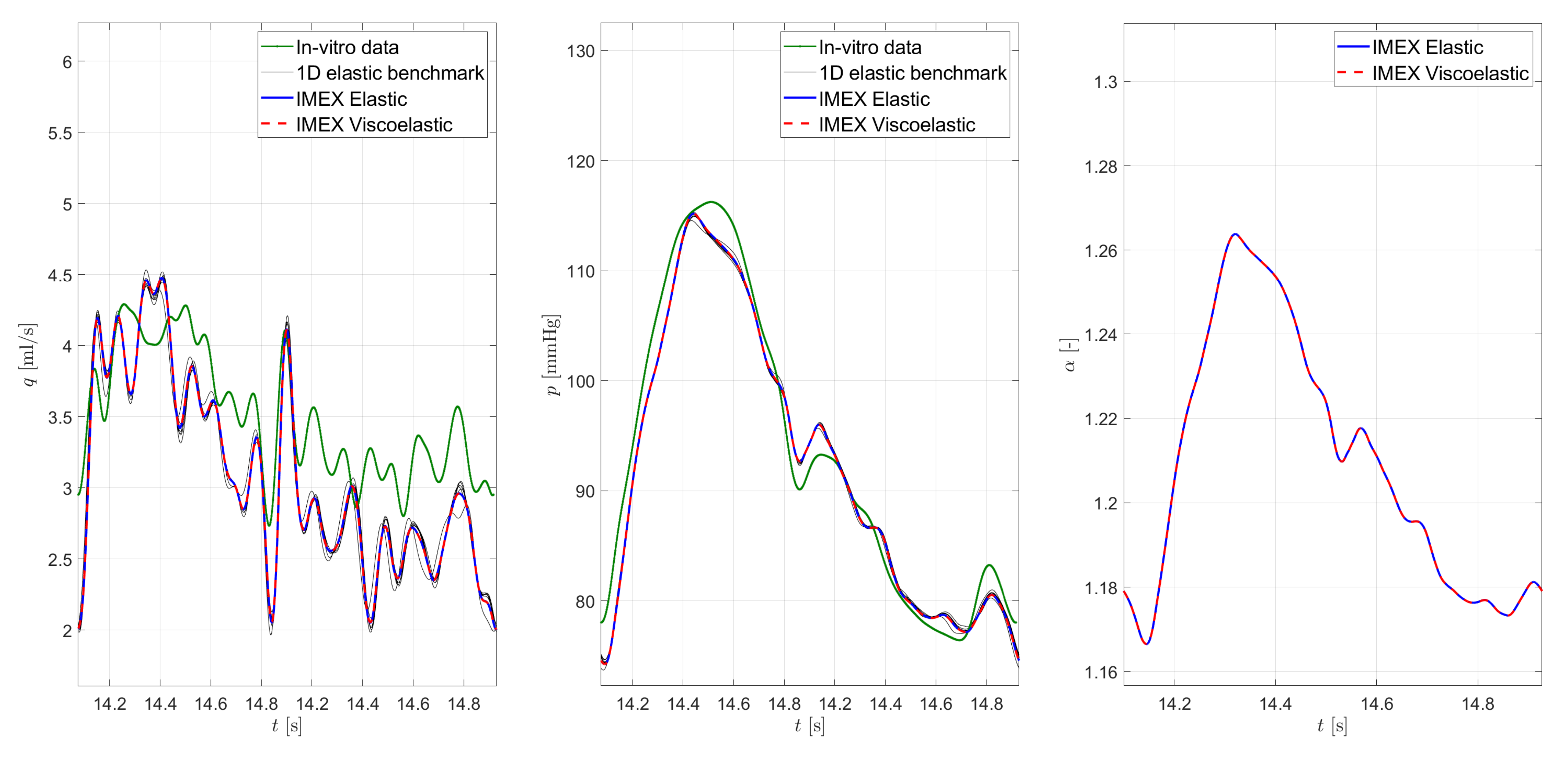}
\vspace*{ - 8mm}
\caption{Left ulnar artery}
\label{37_UL}
\end{subfigure}
\caption{Results of the AN37 network for 2 selected arteries. Results are obtained with the IMEX RK FV scheme considering both the elastic and the viscoelastic tube law to characterize the mechanical behavior of vessel walls. Comparison with \textit{in-vitro} experimental data and \revised{six different} numerical elastic benchmark are presented in terms of flow rate (left column) and pressure (center column), while for the dimensionless cross-sectional area (right column) reference results are not available.}
\label{fig:37AN_3}
\end{figure}
\begin{figure}[t!]
\begin{subfigure}{1.0\textwidth}
\centering
\includegraphics[width=1\linewidth]{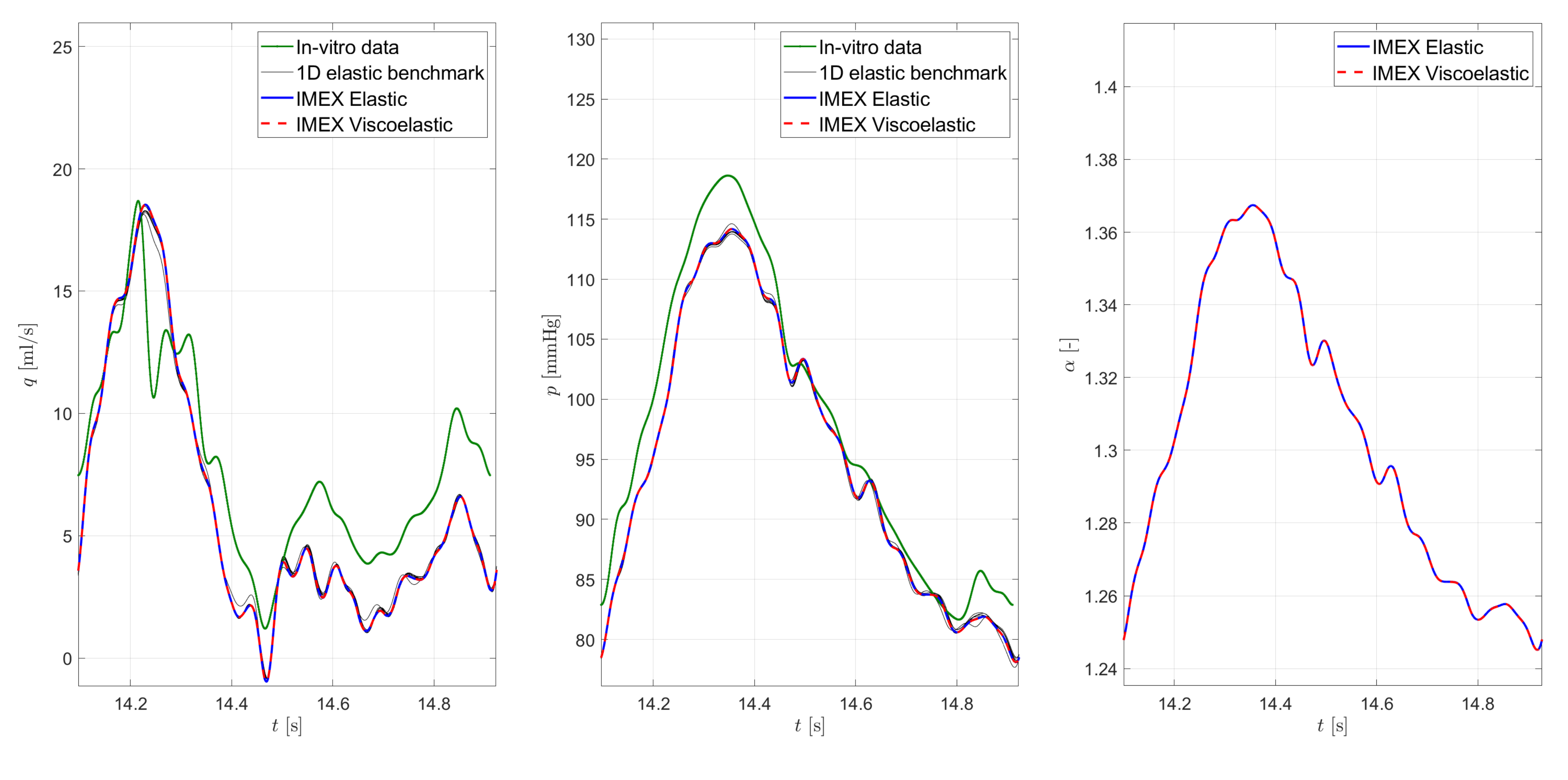}
\vspace*{ - 8mm}
\caption{Left subclavian artery I}
\label{37_SLI}
\end{subfigure}
\begin{subfigure}{1.0\textwidth}
\centering
\includegraphics[width=1\linewidth]{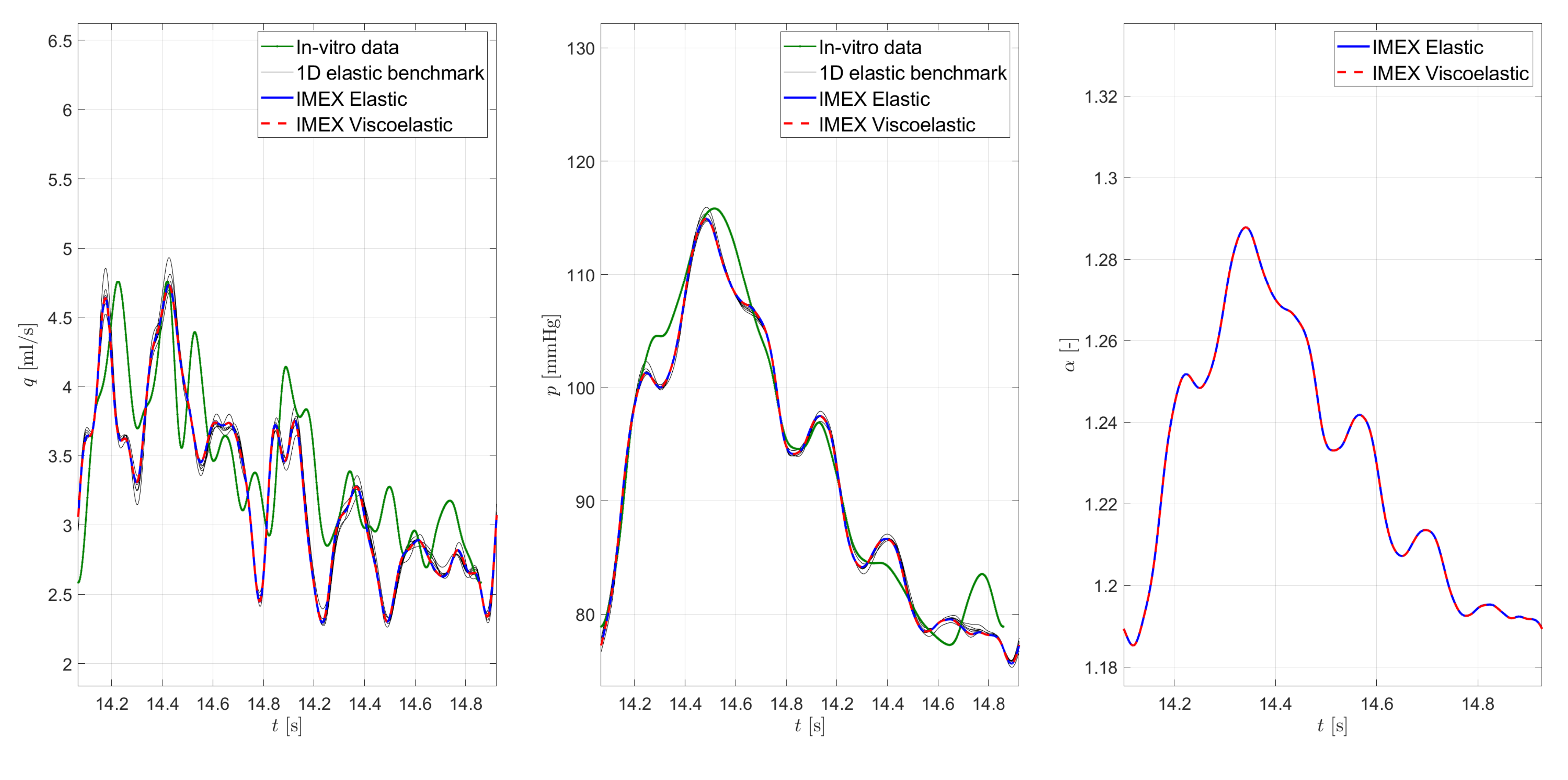}
\vspace*{ - 8mm}
\caption{Splenic artery}
\label{37_S}
\end{subfigure}
\caption{Results of the AN37 network for 2 selected arteries. Results are obtained with the IMEX RK FV scheme considering both the elastic and the viscoelastic tube law to characterize the mechanical behavior of vessel walls. Comparison with \textit{in-vitro} experimental data and \revised{six different} numerical elastic benchmark are presented in terms of flow rate (left column) and pressure (center column), while for the dimensionless cross-sectional area (right column) reference results are not available.}
\label{fig:37AN_4}
\end{figure}
\begin{figure}
\centering
\includegraphics[width=.5\linewidth]{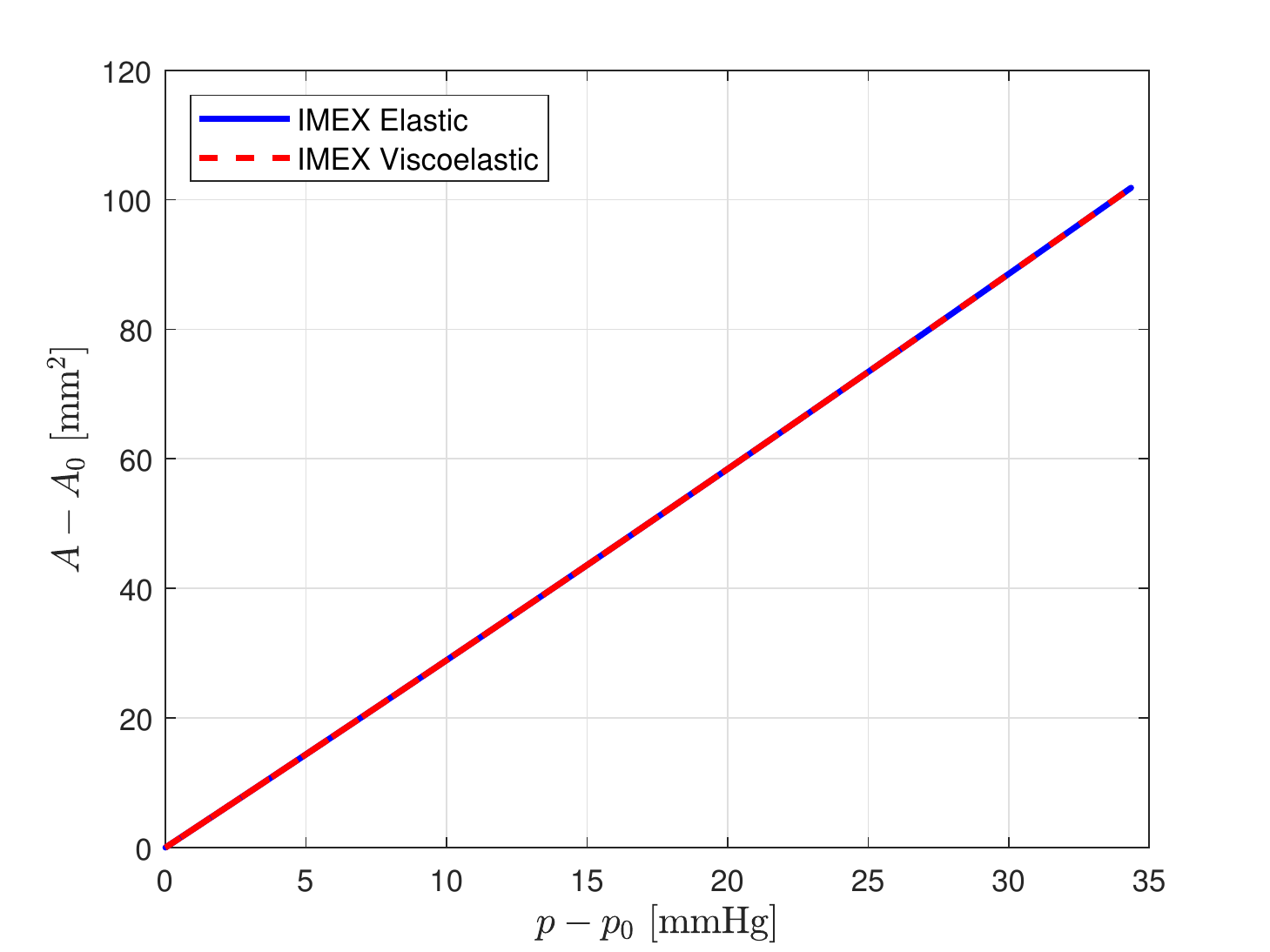}
\caption{Hysteresis loop obtained for the aortic arch II in the AN37 network. The viscoelastic loop \revised{essentially} coincides with the elastic line because of the parameters of the SLSM fixed for this specific test. Since $\tau_r \approx \mathcal{O}(10^{-4}) \to 0$, the material does not have enough time to develop its viscoelastic response. $A_0$ and $p_0$ are the equilibrium cross-sectional area and pressure, respectively, whose value coincides with those of the diastolic phase.}
\label{fig:37AN_HL}
\end{figure}

\subsubsection{AN37}
\label{37AN}
The AN37 arterial tree recalls the one presented in \cite{Alastruey2011a,Matthys2007a}, for which \textit{in-vitro} pressure and flow rate measurements were acquired at multiple locations. The network is composed by 37 silicon vessels representing the largest central systemic arteries of the human vascular system. At the inlet of the ascending aorta, the condition $q_{\mathrm{in}}(t)$ is imposed, which corresponds to the measured \textit{in-vitro} flow rate, based on a human cardiac output. Downstream BC is simulated coupling to each 1-D peripheral vessels a single-resistance ($R$) lumped-parameter model. All these $R$ boundary parameters and the inlet flow rate are supplied by \cite{Boileau2015}. The network is simulated through the a-FSI blood flow model solved with the AP-IMEX RK FV scheme, with initial conditions such that $A_{IC} = A_0$, $p_{IC} = p_{ext}=0$ and $u_{IC} = 0$. 
The reader is referred to \cite{Alastruey2011a,Boileau2015,Matthys2007a} for the topological, geometrical and mechanical parameters of the AN37 network. \revised{To simulate the viscoelastic case, for consistency with the corresponding elastic case, $E_{\infty}$ is set equal to the elastic Young modulus given in the references,} $\eta$ is gathered, for every silicon tube, from \cite{Alastruey2011a} (from which experimental results are also taken for comparison), and finally $E_0$ is computed with \revised{Eq. \eqref{eq:Einf2E0}}. The velocity profile is characterized by a coefficient $\alpha_c = 1.1$, hence $\zeta = 9$. \revised{The spatial discretization is performed setting the cell width $\Delta x$ = 5 mm, imposing that at least two cells are used to discretize the shortest vessels.}

Comparisons between numerical results, elastic benchmark and experimental data, in terms of pressure and flow rate, are presented in Figs.~\ref{fig:37AN_1}--\ref{fig:37AN_4} for selected vessels, as well as numerical results in terms of dimensionless cross-sectional area, for which a reference solution is not available in literature. The simulation is run for 15 heartbeats in order to \revised{converge to a periodic state}, but only the last cardiac cycle is shown. 
{It can be observed that IMEX numerical results are in \revised{very good} agreement with elastic benchmarks, which consist in six 1-D numerical results obtained with different methods \cite{Boileau2015}. \revised{Moreover, IMEX results} correctly captures the main features of flow rate and pressure experimental data at the 8 investigated arterial sites, \revised{suggesting that the proposed methodology permits} to replicate the behavior of an \textit{in-vitro} arterial tree, considering the uncertainties in the experimental measurements and the simplifications of the \revised{1-D} formulation, which cause the observable discrepancies.}

As visible from Figs.~\ref{fig:37AN_1}--\ref{fig:37AN_4}, there is no significant distinction between elastic and viscoelastic results. The viscoelastic contribution appears weak in this simulation because\revised{, with the fixed SLSM parameters,} the relaxation time results very small: $\tau_r \approx \mathcal{O}(10^{-4})~ \mathrm{s}$, \revised{thus approaching} the elastic equilibrium limit ($\tau_r \to 0$), as discussed in Section \ref{sect:asymptotic_lim}. 
This is consistent with Fig.~\ref{fig:37AN_HL}, where the hysteresis loop of the second tract of the aortic arch (aortic arch II) is reported as an example. It can be noticed that the loop is extremely narrow, basically coincident with the elastic line. The instantaneous Young modulus $E_0$ is the slope of the systolic tract of the loop, namely the loop lower path. For $\tau_r \to 0$, $E(t) \to E_{\infty}$, hence coinciding, even from the very first instants, with the slope of the diastolic tract, $E_{\infty}$, which is the loop upper path. Therefore, consistently with the asymptotic limit of the model, with this setting of viscoelastic parameters the loop remains totally thin.

It can be questioned if the model actually \revised{reproduces} the viscoelastic contribution. The answer is positive and it can be verified computing the specific damping parameters as presented in \cite{Wang2012}, which results to be greater than zero, although with a small order of magnitude, $\mathcal{O}(10^{-4})$. \revised{We highlight once more that} the reason behind these results lays in the specific viscoelastic parameters fixed \revised{for the silicon vessels that constitute the AN37 network. To make differences between viscoelastic and elastic results evident in this test, we should have increased the vessel wall viscosity $\eta$ (and therefore the relaxation time $\tau_r$) with respect to the value given in the reference \cite{Alastruey2011a}. However, for consistency with the rest of this work, the parameters given in literature are not changed}.

It is here stressed that the SLSM, besides being more accurate than other simpler viscoelastic models, is defined by three viscoelastic parameters, hence presenting an additional degree of freedom with respect, for instance, to the widely used KV model. \revised{This leads the SLSM to define a richer variety of dynamics and of configurations that permit the convergence towards the asymptotic elastic limit.} In this context, it is neither trivial nor predictable to have a viscoelastic waveform that appears clear and distinct from the elastic corresponding case. These three parameters cannot be considered separately but it is their interaction that truly affects the vessel wall behavior.

\begin{figure}[t!]
\begin{subfigure}{1.0\textwidth}
\centering
\includegraphics[width=1\linewidth]{ 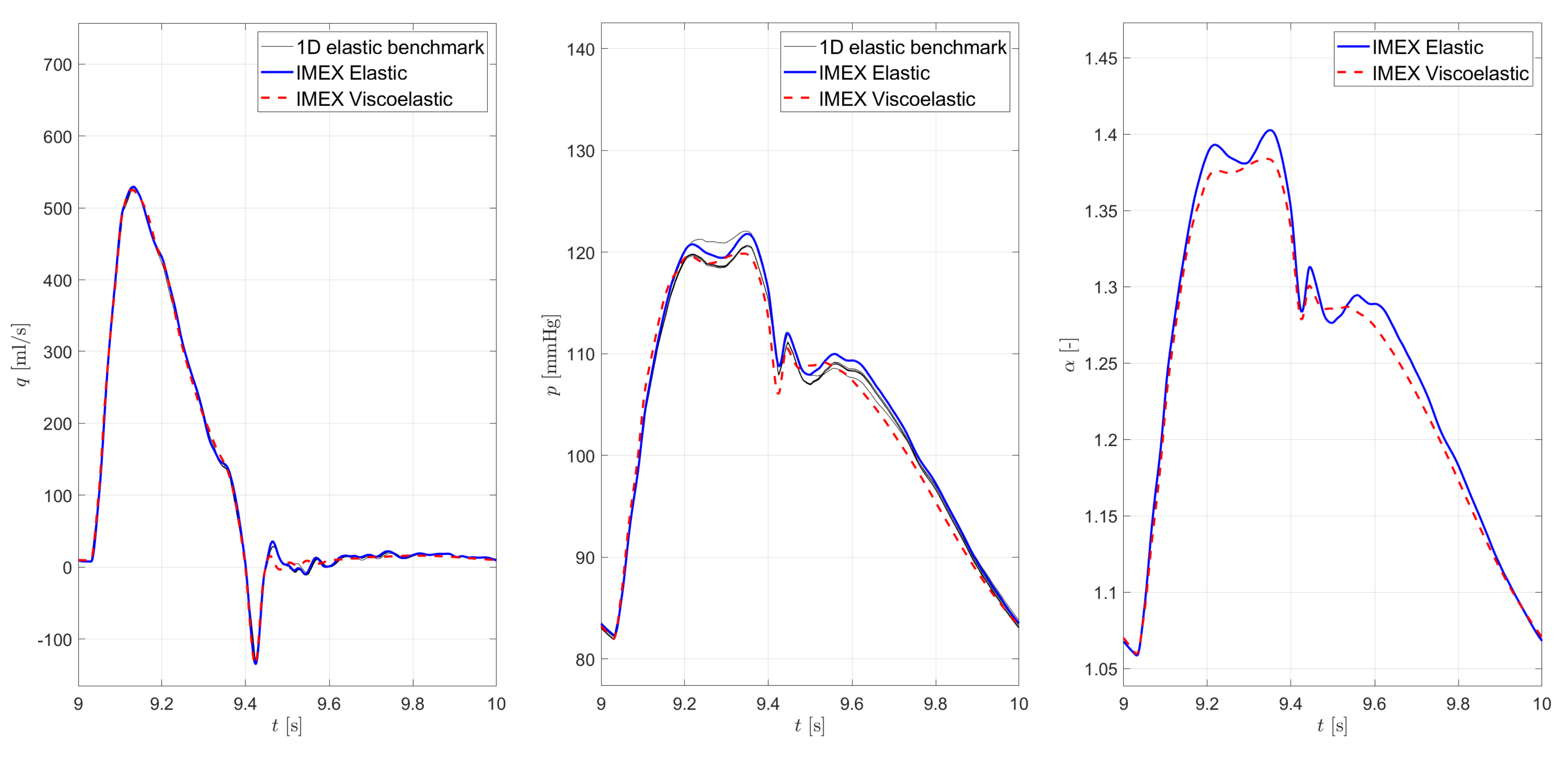}
\vspace*{ - 8mm}
\caption{Aortic arc I}
\label{56_AAI}
\end{subfigure}
\begin{subfigure}{1.0\textwidth}
\centering
\includegraphics[width=1\linewidth]{ 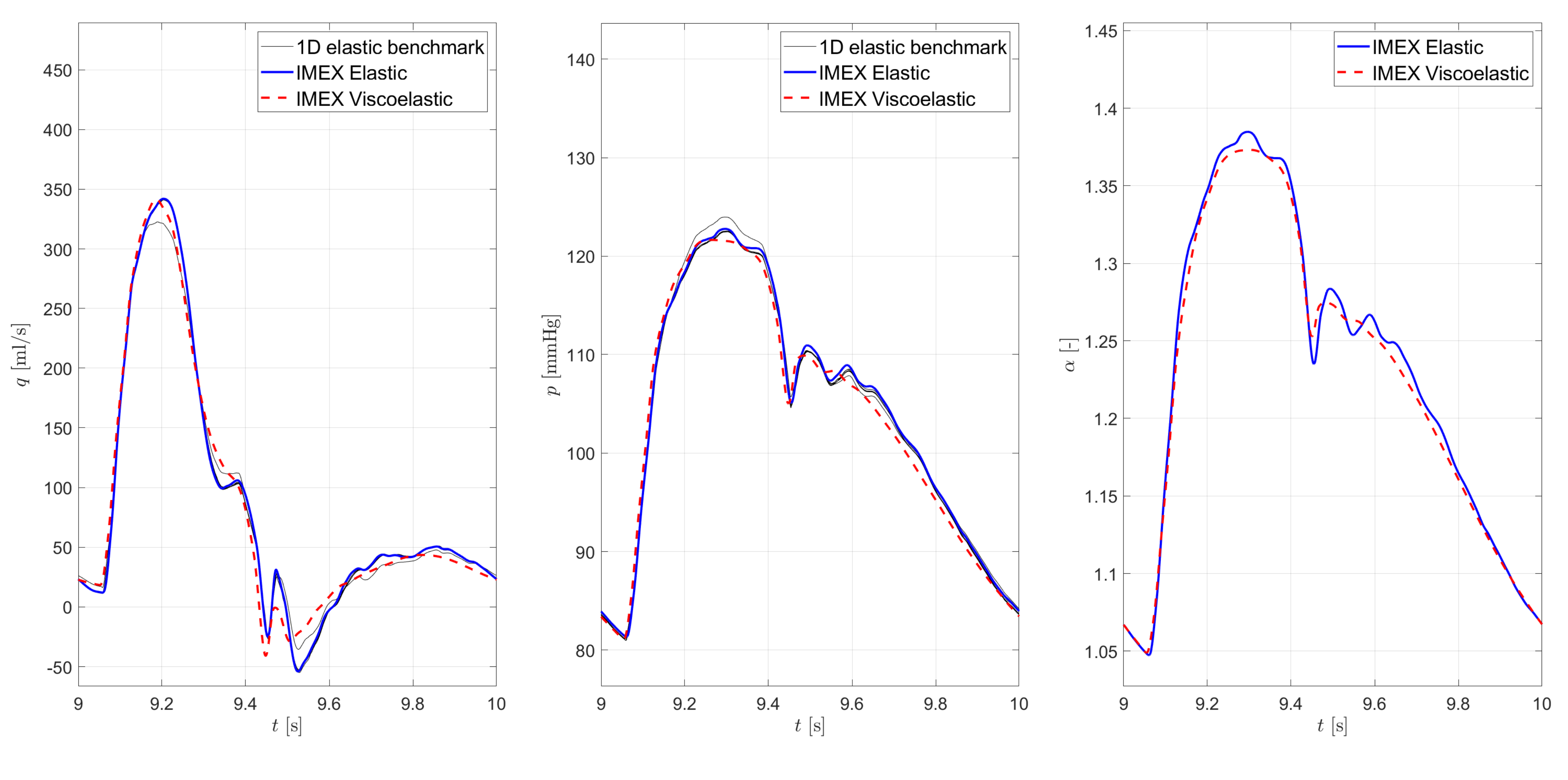}
\vspace*{ - 8mm}
\caption{Thoracic aorta III}
\label{56_TAIII}
\end{subfigure}
\caption{Results of the ADAN56 network for 2 selected arteries. Results are obtained with the IMEX RK FV scheme considering both the elastic and the viscoelastic tube law to characterize the mechanical behavior of vessel walls. Comparison with \revised{six different} 1-D elastic benchmark solutions \revised{\cite{Boileau2015}} are presented in terms of flow rate (left column) and pressure (center column), while for the dimensionless cross-sectional area (right column) reference results are not available.}
\label{fig:56AN_1}
\end{figure}
\begin{figure}[t!]
\begin{subfigure}{1.0\textwidth}
\centering
\includegraphics[width=1\linewidth]{ 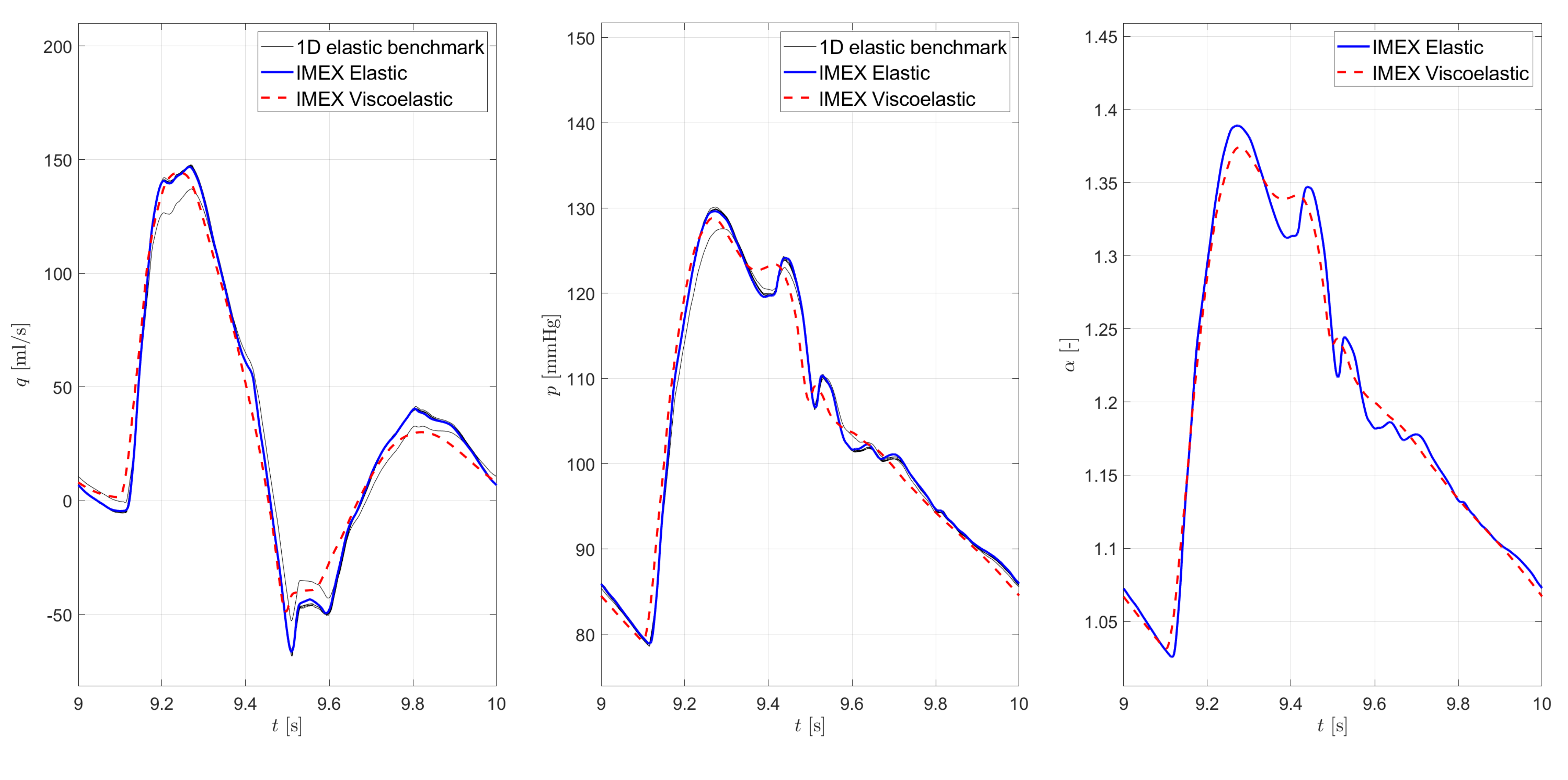}
\vspace*{ - 8mm}
\caption{Abdominal aorta V}
\label{56_AAV}
\end{subfigure}
\begin{subfigure}{1.0\textwidth}
\centering
\includegraphics[width=1\linewidth]{ 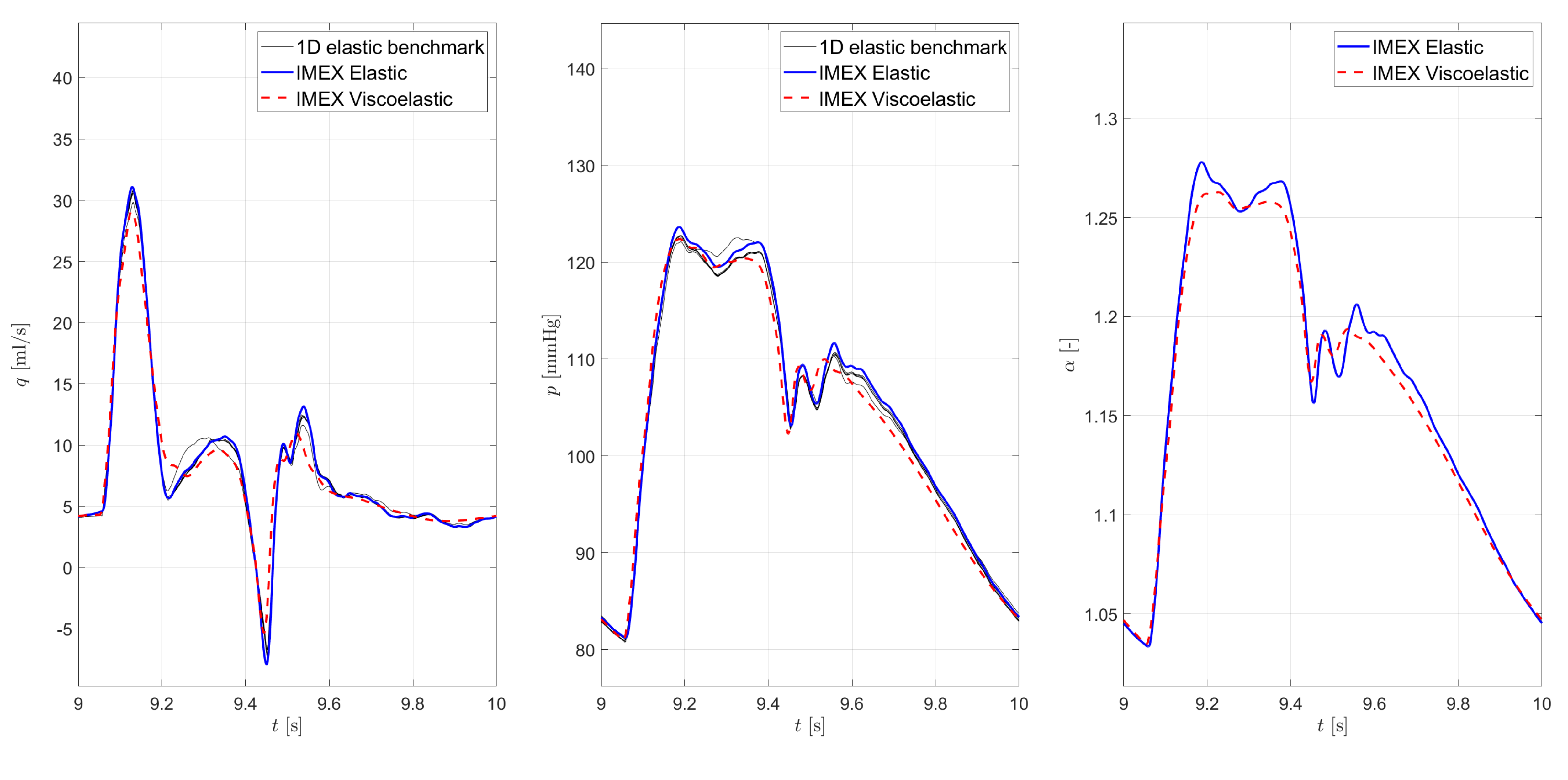}
\vspace*{ - 8mm}
\caption{Right common carotid artery}
\label{56_RCCA}
\end{subfigure}
\caption{Results of the ADAN56 network for 2 selected arteries. Results are obtained with the IMEX RK FV scheme considering both the elastic and the viscoelastic tube law to characterize the mechanical behavior of vessel walls. Comparison with \revised{six different} 1-D elastic benchmark solutions \revised{\cite{Boileau2015}} are presented in terms of flow rate (left column) and pressure (center column), while for the dimensionless cross-sectional area (right column) reference results are not available.}
\label{fig:56AN_2}
\end{figure}
\begin{figure}[t!]
\begin{subfigure}{1.0\textwidth}
\centering
\includegraphics[width=1\linewidth]{ 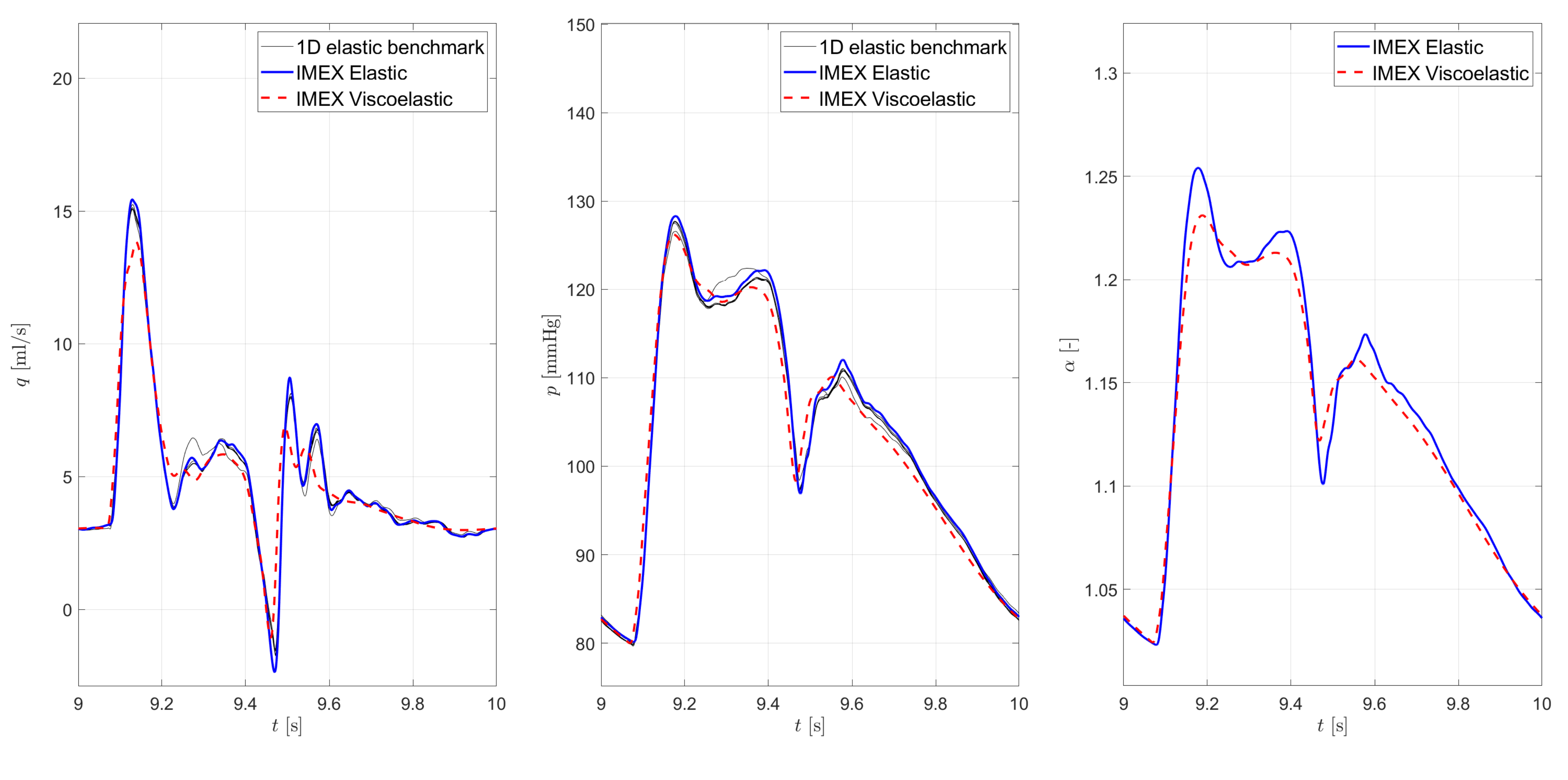}
\vspace*{ - 8mm}
\caption{Right internal carotid artery}
\label{56_RIC}
\end{subfigure}
\begin{subfigure}{1.0\textwidth}
\centering
\includegraphics[width=1\linewidth]{ 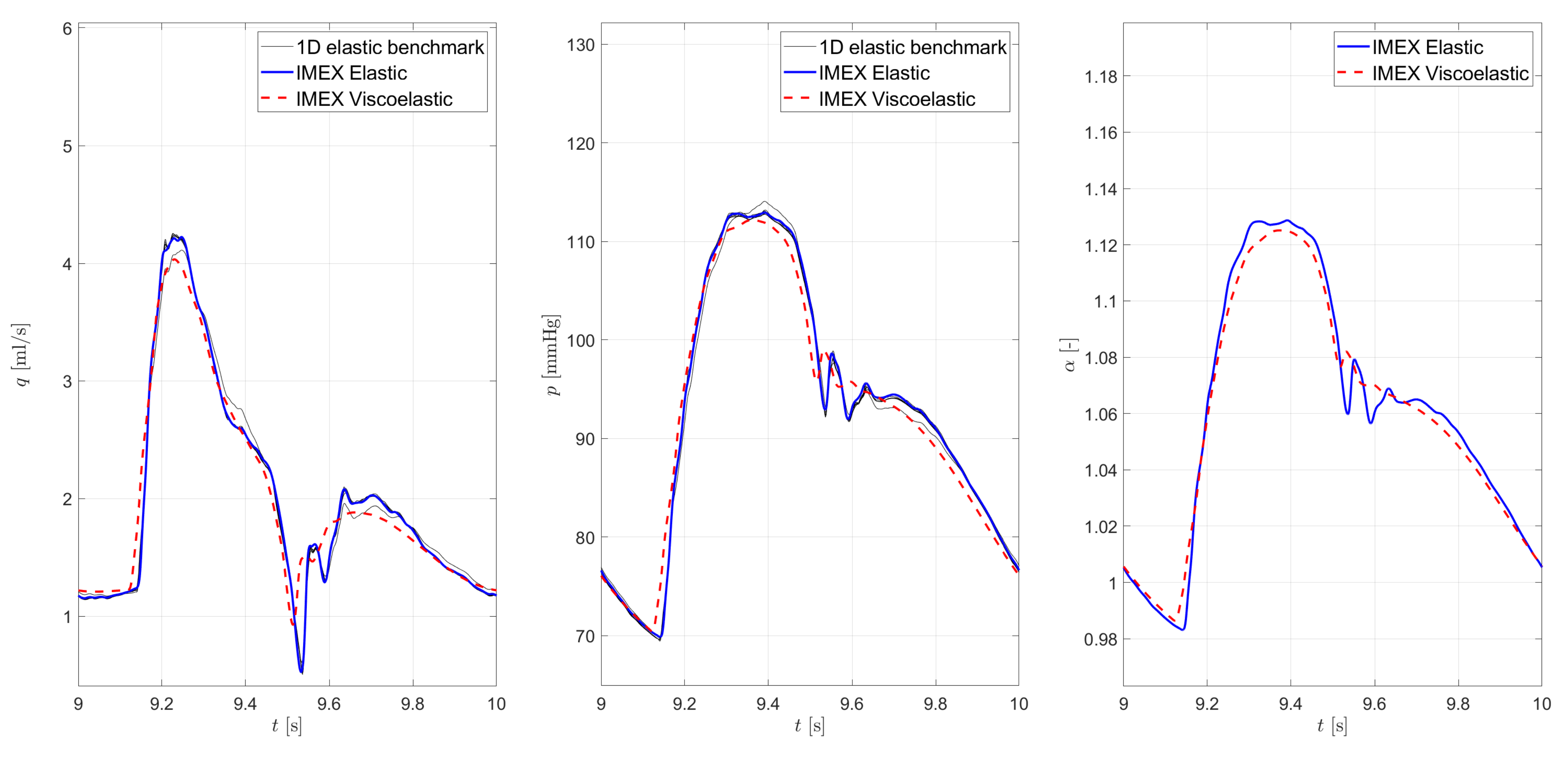}
\vspace*{ - 8mm}
\caption{Right radial artery}
\label{56_RRad}
\end{subfigure}
\caption{Results of the ADAN56 network for 2 selected arteries. Results are obtained with the IMEX RK FV scheme considering both the elastic and the viscoelastic tube law to characterize the mechanical behavior of vessel walls. Comparison with \revised{six different} 1-D elastic benchmark solutions \revised{\cite{Boileau2015}} are presented in terms of flow rate (left column) and pressure (center column), while for the dimensionless cross-sectional area (right column) reference results are not available.}
\label{fig:56AN_4}
\end{figure}
\begin{figure}[t!]
\begin{subfigure}{1.0\textwidth}
\centering
\includegraphics[width=1\linewidth]{ 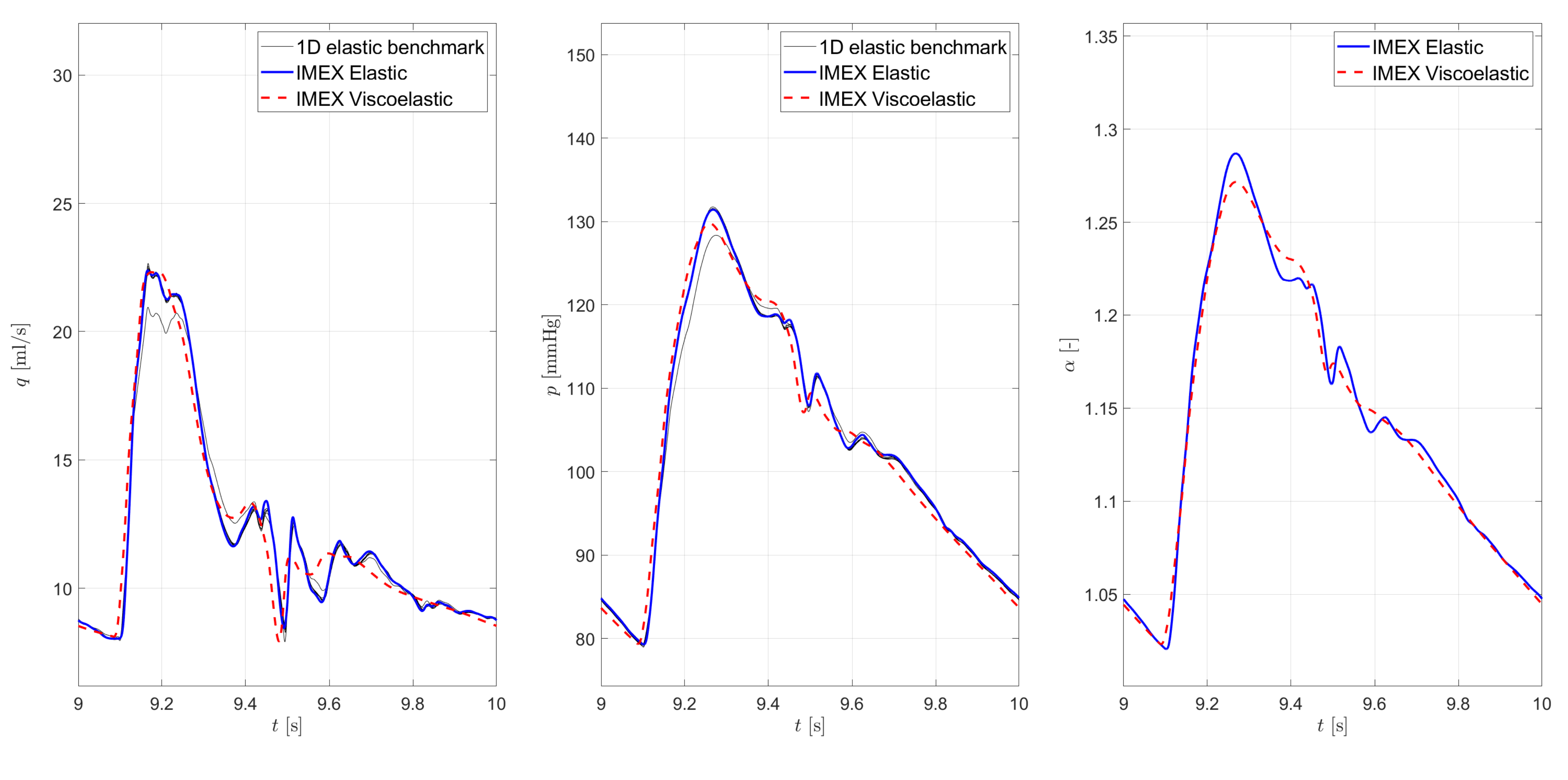}
\vspace*{ - 8mm}
\caption{Right renal artery}
\label{56_RR}
\end{subfigure}
\begin{subfigure}{1.0\textwidth}
\centering
\includegraphics[width=1\linewidth]{ 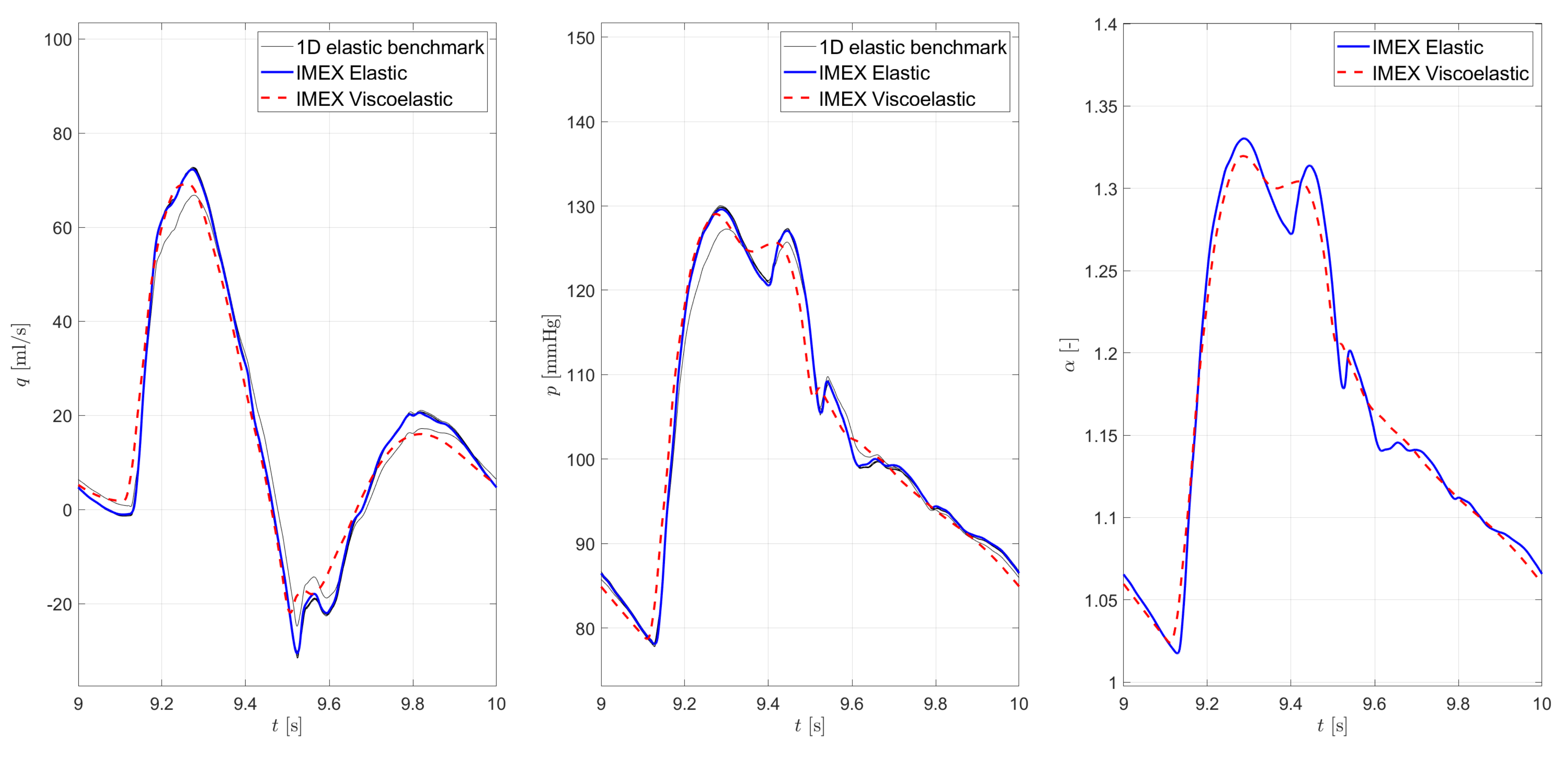}
\vspace*{ - 8mm}
\caption{Right common iliac artery}
\label{56_RCI}
\end{subfigure}
\caption{Results of the ADAN56 network for 2 selected arteries. Results are obtained with the IMEX RK FV scheme considering both the elastic and the viscoelastic tube law to characterize the mechanical behavior of vessel walls. Comparison with \revised{six different} 1-D elastic benchmark solutions \revised{\cite{Boileau2015}} are presented in terms of flow rate (left column) and pressure (center column), while for the dimensionless cross-sectional area (right column) reference results are not available.}
\label{fig:56AN_3}
\end{figure}
\begin{figure}[t!]
\begin{subfigure}{1.0\textwidth}
\centering
\includegraphics[width=1\linewidth]{ 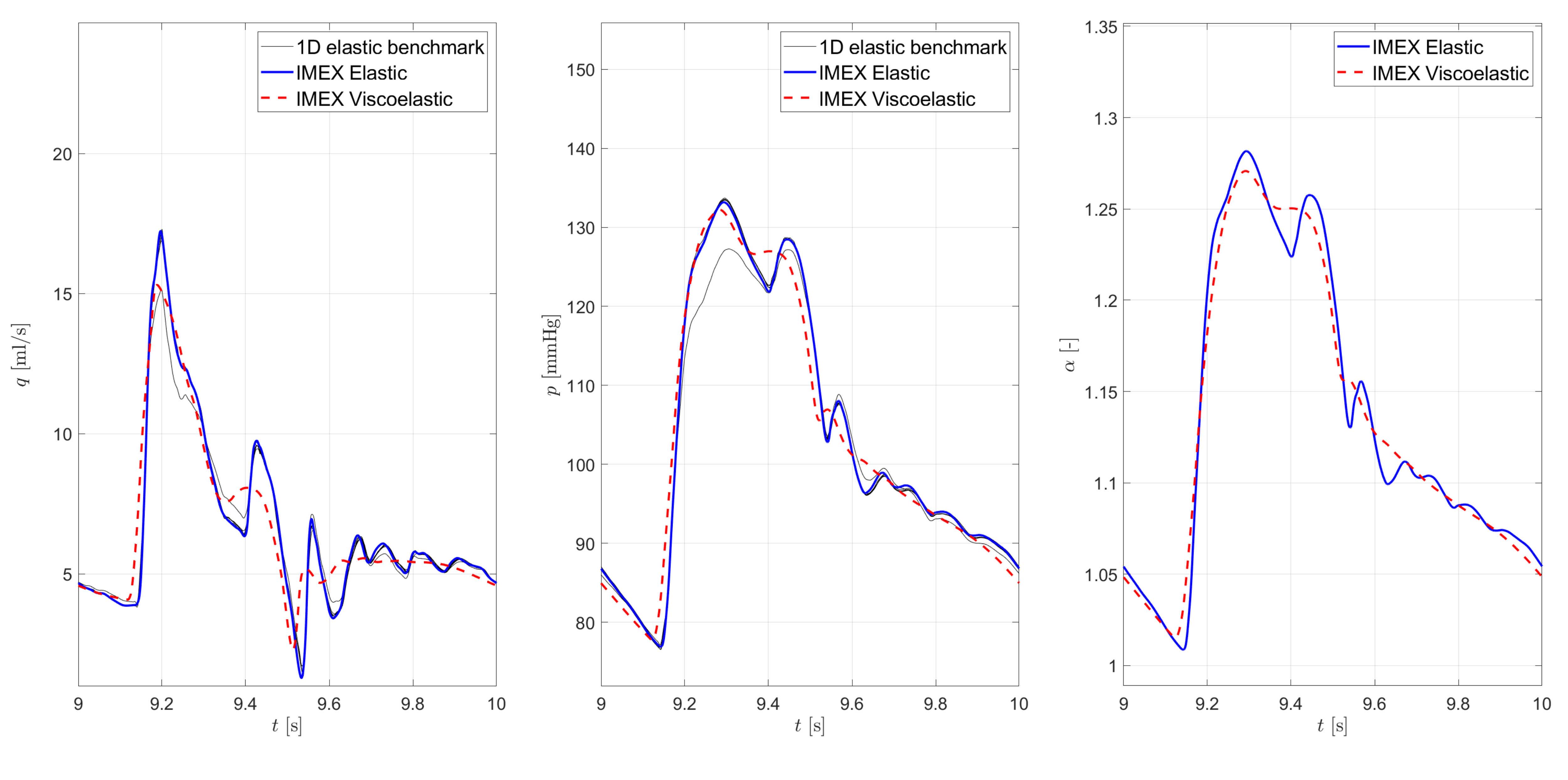}
\vspace*{ - 8mm}
\caption{Right internal iliac artery}
\label{56_RII}
\end{subfigure}
\begin{subfigure}{1.0\textwidth}
\centering
\includegraphics[width=1\linewidth]{ 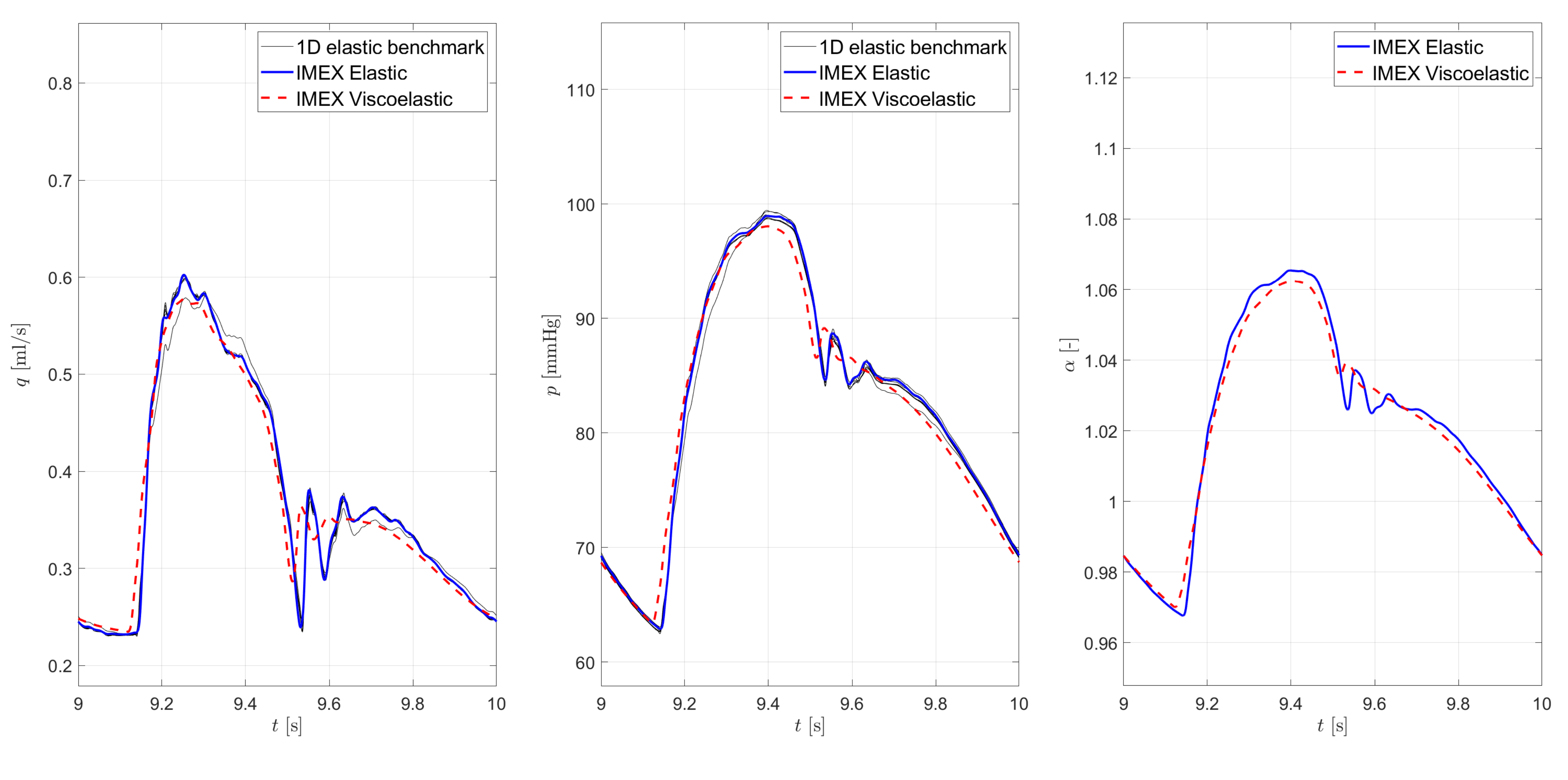}
\vspace*{ - 8mm}
\caption{Right posterior interosseous artery}
\label{56_RPI}
\end{subfigure}
\caption{Results of the ADAN56 network for 2 selected arteries. Results are obtained with the IMEX RK FV scheme considering both the elastic and the viscoelastic tube law to characterize the mechanical behavior of vessel walls. Comparison with \revised{six different} 1-D elastic benchmark solutions \revised{\cite{Boileau2015}} are presented in terms of flow rate (left column) and pressure (center column), while for the dimensionless cross-sectional area (right column) reference results are not available.}
\label{fig:56AN_5}
\end{figure}
\begin{figure}[t!]
\begin{subfigure}{1.0\textwidth}
\centering
\includegraphics[width=1\linewidth]{ 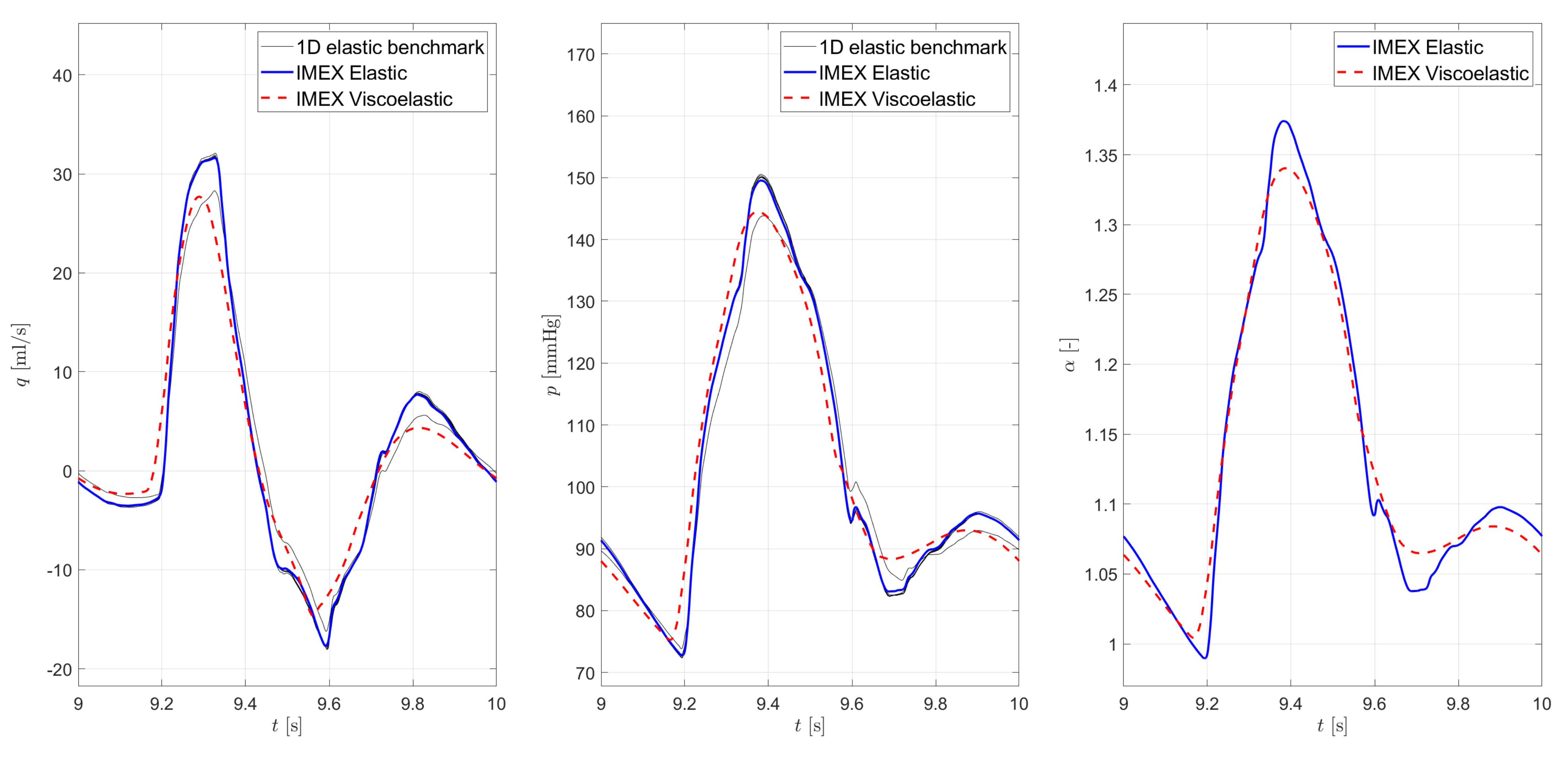}
\vspace*{ - 8mm}
\caption{Right femoral artery II}
\label{56_RFII}
\end{subfigure}
\begin{subfigure}{1.0\textwidth}
\centering
\includegraphics[width=1\linewidth]{ 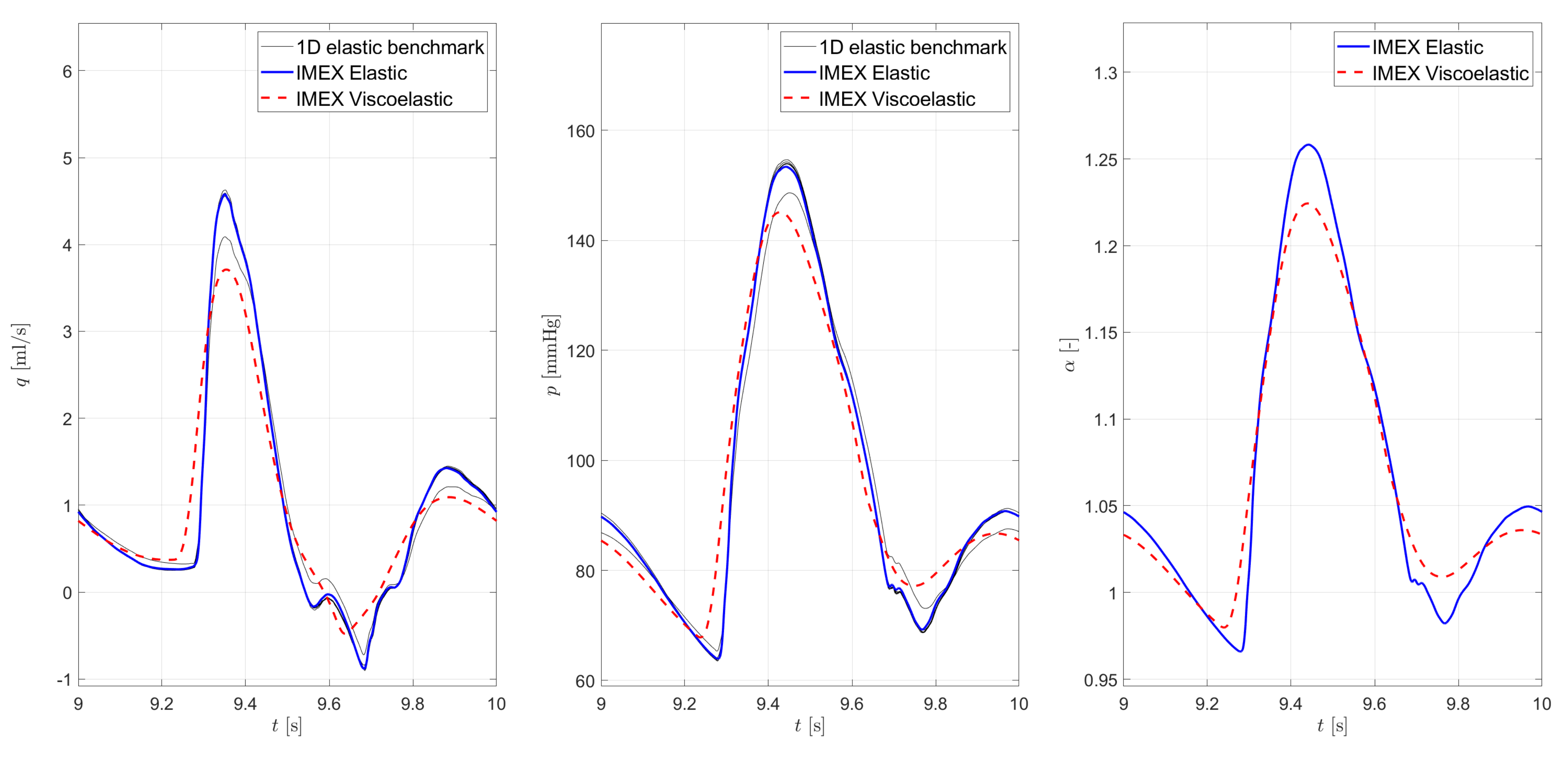}
\vspace*{ - 8mm}
\caption{Right anterior tibial artery}
\label{56_RAT}
\end{subfigure}
\caption{Results of the ADAN56 network for 2 selected arteries. Results are obtained with the IMEX RK FV scheme considering both the elastic and the viscoelastic tube law to characterize the mechanical behavior of vessel walls. Comparison with \revised{six different} 1-D elastic benchmark solutions \revised{\cite{Boileau2015}} are presented in terms of flow rate (left column) and pressure (center column), while for the dimensionless cross-sectional area (right column) reference results are not available.}
\label{fig:56AN_6}
\end{figure}
\subsubsection{ADAN56}
\label{ADAN56}
ADAN56 \revised{represents} a reduced version of 56 vessels of the anatomically detailed arterial network presented in \cite{Blanco2014,Blanco2015}.
The network consists in 77 segments linked by junction \revised{BCs}. At the root of the aortic arch (inlet section), a physiological flow rate, simulating the human cardiac output, is prescribed. Furthermore, there are 31 terminal vessels, in which the outflow and the pressure are simulated via the RCR model. Parameters of the RCR units and the inlet waveform are given by \cite{Boileau2015}, as well as all the topological, geometrical and mechanical characteristics of this test in the elastic configuration. \revised{With the same approach used in the previous test, following \cite{Bertaglia2020a}, for the viscoelastic configuration $E_{\infty}$ is taken from \cite{Boileau2015}, $\eta$ is gathered from \cite{Blanco2014}, and, finally, $E_0$ is computed through Eq. \eqref{eq:Einf2E0}}.
Initial conditions are $A_{IC} = A_0$, $q_{IC} = 0$, $p_{IC} = p_{ext} = 10~\mathrm{kPa}$.
Differently from the previous tests, the velocity profile is assumed to be parabolic in every vessel of the network, hence following the Poiseuille law ($\alpha_c = 4/3$, $\zeta = 2$), as specified by \cite{Boileau2015}.  
\revised{The spatial discretization is performed setting the cell width $\Delta x$ = 2.5 mm, imposing again that at least two cells are used to discretize even the shortest vessels.}
Results computed with the proposed methodology are compared with six reference benchmarks, consisting in numerical results obtained solving the 1-D blood flow model that accounts for a simple elastic tube law for the mechanic characterization of vessel walls with six different numerical methods \cite{Boileau2015}. 

Flow rate and pressure waveforms obtained with the here proposed AP-IMEX RK FV scheme in 12 selected arteries of the network are shown in Figs.~\ref{fig:56AN_1}--\ref{fig:56AN_6}, compared with benchmark solutions. In the same figures, dimensionless cross-sectional area waveforms are presented without a reference solution since not available in literature. To assure the \revised{convergence to a periodic state}, 10 heartbeats are run, but only the last cardiac cycle is here reported. It can be observed that the elastic IMEX simulation is \revised{once more in excellent} agreement with the elastic benchmark, confirming that the proposed methodology is consistent with \revised{the reference elastic configuration}. 

\begin{figure}[p]
\begin{subfigure}{0.5\textwidth}
\centering
\includegraphics[width=1\linewidth]{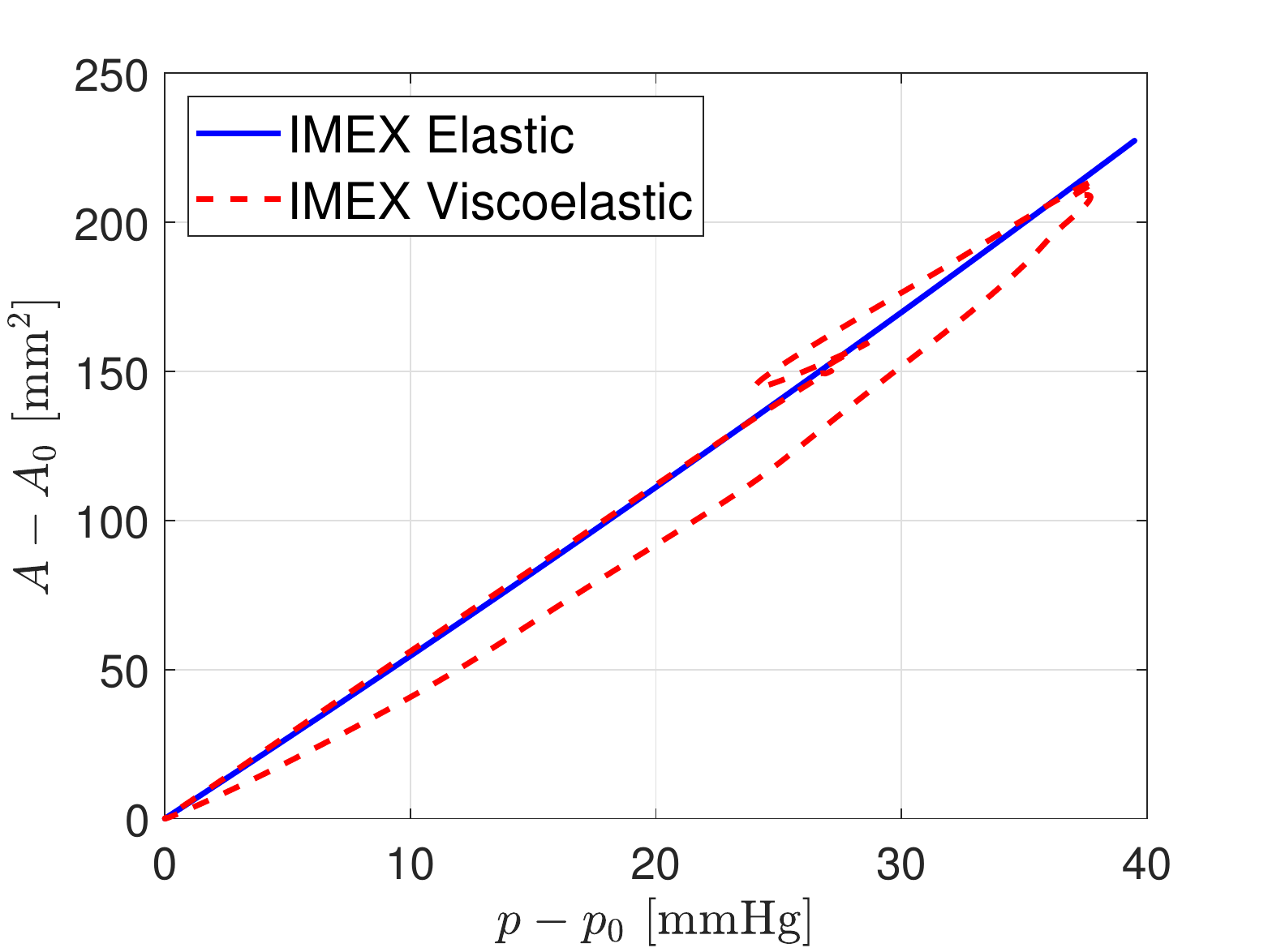}
\vspace*{ - 5mm}
\hspace{2 mm}
\caption{Aortic arch I}
\label{aaI_hl}
\end{subfigure} 
\begin{subfigure}{0.5\textwidth}
\centering
\includegraphics[width=1\linewidth]{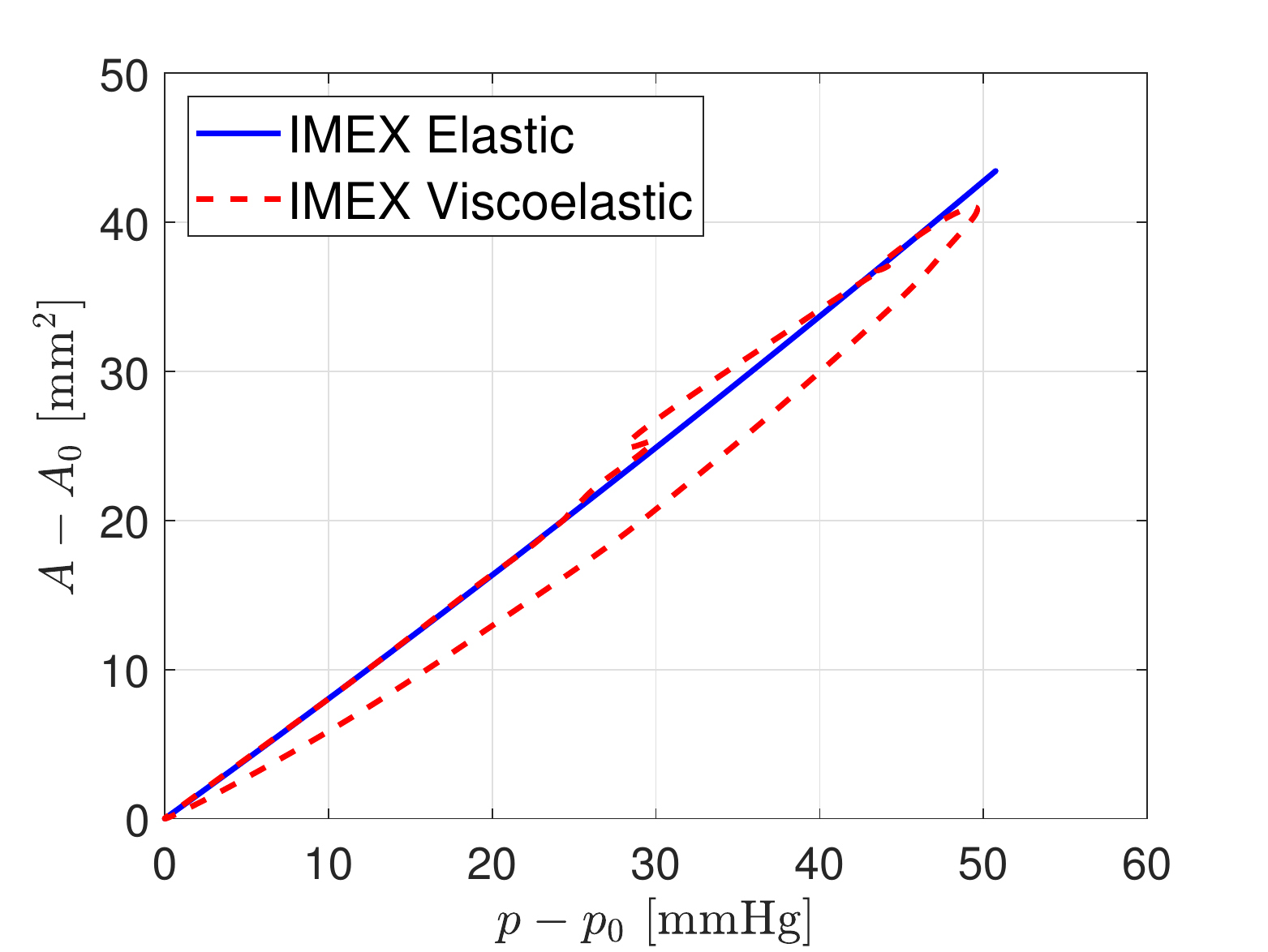}
\vspace*{ - 5mm}
\caption{Abdominal aorta V}
\label{abdominal_hl}
\end{subfigure}
\begin{subfigure}{0.5\textwidth}
\centering
\includegraphics[width=\linewidth]{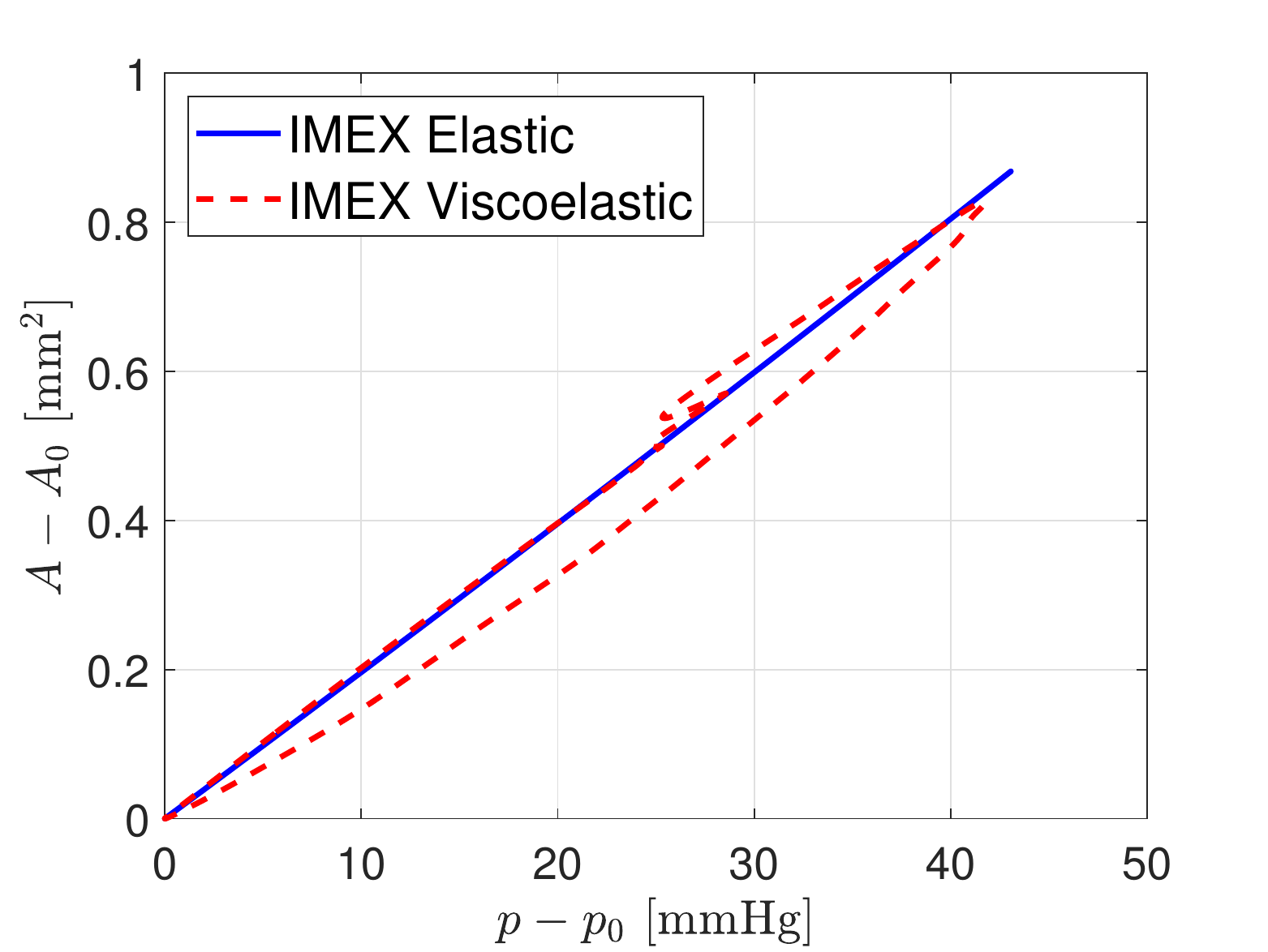}
\vspace*{ - 5mm}
\caption{Right radial artery}
\label{radial_hl}
\end{subfigure} 
\begin{subfigure}{0.5\textwidth}
\centering
\includegraphics[width=\linewidth]{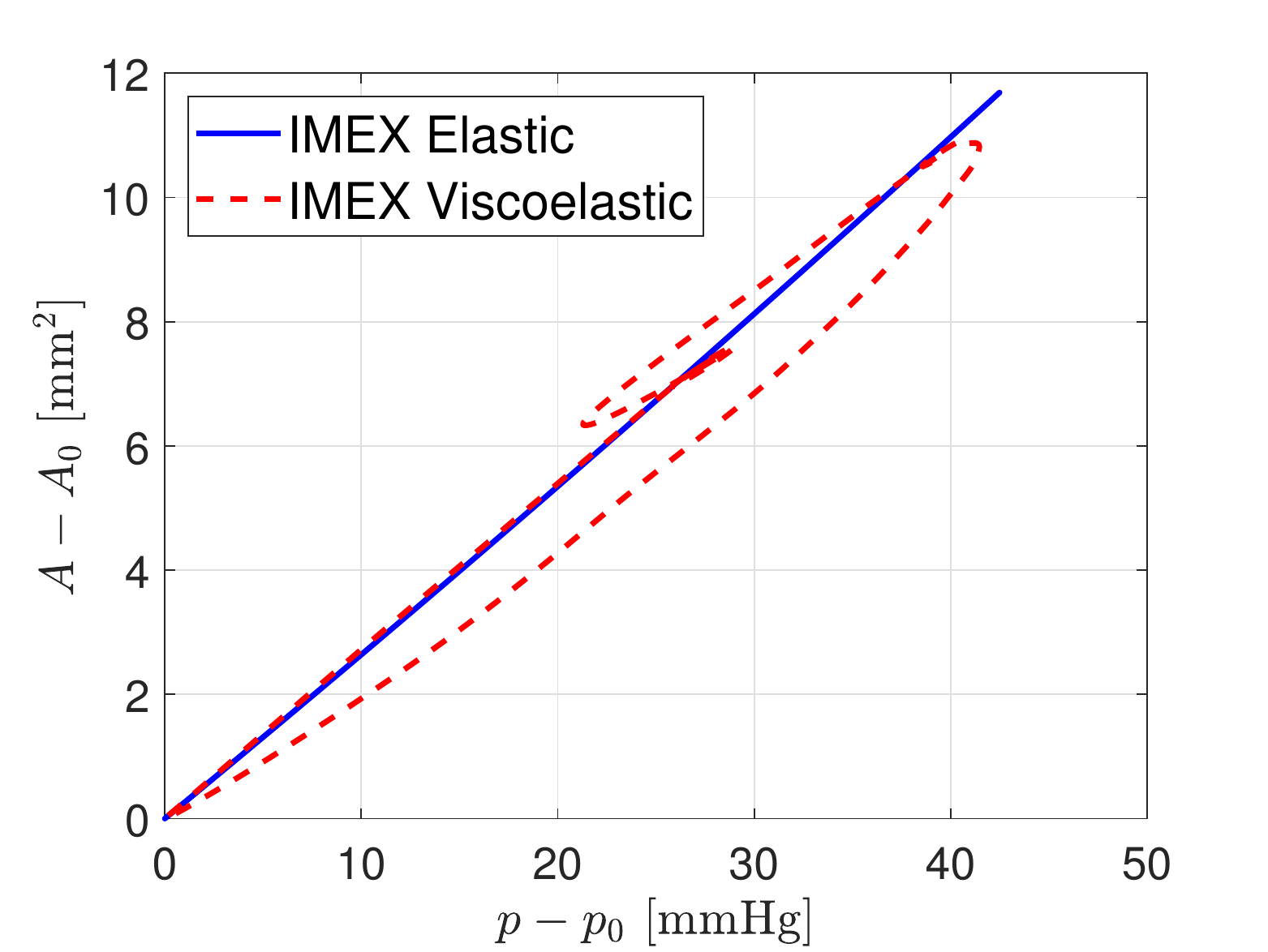}
\vspace*{ - 5mm}
\caption{Right common carotid artery}
\label{cca_hl}
\end{subfigure} 
\begin{subfigure}{0.5\textwidth}
\centering
\includegraphics[width=\linewidth]{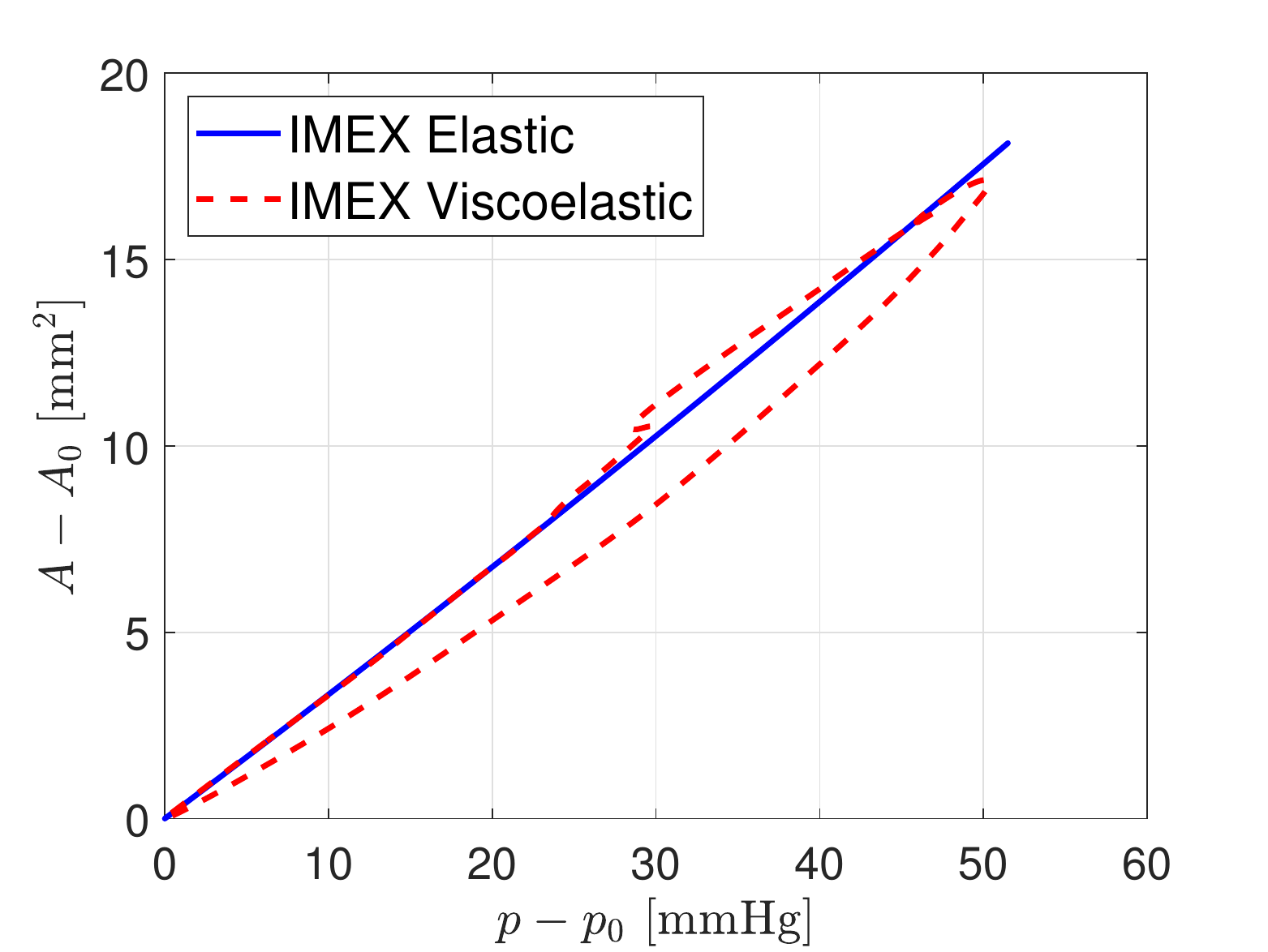}
\vspace*{ - 5mm}
\caption{Right common iliac artery}
\label{iliac_hl}
\end{subfigure} 
\begin{subfigure}{0.5\textwidth}
\centering
\includegraphics[width=\linewidth]{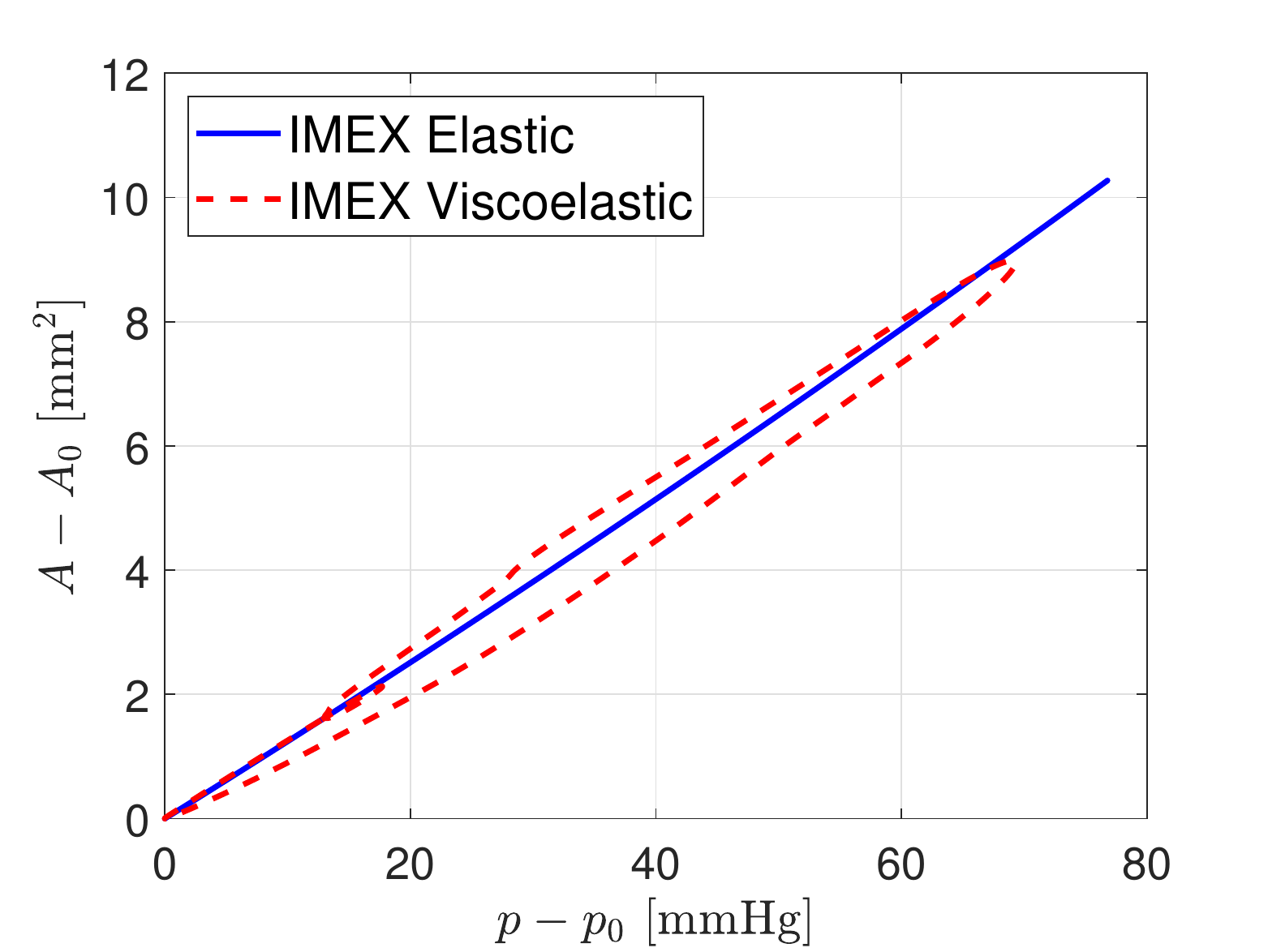}
\vspace*{ - 5mm}
\caption{Right femoral artery II}
\label{femoral_hl}
\end{subfigure} 
\caption{Hysteresis loop obtained for \revised{various arteries} in the ADAN56 network. The elastic lines confirms the absence of dissipative effects, in contrast with the wide viscoelastic curves that shows energy dissipation. In the latter, the curve evolves counterclockwise, showing the characteristic feature connected to the dicrotic notch. $A_0$ and $p_0$ are the equilibrium cross-sectional area and pressure, respectively, whose value coincides with that of the diastolic phase.}
\label{fig:56AN_HL}
\end{figure}

Regarding the viscoelastic simulation, as expected, a pressure peak damping can be observed with respect to elastic results. This network is characterized by a relaxation time $\tau_r \approx \mathcal{O}(10^{-2}) \; \mathrm{s}$, which is two orders of magnitude greater than the AN37 relaxation time. This is the reason why, if compared to AN37 viscoelastic results, ADAN56 results show greater differences between the two mechanical behaviors of vessel walls. The higher relaxation time of this network allows wider hysteresis loops, as shown in Fig.~\ref{fig:56AN_HL} \revised{for various arteries}. Consistently with the viscoelastic model, in  Fig.~\ref{fig:56AN_HL} it can be noticed that, for a fixed area, 
pressure in the systolic tract is higher than the one in the diastolic tract, as the instantaneous Young modulus, which characterizes the systolic phase, is always greater than the asymptotic one (which, instead, characterizes the diastolic phase). \revised{Finally, from the same figure, the ability of the model to capture the variety of viscoelastic mechanical behaviors of each vessel in the network can be appreciated.}

\section{Conclusions}
\label{sect:conclusions}
In the present work, an innovative \revised{numerical} implementation of \revised{a mathematical model representing} cardiovascular networks, with vessel walls characterized by a SLSM viscoelastic behavior, and a consistent treatment of boundary conditions, is presented. The a-FSI blood flow model solved with a second-order accurate AP-IMEX RK FV method, which was previously validated in single--vessel simulations \cite{Bertaglia2021,Bertaglia2020,Bertaglia2020a}, is here adopted to study complete viscoelastic networks, demonstrating that the model is able to accurately simulate even complex circulatory systems. Particular attention is posed on the characterization of junctions, which are geometrical singularities identified by at least two branching vessels. The proposed methodology allows considering the viscoelastic contribution of the model given by the constitutive equation, namely the tube law, even in all boundary sections (external and internal), preserving the prescribed order of accuracy. 
To this aim, to define the Riemann Problem at boundaries, an additional Riemann Invariant is employed, which arise from the study of the eigenstructure of the hyperbolic system of equations that form the a-FSI blood flow model. This additional RI directly derives from the introduction of the viscoelastic constitutive law, in PDE form, inside the system of governing equations. Therefore, the proposed treatment of \revised{BCs} results intrinsically consistent with the mathematical model.

The method has been tested initially on a trivial junction case, named 2--vessels test, to validate it having an exact reference solution and to demonstrate that the second-order of accuracy and the well-balancing of the scheme are preserved even when introducing internal junctions, in both a generic artery and a generic vein.
Furthermore, a simple 3--vessels aortic bifurcation case is analyzed considering the elastic and the viscoelastic tube law. Results (compared with numerical benchmarks in the elastic configuration) in the viscoelastic configuration confirm that the \revised{numerical} implementation of the junction non-linear system is consistent with the viscous contribution produced by the mechanical representation of vessel walls via the SLSM. 
Consequently, two arterial networks are studied. The first one simulates an \textit{in-vitro} network composed of 37 silicon tubes that represent the main human arteries, whereas the second represents a reduced version of an anatomical detailed arterial system, consisting of the 56 largest arteries of the human cardiovascular circulation. Numerical results obtained in the two networks, with the elastic and then the viscoelastic characterization of vessels wall, have been positively tested comparing IMEX results with benchmark solutions (experimental data and numerical elastic benchmark for the AN37 and elastic numerical reference solutions for the ADAN56). 

The viscoelastic \revised{formulation} introduces variations in the system response that are consistent with the mechanical features of the SLSM. Indeed, from the results, the damping effect caused by the viscous contribution is visible \revised{and confirmed by the opening of hysteresis loops}. However, it is worth underlining the strict dependence of the viscoelastic behavior, hence the damping effect, on the viscoelastic parameters defining the SLSM. Concerning especially the definition of the relaxation time of the material, which governs the \textit{stiffness} of the a-FSI blood flow model, \revised{it has been shown that the asymptotic, elastic regime can be approached} either when $\tau_r \to 0$ or $\tau_r \to \infty$. In these cases, viscoelastic features of the material cannot manifest and hysteresis loops coincide with the corresponding elastic straight line in the ($p$--$A$) plane.


Future developments will focus on uncertainty quantification analyses with respect to the geometric and mechanical parameters underlying computational hemodynamics networks and on application works aimed at contributing to the improvement of cardiovascular diagnostics and supporting the study of cardiovascular diseases.
\section*{Aknowledgements}
This work was partially supported by MIUR (Ministero dell’Istruzione, dell’Università e della Ricerca, Italy)
PRIN 2017 for the project “Innovative numerical methods for evolutionary partial differential equations and
applications”, code 2017KKJP4X. GB holds a Research Fellowship from the Italian National Institute of High Mathematics, INdAM (GNCS group) and acknowledges the support by GNCS--INdAM.
AV was also funded by FIR 2020 of the University of Ferrara, while
VC was also funded by MIUR FFABR 2017.
\appendix
\section{Eigenstructure and Riemann Invariants of the a-FSI model}
\label{appendix}
Analyzing the components of system \eqref{a-FSI} in the quasi-linear form \eqref{hypsys2}, we have:
\begin{equation*}
\label{variables_hypsys}
\boldsymbol{Q} = \left( \begin{array}{c} A \\ Au \\ p \\ A_0 \\ E_0 \\ p_{ext} \end{array} \right), \quad
\boldsymbol{A}(\boldsymbol{Q}) =  \left( \begin{array}{cccccc} 
0 & 1 & 0 & 0 & 0 & 0 \\
-u^2 & 2u & \frac{A}{\rho} & 0 & 0 & 0 \\
0 & d & 0 & 0 & 0 & 0 \\
0 & 0 & 0 & 0 & 0 & 0 \\
0 & 0 & 0 & 0 & 0 & 0 \\
0 & 0 & 0 & 0 & 0 & 0 \\ \end{array} \right), \quad
\boldsymbol{S}(\boldsymbol{Q}) = \left( \begin{array}{c} 0 \\ \frac{f}{\rho} \\ S \\ 0 \\ 0 \\ 0 \end{array} \right).
\end{equation*}
It can be demonstrated that system \eqref{a-FSI} is hyperbolic, hence the matrix $\boldsymbol{A}(\boldsymbol{Q})$ is diagonalizable with a diagonal matrix containing all real eigenvalues $\lambda_i$, with $i = 1,\ldots, N$, ($N=6$, number of unknowns of the problem) 
\[ \lambda_1 = u-c \, , \quad \lambda_2 = \lambda_3 = \lambda_4 = \lambda_5 = 0 \, , \quad \lambda_6 = u+c \, , \]
and a complete set of corresponding linearly independent (right) eigenvectors \revised{\cite{Bertaglia2020}}:
\[
\boldsymbol{R}_1 = \gamma_1\left( \begin{array}{c} 1\\u-c\\d\\0\\0\\0 \end{array} \right), \quad
\boldsymbol{R}_2 = \gamma_2\left( \begin{array}{c} 1\\0\\\frac{\rho u^2}{A}\\0\\0\\0 \end{array} \right), \quad
\boldsymbol{R}_3 = \gamma_3\left( \begin{array}{c} 0\\0\\0\\1\\0\\0 \end{array} \right), 
\]
\[
\boldsymbol{R}_4 = \gamma_4\left( \begin{array}{c} 0\\0\\0\\0\\1\\0 \end{array} \right), \quad
\boldsymbol{R}_5 = \gamma_5\left( \begin{array}{c} 0\\0\\0\\0\\0\\1 \end{array} \right), \quad
\boldsymbol{R}_6 = \gamma_6\left( \begin{array}{c} 1\\u+c\\d\\0\\0\\0 \end{array} \right),
\]
where $\gamma_i$ are arbitrary scaling factors.

It is recalled to the reader that to determine the $\lambda_i$--characteristic field of the hyperbolic system it is sufficient to evaluate the quantity $\nabla \lambda_i \cdot \boldsymbol{R}_i$, with $i = 1,\ldots,N$ and see whether it is equal or different from zero \revised{\cite{Toro2013}}. In the former case, the characteristic field is LD and associated to contact discontinuity waves, whereas, in the latter case, the field is genuinely non-linear, hence related to elementary waves, i.e.\ rarefactions and shock waves \cite{Toro2009}. 
The $\lambda_i$--characteristic fields with $i = 2,3,4,5$ are LD, whereas the remaining two fields, on waves $\lambda_1$ and $\lambda_6$, are genuinely non-linear \revised{outside the locus: 
\[\mathcal{G}\left(m,n,\frac{A}{A_0}\right) = m(m+2)\alpha^m - n(n+2)\alpha^n, \]
in the $m$-$n$-$\alpha$ space.
The proof is omitted since it results equivalent to the one presented in \cite{Toro2013}. Indeed, the presence of Eq. \eqref{eq:tubelaw} in the a-FSI system here treated does not give rise to additional contributions in the product $\nabla \lambda_i \cdot \boldsymbol{R}_i$, with $i = 1,6$, being $\lambda_1$ and $\lambda_6$ independent from the third component of state variable $\boldsymbol{Q}$, $q_3 = p$. Thus, for the standard cases in which $m\ge0$ and $n\ge-2$ \cite{Toro2013,Shapiro1977} the $\lambda_1$ and $\lambda_6$ characteristic fields are confirmed to be genuinely non-linear.}

To evaluate the Riemann Invariants associated to each characteristic field, the following $(N-1)$ ordinary differential equations, with $k=1, \ldots, N$ in this work, need to be computed \cite{Toro2009}:
\begin{equation*}
\frac{\mathrm{d}q_1}{r_1^{(i)}} = \dots = \frac{\mathrm{d}q_k}{r_k^{(i)}} = \dots = \frac{\mathrm{d}q_N}{r_N^{(i)}}.
\end{equation*}
These equalities relate the rate of change $\mathrm{d}q_k$ of the \textit{k-th} component $q_k$ of state variable $\boldsymbol{Q}$ to the respective component $r_k^{(i)}$ of the right eigenvector $\boldsymbol{R}_i$ corresponding to the $\lambda_i$--wave family.
Considering only the first three equations of the governing hyperbolic system, for the first LD field, namely for $\lambda_2$, it can be written:
\[ \frac{\mathrm{d}A}{1} = \frac{\mathrm{d}(Au)}{0} = \frac{\mathrm{d}p}{\frac{\rho u^2}{A}},\]
resulting:
\begin{equation*} 
\Gamma_1^{LD} = Au, \qquad
\Gamma_2^{LD} = p + \frac{1}{2} \rho u^2.
\end{equation*}
In this way, it is verified that quantities $\Gamma_1^{LD}$ and $\Gamma_2^{LD}$ are constant across contact discontinuity waves. The LD fields associated to eigenvalues $\lambda_3$, $\lambda_4$ and $\lambda_5$ assure jumps of quantities $A_0$, $E_0$ and $p_{ext}$, consistently with the characterization of the a-FSI system. Concerning the genuinely non-linear fields, for the first eigenvalues, $\lambda_1$, the following relationship holds:
\[ \frac{\mathrm{d}A}{1} = \frac{\mathrm{d}(Au)}{u-c} = \frac{\mathrm{d}p}{d},\]
which, recalling Eq.~\eqref{eq:celerity}, returns:
\begin{equation*} 
\Gamma_1^{(1)} = u + \int \frac{c(A)}{A}\mathrm{dA}, \qquad
\Gamma_2^{(1)} = p - \int d(A)\mathrm{dA}.
\end{equation*}
Finally, for the last eigenvalue, $\lambda_6$, the following relationship holds:
\[ \frac{\mathrm{d}A}{1} = \frac{\mathrm{d}(Au)}{u+c} = \frac{\mathrm{d}p}{d} ,\]
which gives:
\begin{equation*} 
\Gamma_1^{(6)} = u - \int \frac{c(A)}{A}\mathrm{dA}, \qquad
\Gamma_2^{(6)} = p - \int d(A)\mathrm{dA}.
\end{equation*}
Since the second Riemann Invariant results the same in the two genuinely non-linear fields, it is considered just once and labeled as $\Gamma_3$, whereas the first Riemann Invariants associated to $\lambda_1$ and $\lambda_6$ are labeled as $\Gamma_1$ and $\Gamma_2$, respectively.

\section{Viscoelastic SLSM parameters of numerical tests}
\label{appendixB}
\setcounter{table}{0}

Tables \ref{tbl:SLSM_AB_par}, \ref{tbl:SLSM_ADAN56_par} and \ref{tbl:SLSM_AN37_par} report the Standard Linear Solid Model (SLSM) parameters for every bifurcation/network test implemented in this work, namely for the 3-vessels test (aortic bifurcation), the AN37 and the ADAN56. It is reminded to the reader that ADAN56 is characterized by a unique set of viscoelastic parameters for the entire network. Tables are reported here to facilitate the assessment of further analysis.
\vspace{2cm}
\begin{table}[ht!]
\caption{Viscoelastic SLSM parameters for the aortic bifurcation test: instantaneous Young modulus $E_0$, asymptotic Young modulus $E_{\infty}$, ratio $z = E_{\infty} / E_0$, viscosity coefficient $\eta$  and relaxation time $\tau_r$.}
\label{tbl:SLSM_AB_par}
\centering
\begin{tabular}{l c c c c c}
\hline
\rule{0pt}{3.5ex}
Vessel name & 
\makecell{$E_0$  \\{}  [MPa]} & 
\makecell{$E_{\infty}$ \\{} [MPa]} & 
\makecell{$z$ \\{} [-]} & 
\makecell{$\eta$ \\{} [\revised{k}Pa$\cdot$s]} &
\makecell{$\tau_{r}$ \\{}  [ms]} \\
\hline
\rule{0pt}{2.5ex}
Abdominal aorta & 1.4293 & 0.6667 & 0.466 & 58.664 & 21.89 \\
\rule{0pt}{3ex}
Iliac artery & 3.8986 & 0.9333 & 0.239 & 109.97 & 21.45  \\
\hline
\end{tabular}
\end{table}
\vspace{2cm}
\begin{table}[ht!]
\caption{Viscoelastic SLSM parameters for the ADAN56 network: instantaneous Young modulus $E_0$, asymptotic Young modulus $E_{\infty}$, ratio $z = E_{\infty} / E_0$, viscosity coefficient $\eta$  and relaxation time $\tau_r$.}
\label{tbl:SLSM_ADAN56_par}
\centering
\begin{tabular}{l c c c c c}
\hline
\rule{0pt}{3.5ex}
Vessel name & 
\makecell{$E_0$  \\{}  [MPa]} & 
\makecell{$E_{\infty}$ \\{} [MPa]} & 
\makecell{$z$ \\{} [-]} & 
\makecell{$\eta$ \\{} [\revised{k}Pa$\cdot$s]} &
\makecell{$\tau_{r}$ \\{}  [ms]} \\
\hline
\rule{0pt}{2.5ex}
$\forall$ vessels & 0.4431 & 0.3000 & 0.677 & 30.00 & 21.87 \\
\hline
\end{tabular}
\end{table}
\clearpage
\begin{longtable}[t!]{l c c c c c}
\caption{Viscoelastic SLSM parameters for the AN37 network: instantaneous Young modulus $E_0$, asymptotic Young modulus $E_{\infty}$, ratio $z = E_{\infty} / E_0$, viscosity coefficient $\eta$  and relaxation time $\tau_r$. In vessel names, R. stands for the right artery and L. stands for the left artery.}\\
\label{tbl:SLSM_AN37_par}
\endfirsthead
\endhead
\hline
\rule{0pt}{3.5ex}
Vessel name & 
\makecell{$E_0$  \\{}  [MPa]} & 
\makecell{$E_{\infty}$ \\{} [MPa]} & 
\makecell{$z$ \\{} [-]} & 
\makecell{$\eta$ \\{} [\revised{k}Pa$\cdot$s]} &
\makecell{$\tau_{r}$ \\{}  [ms]} \\
\hline
Ascending aorta \rule{0pt}{2.5ex} & 1.6860 & 1.6 & 0.949 & 4.0268 &  0.12\\
Innominate & 1.6846 & 1.6  & 0.950 & 3.9654 & 0.12 \\
R. Carotid  & 1.6869 & 1.6 & 0.948 & 4.0702 & 0.12 \\
R. Subclavian I & 1.6857 & 1.6 & 0.949 & 4.0120 & 0.12 \\
R. Subclavian II &1.6843  & 1.6 & 0.950 & 3.9493 & 0.12 \\
R. Radial & 1.6851 & 1.6 & 0.949 & 3.9869 & 0.12 \\
R. Ulnar & 1.6872 & 1.6 & 0.948 & 4.0837 & 0.13 \\
Aortic arch I & 1.6853 & 1.6 & 0.949 & 3.9945 & 0.12 \\
L. Carotid & 1.6863 & 1.6 & 0.949 & 4.0403 & 0.12 \\
Aortic arch II &1.6846  & 1.6 & 0.950 & 3.9631 & 0.12 \\
L. Subclavian I &1.6865  & 1.6 & 0.949 & 4.0519 & 0.12 \\
L. Subclavian II & 1.6879 & 1.6 & 0.948 & 4.1153  & 0.13 \\
L. Radial & 1.6861 & 1.6 &0.949  & 4.0299 & 0.12 \\
L. Ulnar & 1.6874 & 1.6 & 0.948 & 4.0904 & 0.13 \\
Thoracic aorta I & 1.6857 & 1.6 & 0.949 & 4.0149 & 0.12 \\
Intercostals & 1.6857 & 1.6 & 0.949 &  4.0120 & 0.12 \\
Thoracic aorta II & 1.6843 & 1.6 &0.950  & 3.9493 & 0.12 \\
Celiac I & 1.6855 & 1.6 & 0.949 & 4.0057  & 0.12 \\
Celiac II & 1.6854 & 1.6 &  0.949& 3.9990  & 0.12 \\
Splenic & 1.6833 & 1.6 & 0.950 & 3.9059  & 0.11 \\
Gastric & 1.6877 & 1.6 & 0.948 & 4.1032  & 0.13 \\
Hepatic & 1.6849 & 1.6 & 0.950 & 3.9762  & 0.12 \\
Abdominal aorta I & 1.6847 & 1.6 & 0.950 & 3.9664 & 0.12 \\
L. Renal & 1.6836 & 1.6 & 0.950 & 3.9196  & 0.12\\
Abdominal aorta II & 1.6861 & 1.6 & 0.949 & 4.0299  & 0.12 \\
R. Renal & 1.6858 & 1.6 &  0.949& 4.0199 &  0.12 \\
Abdominal aorta III & 1.6851 & 1.6 &0.949  & 3.9869 & 0.12 \\
R.  Iliac femoral I & 1.6872 & 1.6 & 0.948 & 4.0837  & 0.13 \\
R. Iliac femoral II & 1.6884 & 1.6 & 0.948 & 4.1374  & 0.13 \\
Iliac femoral R. III & 1.6831 & 1.6 & 0.951 & 3.8929  & 0.11 \\
L.  Iliac femoral I & 1.6868 & 1.6 & 0.949 &  4.0622 & 0.12 \\
L. Iliac femoral II & 1.6874 & 1.6 & 0.948 & 4.0904 & 0.13 \\
L. Iliac femoral III & 1.6871 & 1.6 & 0.948 & 4.0795  & 0.12  \\
R. Anterior tibial & 1.6835 &1.6 & 0.950 & 3.9117  & 0.12 \\
R. Posterior tibial &1.6864  &1.6  & 0.949 &  4.0434 & 0.12 \\
L. Posterior tibial & 1.6832 & 1.6 & 0.951 & 3.8980  & 0.11 \\
L. Anterior tibial & 1.6843 & 1.6 & 0.950 &  3.9493 & 0.12 \\
\hline
\end{longtable}


\end{document}